\documentclass[graybox]{svmult}

\usepackage{mathptmx}       
\usepackage{helvet}         
\usepackage{courier}        
\usepackage{type1cm}        
\usepackage{makeidx}         
\usepackage{graphicx}        
\usepackage{multicol}        
\usepackage[bottom]{footmisc}

\makeindex             

\def\reff@jnl#1{{\rm#1\/}}
\def\aj{\reff@jnl{AJ}}                  
\def\araa{\reff@jnl{ARA\&A}}            
\def\apj{\reff@jnl{ApJ}}                
\def\apjl{\reff@jnl{ApJ}}               
\def\apjs{\reff@jnl{ApJS}}              
\def\apss{\reff@jnl{Ap\&SS}}            
\def\aap{\reff@jnl{A\&A}}               
\def\aapr{\reff@jnl{A\&A~Rev.}}         
\def\aaps{\reff@jnl{A\&AS}}             
\def\mnras{\reff@jnl{MNRAS}}            
\def\prd{\reff@jnl{Phys.Rev.D}}         
\def\prl{\reff@jnl{Phys.Rev.Lett}}      
\def\pasp{\reff@jnl{PASP}}              
\def\pasj{\reff@jnl{PASJ}}              
\def\nat{\reff@jnl{Nature}}             
\def\physrep{\reff@jnl{Phys.Rep.}}      
\def\jcap{\reff@jnl{JCAP}}              
\def\procspie{\reff@jnl{Proc.~SPIE}}  

\usepackage[square,sort,comma]{natbib}

\usepackage{graphicx}	
\usepackage{amsmath}	
\usepackage{amssymb}	
\usepackage{xspace}

\newcommand{\bd}{\begin{displaymath}}
\newcommand{\ed}{\end{displaymath}}
\newcommand{\be}{\begin{equation}}
\newcommand{\ee}{\end{equation}}
\newcommand{\beaa}{\begin{eqnarray*}}
\newcommand{\eeaa}{\end{eqnarray*}}
\newcommand{\bea}{\begin{eqnarray}}
\newcommand{\eea}{\end{eqnarray}}

\begin{document}

\title*{Cosmological distance indicators}
\author{Sherry H.~Suyu, Tzu-Ching Chang, Fr{\'e}d{\'e}ric Courbin and Teppei Okumura}
\authorrunning{Suyu, Chang, Courbin \& Okumura}
\institute{Sherry H.~Suyu \at Max-Planck-Institut f{\"u}r Astrophysik, Karl-Schwarzschild-Str.~1, 85748 Garching, Germany;\\
Institute of Astronomy and Astrophysics, Academia Sinica, 11F of ASMAB, No.1, Section 4, Roosevelt Road, Taipei 10617, Taiwan;\\
Physik-Department, Technische Universit\"at M\"unchen, James-Franck-Stra\ss{}e~1, 85748 Garching, Germany\\ 
\email{suyu@mpa-garching.mpg.de}
\and Tzu-Ching Chang \at Jet Propulsion Laboratory, 4800 Oak Grove Dr,
MS~169-237, Pasadena, CA 91109, USA; \\
Institute of Astronomy and Astrophysics, Academia Sinica, 11F of ASMAB, No.1, Section 4, Roosevelt Road, Taipei 10617, Taiwan;\\
\email{tzu-ching.chang@jpl.nasa.gov}
\and Fr{\'e}d{\'e}ric Courbin \at Institute of Physics, Laboratoire d'Astrophysique, Ecole Polytechnique
F{\'e}d{\'e}rale de Lausanne (EPFL), Observatoire de Sauverny, CH-1290
Versoix, Switzerland\\ 
\email{frederic.courbin@epfl.ch}
\and Teppei Okumura \at Institute of Astronomy and Astrophysics, Academia Sinica, 11F of ASMAB, No.1, Section 4, Roosevelt Road, Taipei 10617, Taiwan;\\
Kavli Institute for the Physics and Mathematics of the Universe (WPI), UTIAS, The University of Tokyo, Kashiwa, Chiba 277-8583, Japan\\
\email{tokumura@asiaa.sinica.edu.tw}
}

%
%
\maketitle

\abstract*{ We review three distance measurement techniques beyond the
  local universe: (1) gravitational lens time delays, (2) baryon
  acoustic oscillation (BAO), and (3) HI intensity mapping.  We
  describe the principles and theory behind each method, the
  ingredients needed for measuring such distances, the current
  observational results, and future prospects.  Time-delays from
  strongly lensed quasars currently provide constraints on $H_0$ with
  $<4\%$ uncertainty, and with $1\%$ within reach from ongoing surveys
  and efforts.  Recent exciting discoveries of strongly lensed
  supernovae hold great promise for time-delay cosmography. BAO
  features have been detected in redshift surveys up to $z\lesssim0.8$
  with galaxies and $z\sim2$ with Ly-$\alpha$ forest, providing
  precise distance measurements and $H_0$ with $<2\%$ uncertainty in
  flat $\Lambda$CDM. Future BAO surveys will probe the distance scale
  with percent-level precision. HI intensity mapping has great
  potential to map BAO distances at $z\sim0.8$ and beyond with
  precisions of a few percent. The next years ahead will be exciting
  as various cosmological probes reach $1\%$ uncertainty in determining $H_0$, to assess the current tension in $H_0$ measurements 
  that could indicate new physics.  }




We review three distance measurement techniques beyond the local
universe: (1) gravitational lens time delays, (2) baryon acoustic
oscillation (BAO), and (3) HI intensity mapping.  We describe the
principles and theory behind each method, the ingredients needed for
measuring such distances, the current observational results, and
future prospects.  Time-delays from strongly lensed quasars currently
provide constraints on $H_0$ with $<4\%$ uncertainty, and with $1\%$
within reach from ongoing surveys and efforts.  Recent exciting
discoveries of strongly lensed supernovae hold great promise for
time-delay cosmography. BAO features have been detected in redshift
surveys up to $z\lesssim0.8$ with galaxies and $z\sim2$ with
Ly-$\alpha$ forest, providing precise distance measurements and $H_0$
with $<2\%$ uncertainty in flat $\Lambda$CDM. Future BAO surveys will
probe the distance scale with percent-level precision. HI intensity
mapping has great potential to map BAO distances at $z\sim0.8$ and
beyond with precisions of a few percent. The next years ahead will be
exciting as various cosmological probes reach $1\%$ uncertainty in
determining $H_0$, to assess the current tension in $H_0$ measurements
that could indicate new physics.

\section{Gravitational Lens Time Delays}
\label{sec:gl}

\def\Imvec{\boldsymbol{\theta}}
\def\Srvec{\boldsymbol{\beta}}
\def\tdist{D_{\rm \Delta t}}
\def\zd{z_{\rm d}}
\def\zs{z_{\rm s}}
\def\Dd{D_{\rm d}}
\def\Ds{D_{\rm s}}
\def\Dds{D_{\rm ds}}
\def\vdisp{\sigma_{\rm v}}
\def\hst{{\it HST}}
\def\kext{\kappa_{\rm ext}}

\subsection{Principles of gravitational lens time delays}
\label{sec:gl:principles}

Strong gravitational lensing occurs when a foreground mass distribution is located along the line of sight to a background source such that multiple images of the background source appear around the foreground lens.  In cases where the background source intensity varies, such as an active galactic nucleus (AGN) or a supernova (SN), the variability pattern manifests in each of the multiple images and is delayed in time due to the different light paths of the different images.  The time delay of image $i$, relative to the case of no lensing, is 
\be
\label{eq:delays_Dt}
t(\Imvec_i;\Srvec) = \frac{\tdist}{c} \phi(\Imvec_i;\Srvec), 
\ee
up to an additive constant\footnote{The Fermat potential, being a potential, is defined only up to an additive constant that has no physical consequence.  Furthermore, a ``mass-sheet transformation'' (explained later in Section \ref{sec:gl:adv:modeling}) can also add a term that is independent of $\Imvec_i$ to the Fermat potential.}, where $\Imvec_i$ is the position of the lensed image $i$, $\Srvec$ is the position of the source, $\tdist$ is the so-called ``time-delay distance'', $c$ is the speed of light, and $\phi$ is the ``Fermat potential'' related to the lens mass distribution.  The time-delay distance for a lens at redshift $\zd$ and a source at redshift $\zs$ is
\be
\label{eq:tdist}
\tdist = (1+\zd)\frac{\Dd \Ds}{\Dds},
\ee
where $\Dd$ is the angular diameter distance to the lens, $\Ds$ is the angular diameter distance to the source, and $\Dds$ is the angular diameter distance between the lens and the source.  In the $\Lambda$CDM cosmology with density parameters $\Omega_{\rm m}$ for matter, $\Omega_{\rm k}$ for spatial curvature, and $\Omega_{\Lambda}$ for dark energy described by the cosmological constant $\Lambda$, the angular diameter distance between two redshifts $z_1$ and $z_2$ is
\be
\label{eq:DA}
D(z_1,z_2) = \frac{1}{1+z_2}f_{K}[\chi(z_1,z_2)]
\ee
where
\be
\label{eq:chi}
\chi(z_1,z_2) = \frac{c}{H_0}\int_{z_1}^{z_2} {\rm d}z' \left(\Omega_{\rm m}(1+z')^3 + \Omega_{\rm k} (1+z')^2 + \Omega_{\Lambda} \right)^{-1/2},
\ee
and 
\be
\label{eq:fk}
f_{K}(\chi) = \left\{ \begin{array}{ll}
 K^{-1/2} \sin\left(K^{1/2}\chi\right) & \textrm{for $K>0$}\\
 \chi & \textrm{for $K=0$} \\
 (-K)^{-1/2}\sinh\left[(-K)^{1/2}\chi \right] & \textrm{for $K<0$}
  \end{array} \right. , 
\ee
$K=-\Omega_{\rm k}H_0^2/c^2$ is the spatial curvature, and $H_0$ is the Hubble constant.

By monitoring the variability of the multiple images, we can measure the time delay between the two images $i$ and $j$:
\be
\label{eq:tdiff}
\Delta t_{ij} = t(\Imvec_i;\Srvec)- t(\Imvec_j;\Srvec) = \frac{\tdist}{c}\Delta \phi_{ij}.
\ee
The Fermat potential $\phi$ can be determined by modeling the lens mass distribution using observations of the lens system such as the observed lensed image positions, shapes and fluxes.
Therefore, with $\Delta t$ measured and $\Delta \phi$ determined, we can use equation (\ref{eq:tdiff}) to infer the value of $\tdist$, which is inversely proportional to $H_0$ ($\tdist \propto H_0^{-1}$) through equations (\ref{eq:tdist}) and (\ref{eq:DA}).  Being a combination of three angular diameter distances, $\tdist$ is mainly sensitive to the Hubble constant $H_0$ and weakly depends on other cosmological parameters \citep[e.g.,][]{Refsdal64, SchneiderEtal06, SuyuEtal10}.   

One can further measure $\Dd$ from a lens system by measuring the velocity dispersion of the foreground lens, $\vdisp$, and combining it with the time delays \citep{ParaficzHjorth09, JeeEtal15}.  The measurement of $\Dd$ provides additional constraints on cosmological models \citep{JeeEtal16}.

In order to measure $\tdist$ and $\Dd$ from a time-delay lens system for cosmography, we need the following 
\begin{enumerate}
\item spectroscopic redshifts of the lens $\zd$ and source $\zs$
\item time delays between the multiple images
\item lens mass model to determine the Fermat potential
\item lens velocity dispersion, which is not only required for $\Dd$ inference, but also provides additional constraints in breaking lens mass model degeneracies
\item lens environment studies to break lens model degeneracies, such as the mass-sheet degeneracy 
\end{enumerate}
In the next sections, we describe the history behind this approach, and detail the advances in recent years in acquiring these ingredients before presenting the latest cosmographic inferences from this approach.

\subsection{A brief history}
\label{sec:gl:history}

In his original paper \citet{Refsdal64} proposed to use gravitationally lensed supernovae to measure the time delays: the light curves associated to each lensed image of a supernova are expected to be seen shifted in time by a value that depends on the potential well of the lensing object and on cosmology. However, due to the shallow limiting magnitude of the telescopes available at the time and due to their restricted field of view, discovering faint and distant supernovae right behind a galaxy or a galaxy cluster was completely out of reach. But Refsdal's idea came right when quasars were discovered \citep{Schmidt1963, Hazard1963}. These bright, distant and photometrically variable point sources were coming timely, offering a new opportunity to implement the time-delay method: light curves of lensed quasars are constantly displaying new features that can be used to measure the delay. 

With the increasing discovery rate of quasars, the first cases of multiply imaged ones also started to grow. The first lensed quasar, Q~0951+567, was found by \citet{Walsh1979}, displaying two lensed images. This was followed by the quadruple PG~1115+080 \citep{Young1981}, and a few years later by the discovery of the ``Einstein Cross'' \citep{Huchra1985} and of the ``cloverleaf'' \citep{Magain1988}. The first time-delay measurement became available only in the late 80s with the optical monitoring of Q~0951+567 by \citet{Vanderriest1989} and the radio monitoring of the same object by \citet{Lehar1992}. Unfortunately, given the two data sets and methods of analysis to measure the delay, the radio and optical values of the time delay remained in disagreement until new optical data came \citep[e.g.][]{Kundic1997, Oscoz1997}, allowing to confirm and improve the optical delay of \citet{Vanderriest1989}. Further improvement was possible with the ``round-the-clock'' monitoring of \citet{Colley2003}, leading to a time-delay determination to a fraction of a day.

Because of the time and effort it took to solve the ``Q0957 controversy'', astrophysicists quickly limited their interest in the time-delay method as a cosmological probe. But at least two sets of impressive monitoring data revived the field. The first one is the optical monitoring of the quadruple quasar PG~1115+080 by \citet{Schechter1997}, providing time delays to 14\% and the second is the radio monitoring, with the VLA, of the quadruple radio source B1608+656 \citep{Fassnacht1999}, reaching similar accuracy. As the uncertainty on the time delay propagates linearly in the error budget on $H_0^{-1}$ this is still not sufficient for precision cosmology to a few percents.

In a large part thanks to the results obtained for PG~1115+080 and
B1608+656 several monitoring campaigns were put in place by
independent teams in the late 90s and early 2000. Because lensed
quasars were more often discovered in the optical and because their
variability is faster at these wavelengths due to the smaller source size than in the radio, these new monitoring campaigns took place in the optical. The teams involved used 1m class telescopes to measure delays to typical accuracies of 10\% or slightly better, i.e. a 30\% improvement over previous measurements but still too large for cosmological purposes.

Some of the most impressive results were obtained in the years 2000
with the 2.6m Nordic Optical Telescope (NOT) for FBQ~0951+2635
\citep{Jakobsson2005}, SBS~1520+530 \citep{Burud2002}, RX~J0911+0551
\citep{Hjorth2002}, B1600+434 \citep{Burud2000}, at ESO with the 1.54m
Danish telescope for HE~2149$-$2745 \citep{Burud2002b} and at Wise
observatory with the 1m telescope for HE~1104$-$1805
\citep{Ofek2003}. With these new observations and studies it was shown
that ``mass production" of time delays was possible and not restricted
to a few lenses for which the observational situation was
particularly favorable. However, the temporal sampling adopted for
the observations and the limited signal-to-noise per observing epoch
was still limiting the accuracy on the time-delay measurement to 10\%
in most cases hence limiting $H_0^{-1}$ measurements with indivisual lenses to this precision.  

Fifty years after \citet{Refsdal64}'s foresight
on lensed SN, the first strongly lensed SN was discovered by
\citet{Kelly2015} serendipitously in the galaxy cluster
MACS\,J1149.5+2223 with {\it Hubble Space Telescope} (\hst) imaging.
This core-collapse SN was named ``SN Refsdal'', and showed 4 multiple
images at detection.  The predictions \citep{TreuEtal16, GrilloEtal16,
  KawamataEtal16, JauzacEtal16} and subsequent detection
\citep{Kelly2016} of the re-occurrence of the next (time-delayed)
image of SN Refsdal provided a true blind test of our understanding of
lensing theory and mass modeling.  It is reassuring that
some teams predicted accurately the re-occurrence \citep{GrilloEtal16,
  KawamataEtal16}, and the modeling software {\sc Glee}\footnote{{\sc Glee} (Gravitational Lens Efficient Explorer) is a gravitational lens modeling software developed by A.~Halkola and S.~H.~Suyu \citep{SuyuHalkola10, SuyuEtal12a}} used by \citet{GrilloEtal16} was also the software employed for cosmography with lensed quasars \citep[e.g.,][]{SuyuEtal13, SuyuEtal14, WongEtal17}. 

In the fall of 2016, the first spatially-resolved multiply-imaged Type
Ia SN, iPTF16geu, was discovered by \citet{Goobar2017} in the
intermediate Palomar Transient Factory survey.  \citet{More2017}
independently modeled a single-epoch \hst\ image of the system,
finding short model-predicted time delays ($<$1 day) between the
multiple images.  Furthermore, \citet{More2017} found anomalous flux
ratios of the SN compared to the smooth model prediction, indicating
possible microlensing effects, although \citet{YahalomiEtal17} showed
that microlensing is unlikely to be the sole cause of the anomalous
flux ratios.

Both SN Refsdal and iPTF16geu have been monitored for time-delay
measurements \citep{Rodney2016, Goobar2017}, opening a new window to
study cosmology with strongly-lensed SN.  Recently
  \citet{GrilloEtal18} estimated the time delay of the image SX of SN
  Refsdal based on the detection presented in \citet{KellyEtal16}
  (image SX has the longest delay compared to other images of SN
  Refsdal, so image SX will ultimately provide the most precise
  time-delay measurement for cosmography from this system), and modeled the mass
  distribution of the galaxy cluster MACS\,J1149.5+2223 to infer
  $H_0$.  This feasibility study shows that $H_0$ can be measured with
  $\sim7\%$ statistical uncertainty, despite the complexity in
  modeling the cluster lens mass distribution.  The full analysis
  including various systematic uncertainties is forthcoming, after the
  time delays are measured from the monitoring data.  As lensed SNe
are only being discovered/observed recently and their utility
  as a time-delay cosmological probe is just starting, we focus in
the rest of the review on the more common lensed quasars as time-delay
lenses, but there is a wealth of information to gather with lensed
supernovae, both on a cosmological and stellar physics point of view.

\subsection{Recent advances}
\label{sec:gl:adv}

A 10\% error bar on the time delay translates to a similar error on
$H_0$. Improving this further and obtaining $H_0$ measurements
competitive with other techniques, i.e. of 3-4\% currently and 1-2\%
in the near future, requires several ingredients. Time-delay
measurements of individual lenses must reach 5\% at most and many more
systems must be measured. With such precision on individual time
delays, and under the assumption that all sources of systematic errors
are controlled or negligible, measuring $H_0$ to 2\% is possible with
only a handful of lenses and a 1\% measurement is not out of reach with
of the order of 50 lenses!  
However, the lens model for the main lensing galaxy must be well constrained and their systematics evaluated and/or mitigated. Third, the contribution of all objects on the line of sight to the overall potential well must be accounted for. Excellent progress has been made on all three fronts in the recent years and there is still room for further (realistic) improvement.

\subsubsection{Time delays}
\label{sec:gl:adv:delays}

Measuring time delays is hard, but feasible provided telescope time can be guaranteed over long periods of time with stable instrumentation. The main limiting factors are astrophysical, observational or instrumental. 

Astrophysical factors include the characteristics of the intrinsic variability of the source quasar and extrinsic variability of its lensed images due to microlensing by stars in the main lensing galaxy. If the source quasar is highly variable intrinsically, both in amplitude and temporal frequency, the time delay is easier to measure. If microlensing variations are strong and/or comparable in frequency to the intrinsic variations then the time-delay value can be degenerate with the properties of the microlensing variations. In some extreme cases microlensing dominates the observed photometric variations to the point the time delay is hardly measurable \citep[e.g.][]{Morgan2012} even though microlensing itself can be used to infer details properties of the lensed source on micro-arcsec scales, i.e. out of reach of any current and future instrumentation. 

The observations needed to measure time delays must be adapted to the intrinsic and extrinsic variations of the selected quasars and of course to the expected time delay for each target. Not surprisingly the shorter the time delay, the finer the temporal sampling is needed. The position of the target on the sky also influences the results: equatorial targets will hardly be visible more than 6 months in a row, but can be followed both from the North and the South, while circumpolar targets can be seen up to 8-9 consecutive months, hence allowing to measure longer time delays and minimizing the effect of the non-visibility gaps between observing seasons.

Finally, instrumental factors strongly impact the results. A key factor with current monitoring campaigns is the availability of telescopes on good sites and with stable instrumentation, i.e. if possible at all with no camera or filter change and with regular temporal sampling. Long gaps in light curves seriously affect the time-delay values in the sense that they make it more difficult to disentangle the microlensing variations from the quasar intrinsic variations. And since angular separation between the lensed images are small, fairly good seeing is required, typically below 1.2 arcsec even though techniques like image deconvolution \citep[e.g.][]{Magain1998, Cantale2016} and used, e.g. by the COSMOGRAIL collaboration (see below) help dealing with data sets spanning a broad range of seeing values.   

Once long and well-sampled photometric light curves are available, the time delay must be measured. At first glance, this step may be seen as an easy one. However, one has to deal with irregular temporal sampling, gaps in the light curves, variable signal-to-noise and seeing, atmospheric effects (night-to-night calibration) and with microlensing. A number of numerical methods have been devised over the years to carry out the measurement, with different levels of accuracy and precision. They split in different broad categories. Some attempt to cross-correlate the light curves without trying to model/subtract the microlensing variations. Others involve an analytical representation of the intrinsic quasar variations and of the microlensing or involve e.g. Gaussian processes to mimic the microlensing erratic variations. Recent work in this area has been developed in \cite{Tewes2013, Hojjati2013, Hojjati2014, Kumar2015}. 

These methods (and others, so far unpublished) were tested in an objective way using simulated light curves proposed to the community in the context of the ``Time Delay Challenge''  \cite[TDC;][]{Dobler2015}. In the TDC, thousands of light curves of different lengths, sampling rate, signal-to-noise, and visibility gaps are proposed to the participants. Once all participants have submitted their time-delay evaluations to the challenge organizers, the time-delay values are revealed and the results are compared objectively using a metrics common to all participating methods. This metrics was known before the start of the challenge. The results of this TDC are summarized in \citet{Liao2015} as well as in individual papers \citep[e.g.][]{Bonvin2016}. A general conclusion of the TDC was that with current lens monitoring data, curve-shifting technique so far in use are sufficient to extract precise and accurate time delays, given the temporal sampling and signal-to-noise of the data.

\begin{figure}[t!]
\centering
\includegraphics[width=11cm]{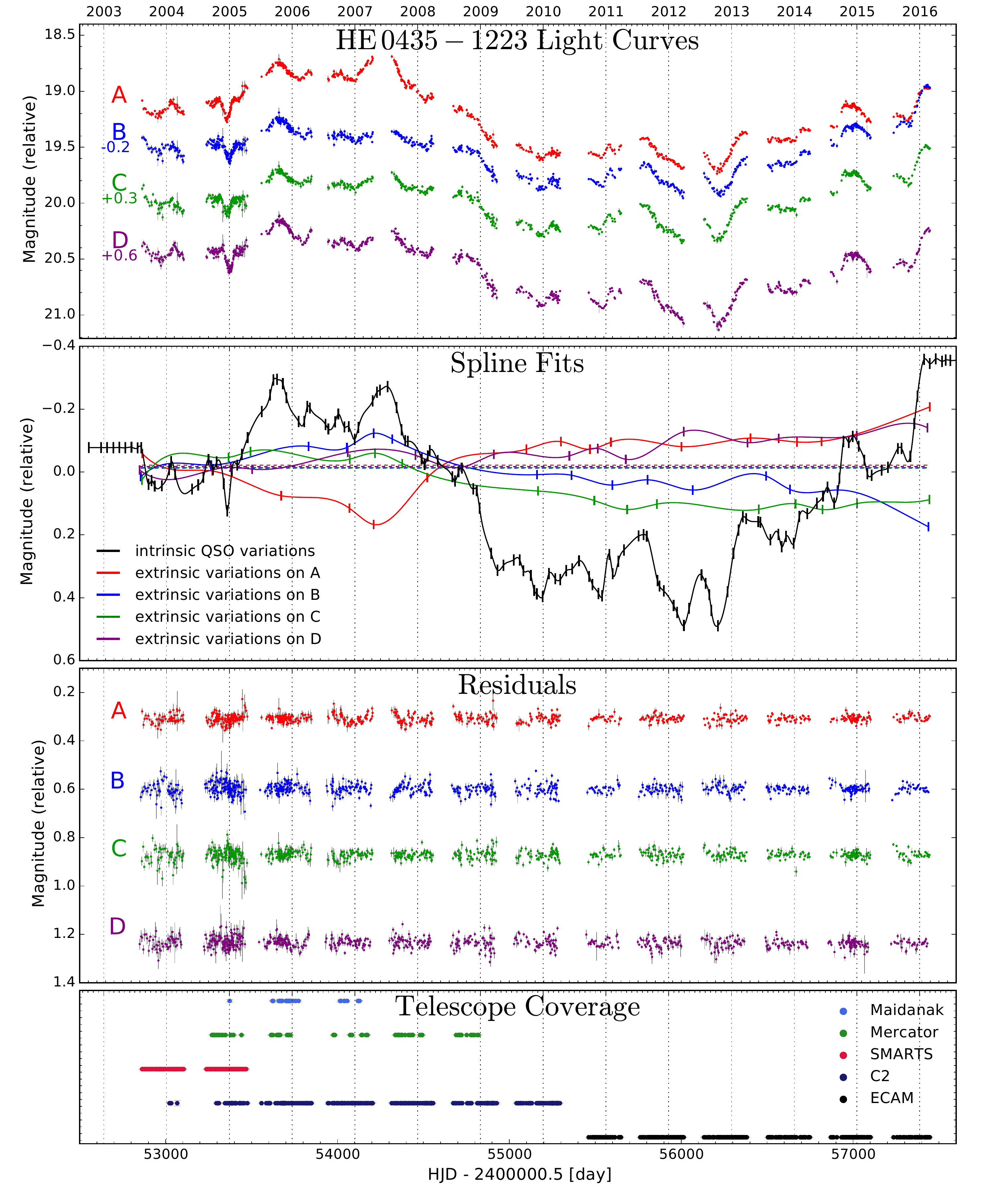}
\caption{From top to bottom: example of light curves produced and exploited by the COSMOGRAIL and H0LiCOW programs, here for the quadruply imaged quasar HE~0435$-$1223. The original light curves are shown on the top. The second panel shows spline fitting to the data including the intrinsic and extrinsic quasar variations. Crucially, long light curves are needed to extract properly the extrinsic variation (microlensing). The residuals to the fit and the journal of the observations with 5 instruments are displayed in the two lower panels \citep[reproduced with permission from][]{Bonvin2017}.}
\label{Fig:he0435}
\end{figure}

Following the encouraging results obtained at NOT, ESO and Wise, long-term monitoring campaigns were organized to measure time delays in a systematic way. Two main teams invested effort in this research:  the OSU group lead by C.S. Kochanek (OSU, USA) and the COSMOGRAIL (COSmological MOnitoring of GRAvItational Lenses) program led by F. Courbin and G. Meylan at EPFL, Switzerland \citep[e.g.][]{Courbin2005, Eigenbrod2005}. Both monitoring programs involve 1-m class telescopes with a temporal sampling of 2 to 3 observing epochs per week and a signal-to-noise of typically 100 per quasar image and per epoch. Both projects started in 2004 and are so far the main (but not only) source of time-delay measurements. Early results from the OSU program were obtained in 2006 for HE~0435$-$1223 \citep{Kochanek2006} while COSMOGRAIL delivered its first results starting in 2007 for SDSS~J1650+4251 \citep{Vuissoz2007}, WFI~J2033$-$4723 \citep{Vuissoz2008} and HE~0435$-$1223 \citep{Courbin2011}. More recent time-delay measurements from COSMOGRAIL were obtained for RX~J1131$-$1231 \citep{Tewes2013b}, HE~0435$-$1223 \citep{Bonvin2017} as well as SDSS~J1206+4332 and HS~2209+1914 \citep{Eulaers2013} and SDSS~J1001+5027 \citep{Kumar2013}. An example of COSMOGRAIL light curve is given in Fig.~\ref{Fig:he0435}. These data are analysed jointly with the H0LiCOW program (see Section~\ref{sec:gl:dist}).

Other recent studies for specific objects include \citet{Giannini2017} for WFI~2033$-$4723 and HE~0047$-$1756, and \citet{Shalyapin2017} reporting a delay for SDSS~J1515+1511. \citet{Hainline2013} measure a tentative time delay for SBS~0909+532, although the curves suffer from strong microlensing. Finally, two (long) time delays have been estimated for two quasars lensed by a galaxy group/cluster: SDSS~J1029+2623 \citep{Fohlmeister2013} and SDSSJ1004+4112 \citep{Fohlmeister2008}. These may not be ideal for cosmological applications though, as a complex lens model for a cluster is harder to constrain than models at galaxy-scale, unless the cluster has additional constraints coming from multiple background sources at different redshifts being strongly lensed.

With the observing cadence of 1 point every 3-4 nights and an SNR of 100 per epoch, the current data can catch quasar variations of the order of 0.1 mag in amplitude, arising on time-scales of months. These time scales are unfortunately of the same order of magnitude as the microlensing variations (see 2nd panel of Fig.~\ref{Fig:he0435}) making it hard to disentangle between intrinsic and extrinsic variations. For this reason, lensed quasars must be monitored for extended periods of time, typically a decade, to infer any reliable time-delay measurement. 

\begin{figure}[h!]
\centering
\includegraphics[width=8cm, angle=-90]{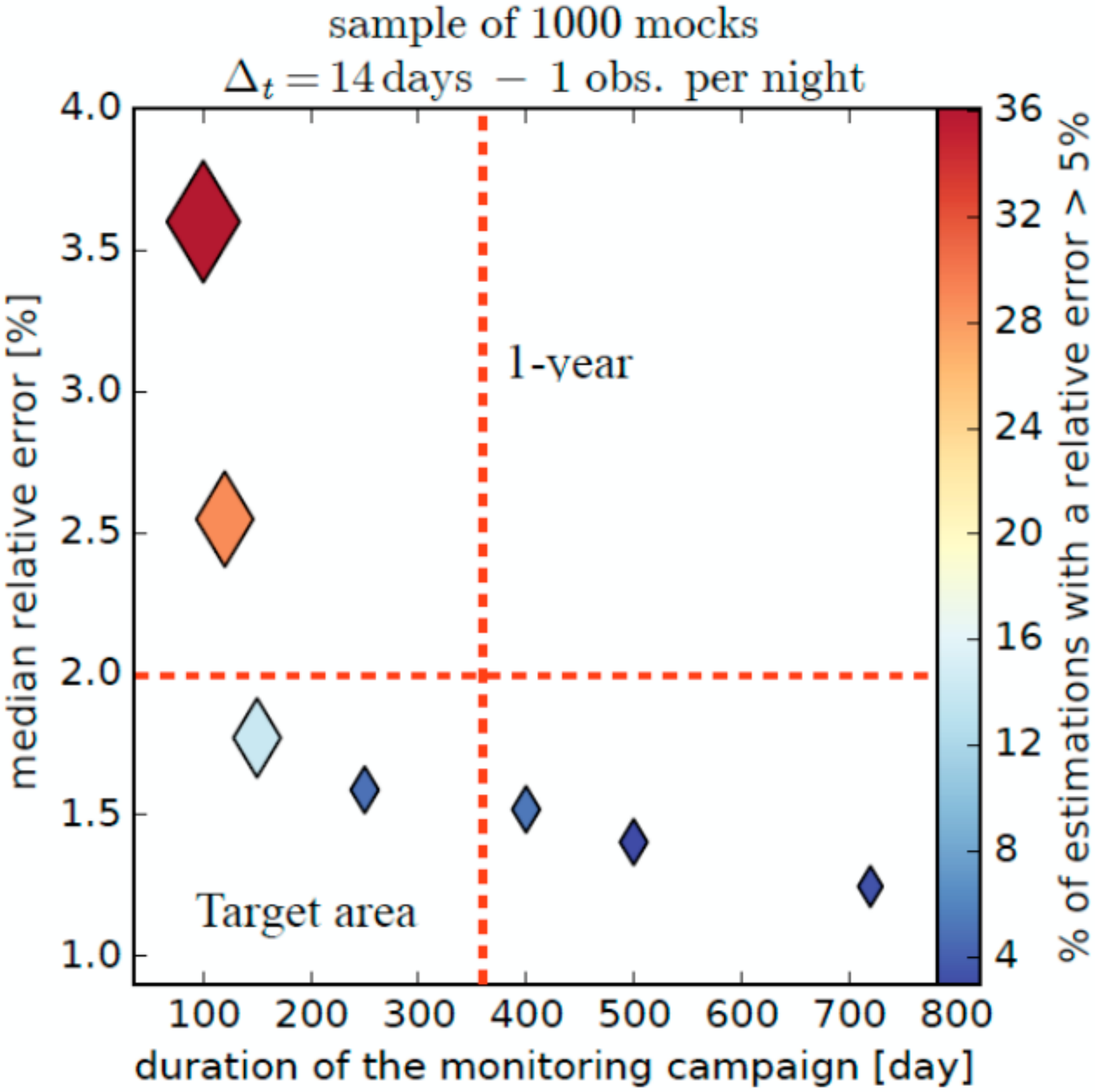}
\caption{Expected relative precision on a time delay measurement as a function of the length of the campaign. High-cadence (1 day$^{-1}$) monitoring is assumed and the fiducial delay in this simulations mimics the longer delay of HE~0435-1223, i.e. 14 days. Clearly, 2\% precision can be reached in only 1 observing season. The color code shows the catastrophic failure rate, i.e. the probability of getting a measurement wrong by more than 5\%. This probability is about 10\% for a 1-season campaign and 3\% for a 2-season campaign. (Courtesy: Vivien Bonvin)}
\label{Fig:highcandence}
\end{figure}

Going beyond current monitoring campaigns like COSMOGRAIL and others
is possible, but measuring massively time delays for dozens of lensed
quasars requires a new observing strategy to minimize the effect of
microlensing and to measure time delays in individual objects in less
than 10 years! One solution is to observe at high cadence (1
day$^{-1}$) and high SNR, of the order of 1000. In this way, very
small quasar variations can be caught, on time scales much shorter
than microlensing hence allowing the separation of the two signals in
frequency. We show with 1000 mock light curves that mimic those of
HE~0425$-$1223 (Fig.~\ref{Fig:he0435}) that time delays can be measured
precisely in only 1 observing season. In doing the simulation, we
include realistic microlensing and fast quasar variations with a few
mmag amplitude. We then run {\tt PyCS}, the COSMOGRAIL curve-shifting
algorithm \citep{Tewes2013, Bonvin2016}, to recover the fiducial time
delay of 14 days.  Fig.~\ref{Fig:highcandence} summarizes our results
and provides the length of the monitoring campaign needed to reach a
desired relative precision on the time delay, assuming daily
observations and an SNR of 1000 per epoch. It appears that a typical
2\% precision is achievable in 1 observing season with a 10\% failure
rate. Doing two seasons allows one to reach the percent precision and a
failure rate below 3\%. At the time this paper is being written, an
intensive lens monitoring program has been started at the 2.2m MPI
telescope at La Silla Observatory, with the above
characteristics. Three targets have been observed for 1 semester and
time delays have been measured to a few percents for all three! The first of these is presented in \citet{Courbin2017} and features a 1.8\% measurement of one of the delays in the newly discovered quadruple lens DES~J0408$-$5354 \citep{Lin2017, Agnello2017}.

Finally, we note that although quasars have been used so far to
implement the time delay method, the original idea of Refsdal was to
use lensed supernovae. The first systems have finally be found, as
mentioned in Section \ref{sec:gl:history}: SN Refsdal
\citep[e.g.][]{Kelly2015, Kelly2016, Rodney2016} and iPTF16geu
\citep[e.g.][]{Goobar2017, More2017}.  With the advent of large
imaging surveys such as the Zwicky Transient Facility and the Large
Synoptic Survey Telescope, prospects to find lensed supernovae
are excellent \citep[e.g.][]{Goldstein2017}. As supernovae have known
light curves, one can measure the time delay by fitting a template to
the observed light curves in the lensed images, hence giving much more
constraining power than quasars whose photometric variations are close
to a random walk. In addition, if the lensed supernova is a Type Ia,
then two cosmological probes are available in the same object, hence
provide a fantastic cross-check of otherwise completely different
methods: standard candles and a geometrical method, provided
microlensing effects could be corrected \citep[][]{DoblerKeeton06, YahalomiEtal17, GoldsteinEtal18}.  
  
For all the above reasons, we believe that the
future of time-delay cosmography resides in lensed supernovae and in
high-cadence monitoring of lensed quasars. However, \cite{TK18} recently pointed out that microlensing by stars in the lensing galaxy can introduce a bias in the time-delay measurements. This is due to a combination of differential magnification of  different parts of the source and the source geometry itself. The net result is that cosmological time delays can be affected both in a statistical and a systematic way by microlensing. The effect is absolute, with biases on time delays of the order of a day for lensed quasars and tenths of a day for lensed supernovae. Mitigation strategies have been successfully devised \citep{Chen2018} for lensed quasars and the effect seems less pronounced in lensed supernovae than in lensed quasars \citep{GoldsteinEtal18, FM18, Bonvin2018}, but clearly this new effect must be accounted for in any future work in the field.

\subsubsection{Lens mass modeling}
\label{sec:gl:adv:modeling}

To convert the time delays into a measurement of the time-delay
distance via equation (\ref{eq:delays_Dt}), one needs to determine the
Fermat potential $\phi(\Imvec_i;\Srvec)$, which depends both on the
mass distribution of the main strong-lens galaxy and the mass
distribution of other galaxies along the line of sight.  

The mass distribution of the main strong-lens galaxy can be modeled
using either simply parametrized profiles
\citep[e.g.,][]{KormannEtal94, Barkana98, GolseKneib02} or grid-based
approaches \citep[e.g.,][]{WilliamsSaha00, BlandfordEtal01,
  SuyuEtal09, VegettiKoopmans09}.  The total mass distribution of galaxies
appear to be well described by profiles close to isothermal
\citep[e.g.,][]{KoopmansEtal06, BarnabeEtal11, CappellariEtal15}, even
though neither the baryons nor the dark matter distribution follow
isothermal profiles.  Even in the complex case of the gravitational
lens B1608+656 with two interacting lens galaxies, simply parametrized
profiles provide a remarkably good description of the galaxies when
compared to the pixelated lens potential reconstruction
\citep{SuyuEtal09}.  Therefore, most of the current mass modeling for
time-delay cosmography use simply parametrized profiles, either for
the total mass distribution \citep[e.g.,][]{KoopmansEtal03, FadelyEtal10,
  SuyuEtal10, SuyuEtal13, BirrerEtal16} or for
separate components of baryons and dark matter
\citep[e.g.,][]{CourbinEtal11, SchneiderSluse13, SuyuEtal14, WongEtal17}.

The source (quasar) properties need to be modeled simultaneously with
the lens mass distribution to predict the observables.  In particular,
source position and intensity are needed to predict the positions, fluxes
and time delays of the lensed quasar images, whereas the source
surface brightness distribution (of the quasar host galaxy) is needed
to predict the lensed arcs.  These observables (image positions,
fluxes and delays of the multiple quasar images, and lensed arcs) are
then used to constrain the parameters of the lens mass model and the
source.  Several softwares are available publicly for modeling lens
systems, including {\sc Gravlens} \citep{Keeton01}, {\sc lenstool}
\citep{JulloEtal2007}, {\sc glafic} \citep{Oguri10} and {\sc Lensview}
\citep{WaythWebster06}.  

Observed quasar image positions, fluxes and delays provide around a
dozen of constraints for quads (four-image systems) and even fewer
constraints for doubles (two-image systems).  Thus lens mass models
using only these quasar observables are often not precisely
constrained.  In particular, the radial profile slope of the lens
galaxy is strongly degenerate with $\tdist$
\citep[e.g.,][]{Wucknitz02, Suyu12}. The time delays depend primarily
on the average surface mass density between the multiple images, and
thus provide information on the radial profile slope
\citep{Kochanek02}.  Nevertheless, even with multiple time delays from
quad systems, it is difficult to infer the slope precisely to better
than $\sim10\%$ precision\footnote{\label{fn:impact} in terms of
  impact on $\tdist$}.  While mass distribution of massive early-type
galaxies, which are the majority of lens galaxies, are close to being
isothermal, there is an intrinsic scatter in the slope of about
$\sim15\%$$^{\ref{fn:impact}}$ \citep{KoopmansEtal06, BarnabeEtal11}.
For precise and accurate $\tdist$ measurement of a lens system, it is
important to measure directly, at the few percent level, the radial
profile slope of the lens galaxy near the lensed images of the
quasars.  This requires more observations of the lens system, beyond
just the multiple point images of lensed quasars.

\begin{figure}[t!]
\centering
\includegraphics[width=11cm]{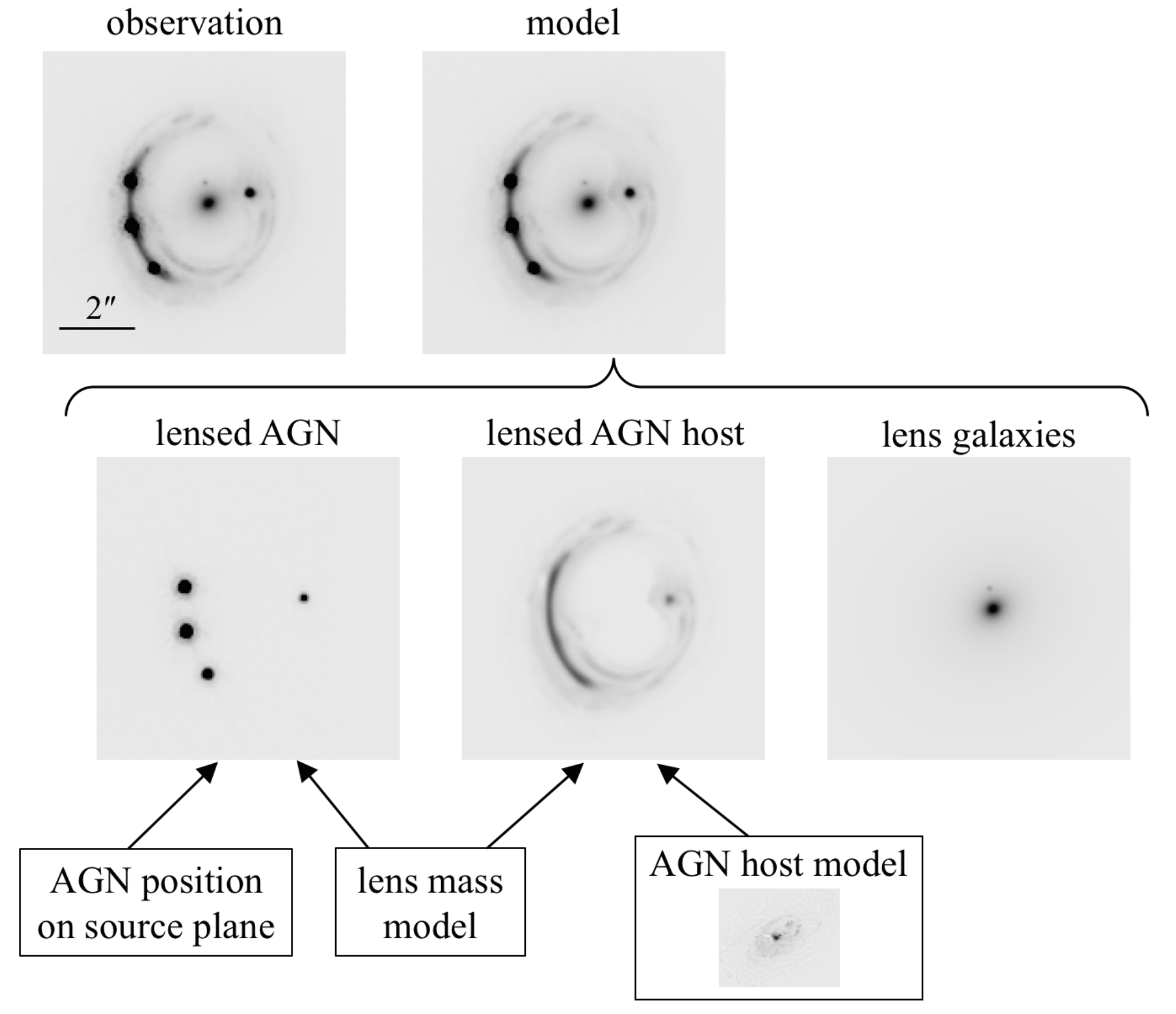}
\caption{Illustration of lens mass modeling of the gravitational lens RXJ1131$-$1231.  Top left is the observed \hst\ image.  Top middle panel is the modeled surface brightness of the lens system, which is composed of three components shown in the second row: lensed AGN images (left), lensed AGN host galaxy (middle), and foreground lens galaxies (right).  The bottom row shows that a mass model is required together with the AGN source position and AGN host galaxy surface brightness, to model the lensed AGN and lensed AGN host images.  See the text and \citet{SuyuEtal13, SuyuEtal14} for more details.}
\label{fig:lensmodel}
\end{figure}

Over the past decade, multiple methods have been developed to make use
of the lensed arcs (lensed host galaxies of the quasars) to constrain
the lens mass distribution.  The source intensity distribution can be
described by simply parametrized profiles, such as Gaussians or Sersic
\citep[e.g.,][]{BrewerLewis08, Oguri10, MarshallEtal07, OldhamEtal17},
or on a grid of pixels, either regular
\citep[e.g.,][]{WallingtonEtal96, WarrenDye03, Koopmans05, SuyuEtal06}
or adaptive \citep[e.g.,][]{DyeWarren05, VegettiKoopmans09,
  TagoreKeeton14, NightingaleDye15}, or based on basis functions
\citep[e.g.,][Joseph et al.~in prep.]{BirrerEtal15}.  These lensed
arcs, when imaged with \hst\ or
ground-based telescopes assisted with adaptive optics, contain
thousands of intensity pixels and thus allow the measurement of the
radial profile slope of the lens galaxies with a precision of a few
percent \citep[e.g.,][]{DyeWarren05, Suyu12, ChenEtal16}, that are
required for cosmography.  In Fig.~\ref{fig:lensmodel}, we show an
example of the mass modeling using the full surface brightness
distributions of quasar host galaxy.

Once a model of the surface mass density $\kappa$ is obtained, lens
theory states that the following family of models $\kappa_{\lambda}$
fits equally well to the observed lensing data:
\be
\label{eq:mst}
\kappa_{\lambda} = \lambda + (1-\lambda) \kappa,
\ee
where $\lambda$ is a constant.  This transformation is analogous to
adding a constant mass sheet $\lambda$ in convergence, and rescaling
the mass distribution of the strong lens (to keep the same mass within
the Einstein radius); it is therefore called the ``mass-sheet
degeneracy'' \citep[][]{FalcoEtal85, SchneiderSluse13,
  SchneiderSluse14, Schneider14}.  Such a transformation corresponds
to a rescaling of the background source coordinate by a factor
$(1-\lambda)$, leaving the observed image morphology and brightness
invariant.  Furthermore, the Fermat potential transforms as
\be
\label{eq:mst:fp}
\phi_{\lambda}(\Imvec;\Srvec) = (1-\lambda) \phi(\Imvec; \Srvec) + {\rm
  constant\ that\ depends\ only\ on\ \Srvec}.
\ee
Therefore, for given observed time delays $\Delta t_{ij}$, equations
(\ref{eq:tdiff}) and (\ref{eq:mst:fp}) imply that the time-delay
distance $\tdist$ would be scaled by $(1-\lambda)$.  The mass-sheet
degeneracy has thus a direct impact on cosmography in measuring
$\tdist$.

While $\lambda$ so far is simply a constant in this mathematical
transformation (equation \ref{eq:mst}), we can identify it with the 
physical external convergence, $\kext$, due to mass structures along
the sight line to the lens system.  By gathering additional data sets beyond
that of the strong lens system, we can infer $\kext$ and thus measure
$\tdist$.  Two practical ways to break the mass-sheet degeneracy are
(1) studies of the lens environment, to estimate $\kext$ based on the
density of galaxies in the strong-lens line of sight in comparison to
random lines of sight \citep[e.g.,][]{MomchevaEtal06, FassnachtEtal06,
  HilbertEtal07, SuyuEtal10, GreeneEtal13, CollettEtal13, RusuEtal17,
  SluseEtal17, McCullyEtal17}, and (2) stellar kinematics of the strong lens galaxy,
which provides an independent mass measurement within the effective
radius to complement the lensing mass enclosed within the Einstein
radius \citep[e.g.,][]{GroginNarayan96a, KoopmansTreu02, BarnabeEtal09,
  SuyuEtal14}.  The time-delay distance can then be inferred via
\be
\label{eq:mst:tdist}
\tdist = \frac{\tdist^{\rm model}}{(1-\kext)},
\ee
where $\tdist^{\rm model}$ is the modeled time-delay distance without
accounting for the presence of $\kext$.  In practice, both lens
environment characterisations and stellar kinematics are employed to
infer $\tdist$ for cosmography \citep[e.g.,][]{SuyuEtal10, SuyuEtal13,
  BirrerEtal16, WongEtal17}.  The stellar kinematic data further help
constrain the strong-lens mass profile \citep[e.g.,][]{SuyuEtal14}.

Lens systems that have massive galaxies close in projection (within
$\sim10''$) to the strong lens, but at a different redshift from the
strong lens, will need to be accounted for explicitly in the strong
lens model.  In such cases, multi-lens plane modeling is needed
\citep[e.g.,][Suyu et al., in preparation]{BlandfordNarayan86,
  SchneiderEtal92, GavazziEtal08, WongEtal17}, but equation
(\ref{eq:delays_Dt}) for single-lens plane is then not directly
valid.  In particular, there is not a single time-delay distance,
but rather there are multiple combination of distances between the
multiple planes.  Nonetheless, for some cases, one could obtain an
effective time-delay distance as if it were a single-lens plane system
\citep[see, e.g., ][for details]{WongEtal17}. 

As noted in Section \ref{sec:gl:principles}, with stellar kinematic and time-delay data, we can infer the angular
diameter distance to the lens, $\Dd$, in addition to $\tdist$
\citep[][]{ParaficzHjorth09, JeeEtal15, BirrerEtal16,
  ShajibEtal17}. Measurement of $\Dd$ is often more sensitive to the dark
energy parameters \citep[for typical lens redshifts $\lesssim 1$, see e.g.,~Fig.~2 of][]{JeeEtal16}, and can also be used as an
inverse distance ladder to infer $H_0$ (Jee et al., submitted).
Currently, the precision in $\Dd$ is limited by the uncertainty in the
single-aperture averaged velocity dispersion measurement and the
unknown anisotropy of stellar orbits \citep{JeeEtal15}.  Nonetheless,
we anticipate that spatially resolved kinematic data will help to
constrain more precisely $\Dd$.

We have focussed here on the advances in getting $\tdist$ and $\Dd$ from
individual lenses with exquisite follow-up data to control the
systematic uncertainties.  Alternatively, one could
analyse a sample of lenses and constrain a global $H_0$ parameter that
is common to all the lenses \citep[e.g.,][]{SahaEtal06, Oguri07,
  SerenoParaficz14}, assuming that the systematic effects for the
lenses average out.  For small samples, this assumption might not be
valid.  Nonetheless, in the future where thousands of lensed quasars
are expected \citep{OguriMarshall10} but most of which will not have
exquisite follow-up observations, this large sample of lenses could
provide information on the population of lens galaxies as a whole for
cosmography (P.~J.~Marshall \& A.~Sonnenfeld, priv.~comm.).  We therefore advocate
getting exquisite follow-up observations of a sample of $\sim40$
lenses to reach an $H_0$ measurement with 1\% uncertainty
\citep{JeeEtal16, ShajibEtal17}, with the other lenses providing
information on the profiles of galaxies to use in the mass modeling.

\subsection{Distance measurements and cosmological inference}
\label{sec:gl:dist}

There are so far only a few lensed quasars for which all required data exist to do time-delay cosmography, i.e., with time-delay measurements to a few percent, deep \hst\ images showing the lensed image of the host galaxy, deep spectra of the lens to measure the velocity dispersion, and multiband data to map the line of sight contribution to the lensing potential. 

Some of the best time-delay measurements available to date include the radio time delay for B1608+656 \citep{Fassnacht1999, FassnachtEtal02} and the two optical measurements of COSMOGRAIL for RX~J1131$-$1231 \citep{Tewes2013b}, and HE~0435$-$1223 \citep{Bonvin2017}. These 3 quadruply imaged quasars, for which all the ancillary imaging and spectroscopic data are also available, gave birth to the H0LiCOW program \citep{SuyuEtal17}, which capitalizes on more than a decade of COSMOGRAIL monitoring as its name reflects: {\bf H0} {\bf L}enses {\bf i}n {\bf CO}MOGRAIL's {\bf W}ellspring. With the precise time-delay measurements of COSMOGRAIL, H0LiCOW addresses what has been so far limiting the effectiveness of strong lensing in delivering reliable $H_0$ measurements: the different systematics at work at each step leading to a value for the Hubble constant. 

\begin{figure}[ht!]
\centering
\includegraphics[width=8.3cm, angle=-90]{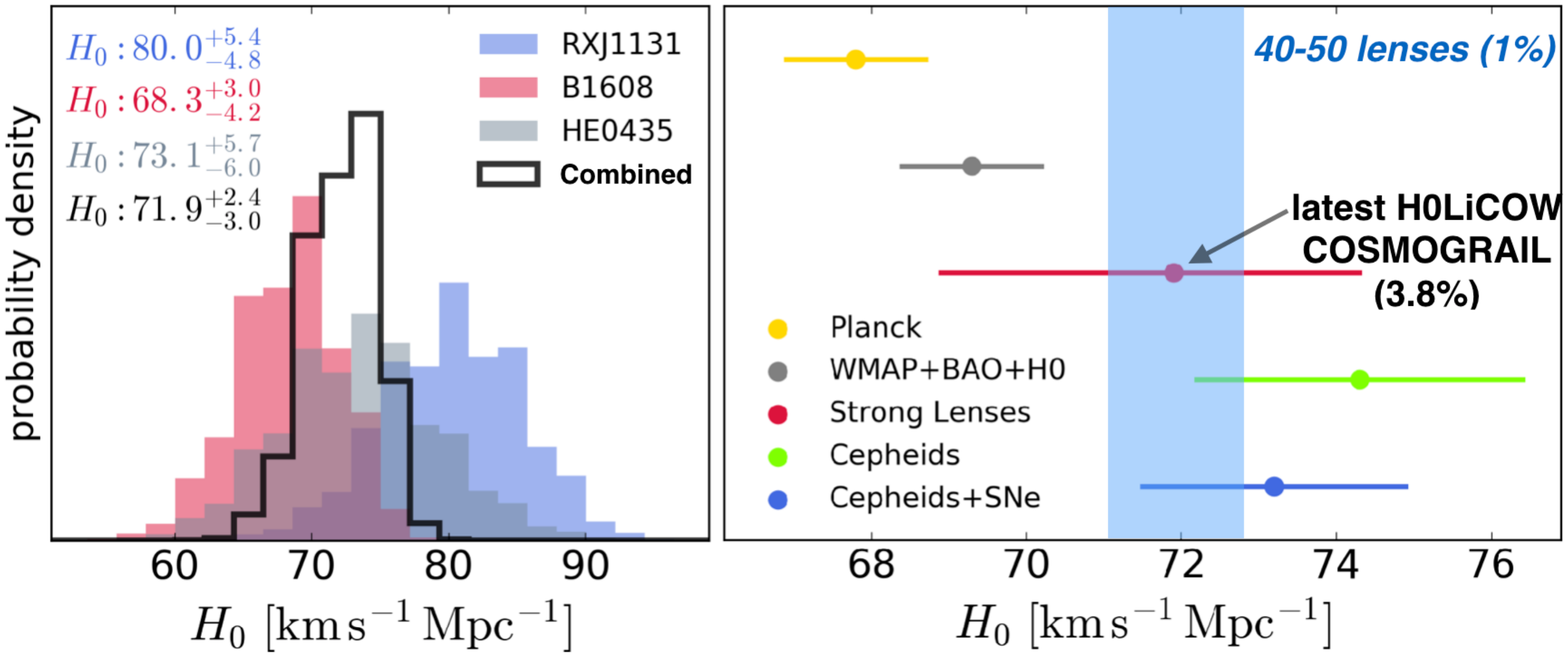}
\vskip -50pt
\caption{{\bf Left:} Latest $H_0$ measurement from quasar time delays from H0LiCOW and COSMOGRAIL for 3 lenses and for their combination in a $\Lambda$CDM Universe \citep[reproduced with permission from][]{Bonvin2017}. {\bf Right:} comparison between time delay $H_0$ measurements and other methods such as CMB shown in yellow \citep[Planck;][]{Planck2016} and gray \citep[WMAP;][]{Bennett2013} or local distance estimators such as Cepheids \citep[green;][]{FreedmanEtal12} and Type Ia supernovae \citep[blue;][]{RiessEtal16}. Quasar time delays are so far in agreement with local estimators but higher than Planck. Measurements for 40-50 new time delays will allow one to confirm (or not) the current tension with Planck to more than 5$\sigma$.}
\label{Fig:stateofart}
\end{figure}

The most recent work by H0LiCOW is summarized in the left panel of Fig.~\ref{Fig:stateofart} \citep{Bonvin2017}, based on state-of-the-art lens mass modeling and characterisations of mass structures along the line of sight \citep{SluseEtal17, RusuEtal17, WongEtal17, TihhonovaEtal17}.  In the right panel of Fig.~\ref{Fig:stateofart}, we compare the value of $H_0$ from H0LiCOW with other fully independent cosmological probes such as Type Ia supernovae, Cepheids, and CMB(+BAO) for a $\Lambda$CDM cosmology. With the current error bars claimed by each probe there exist a tension between local measurements of $H_0$ \citep[e.g.,][]{FreedmanEtal12, Efstathiou14, RiessEtal18} and the value by the Planck team.  When completed, H0LiCOW will feature 5 lenses, with an accuracy on $H_0$ of the order of 3\% \citep{SuyuEtal17}, but reaching close to 1\% precision is possible. This will be enabled by working on several fronts simultaneously, by finding more lenses, measuring up to 50 new time delays, and refining the lens modeling tools to mitigate degeneracies between model parameters. Chapter 8 on ``Towards a self-consistent astronomical distance scale'' provides more details about the (expected) future of quasar time delay cosmography. \\

%
\section{Baryon Acoustic Oscillations}
\label{sec:bao}

\newcommand{\himpc}{{\hbox {$h^{-1}$}{\rm Mpc}}}


\subsection{BAO as a standard ruler}
\label{sec:bao:introduction}
The universe has been expanding, and thus the universe in the earlier stage was
much smaller, denser and hotter than today.
In such an early universe, electrons interacted with photons via Compton scattering and 
with protons via Coulomb scattering. Thus, the three components acted as a mixed fluid \citep{Peebles:1970}. 
They were in the equilibrium state due to the gravity of protons and pressure of photons, and 
oscillated as sound modes. These oscillations are called baryon acoustic oscillations (BAO) (see \cite{Bassett:2010} and \cite{Weinberg:2013} for a comprehensive review). 
It moved with the speed of sound $c_{\rm s} = c\sqrt{\frac{1}{3(1+R)}}$, where the ratio of photon density ($\rho_{\rm r}$) to baryon density ($\rho_{\rm b}$) is defined as
$1/R=4\rho_{\rm r} / 3\rho_{\rm b}$.

At recombination ($z\sim 1100$), photons decouple from the baryons and 
start to free stream. 
We observe the photons as a map of cosmic microwave background (CMB).
The left panel of Fig.~\ref{fig:bao} shows the angular power spectrum of the latest data of the CMB anisotropy probe, 
Planck satellite \citep{Planck-Collaboration:2016a}.
One can see a clear oscillation feature in the power spectrum, which is characterized by the 
sound horizon scale at recombination, expressed as \citep{Hu:1996,Eisenstein:1998}
\be
r_{\rm d}=\int^{\infty}_{z_*} \frac{c_{\rm s}(z)}{H(z)} dz,
\ee
where $z_*$ is the redshift at recombination. $H(z)$ is the Hubble parameter, 
\be
H(z)=  H_0 \left[ \Omega_{\rm m} (1+z)^3 + \Omega_{\rm r}(1+z)^4 +\Omega_{\rm DE}(1+z)^{1+w} \right]^{1/2}, \label{eq:hubblez}
\ee
where $\Omega_{\rm m}$ (previously introduced in Section \ref{sec:gl:principles}), $\Omega_{\rm r}$ and $\Omega_{\rm DE}$ are the matter, radiation and dark energy density parameters, respectively, and $w$ is the equation-of-state parameter of dark energy and the simplest candidate for dark energy, the cosmological constant $\Lambda$, gives $w=-1$.
With standard cosmological models, $r_{\rm d} \simeq 150$\,Mpc.
From the Planck observation, it is constrained to $r_{\rm d}=144.61 \pm 0.49$\,Mpc.

\subsection{Probing BAO in galaxy distribution}\label{sec:1d_bao}

After the recombination, motion of the baryons becomes non-relativistic.
The perturbation of baryons then starts to grow at their locations and interact with 
the perturbation of dark matter. 
Thus the baryon acoustic feature should be imprinted onto the late-time large-scale structure of the Universe.
Theoretically it is predicted to produce the overdensity at the sound horizon scale, $\sim 150$ Mpc. 
It is, however, observationally not easy because the observation of BAO signal requires the number of tracers of matter overdensity field to be large enough at the scale to overcome the cosmic variance. 

\begin{figure*}[tb]
\begin{center}
\includegraphics[width=70mm]{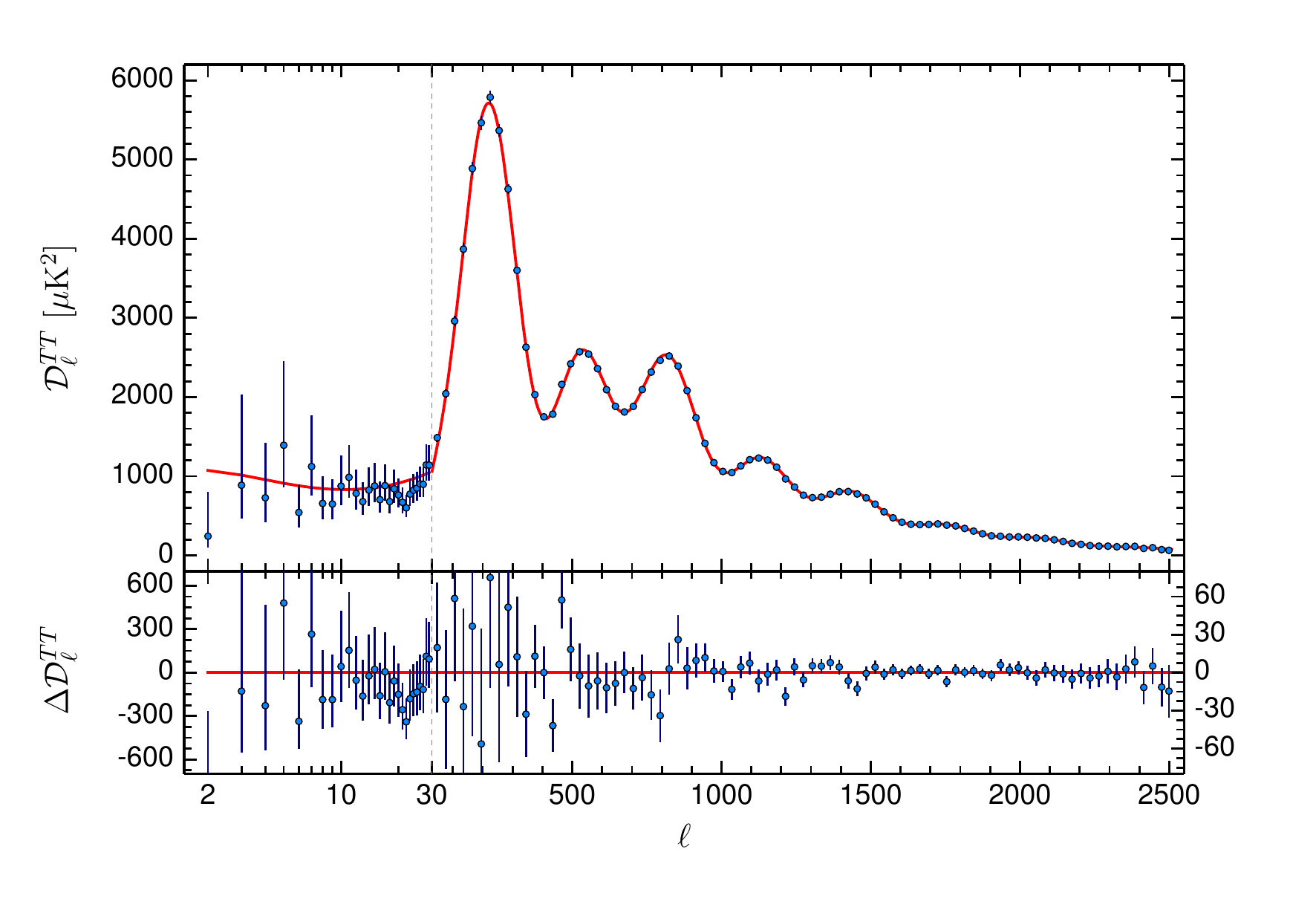}
\includegraphics[width=45mm]{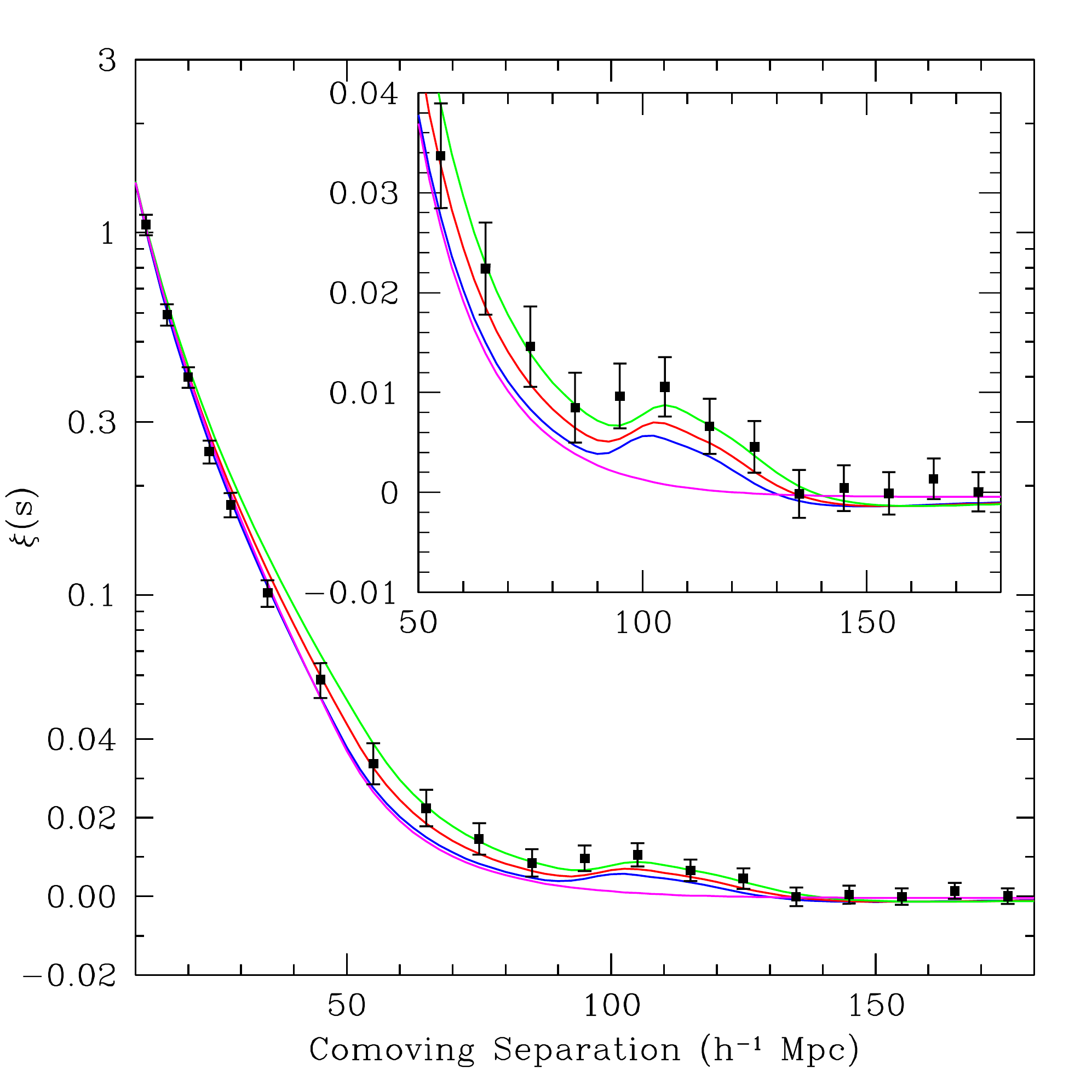}
 \caption{({\it Left}) Angular power spectrum of CMB anisotropies measured from the latest Planck satellite data ($\copyright$ESA and the Planck Collaboration). 
 The wiggles seen in the spectrum are the feature of BAO, and the oscillation scale corresponds to the sound horizon at recombination.
 The best-fitting $\Lambda$CDM theoretical spectrum is plotted as the solid line in the upper panel. 
 Residuals of the measurement with respect to this model are shown in the lower panel.
 ({\it Right}) BAO feature detected in large-scale correlation function of the galaxy distribution of the SDSS \citep[reproduced with permission from][]{Eisenstein:2005}. The bump seen at $105\himpc ~(\simeq 150 {\rm Mpc})$ corresponds to the sound horizon scale at recombination. The solid lines are the theoretical models with $\Omega_{\rm m}h^2=0.12$ (top line), $0.13$ (second line) and $0.14$ (third line). The bottom line shows a pure CDM model with $\Omega_{\rm m}h^2=0.105$, which lacks the acoustic peak. The inset zooms into the BAO peak position. 
 \label{fig:bao}}
\end{center}
\end{figure*}


In 2005, detection of the BAO was reported almost simultaneously by two independent groups  
using the Sloan Digital Sky Survey (SDSS) \citep{Eisenstein:2005} and the 2dF Galaxy Redshift Survey (2dFGRS) \citep{Cole:2005}. 
The right panel of Fig.~\ref{fig:bao} shows the two-point correlation function obtained from the SDSS galaxy sample by \cite{Eisenstein:2005}. 
The 2-point correlation function $\xi(s)$ is defined as an excess of the probability that one can find pairs of galaxies at a given scale $s$  from the case of a random distribution. 
Thus the scales where $\xi >0$ and $\xi <0$ correspond to the statistically overdense and underdense regions respectively.
The bump seen around $s\simeq 105\himpc$ ($\simeq 150$ Mpc) is the feature of BAO, and the scale of the bump
corresponds to the sound horizon scale at recombination. 
The inset of the right panel of Fig.~\ref{fig:bao} zooms into the feature. 
Unlike observations of the CMB, galaxy redshift surveys are the observation of 3 dimensional space.  
The peak scale of BAO should be isotropic because the scale corresponds to the sound horizon at recombination.
On the other hand, the BAO scale along the line of sight depends on $H^{-1}(z)$ while the BAO scale perpendicular to the line of sight depends on the comoving angular diameter distance $D_{\rm M}=(1+z)D_{\rm A}(z)$, 
where $D_{\rm A}(z)$ is expressed in flat universe as equation (\ref{eq:DA}) with $z_1=0$, or, 
\be
D_{\rm A}(z)=\frac{1}{1+z}\int^z_0  \frac{c{\rm d}z'}{H(z')}. \label{eq:angular_diameter_z}
\ee
Thus, the scale of BAO probed by a galaxy correlation function has a cosmology dependence of 
\be
D_{\rm V}(z)=\left[ (1+z)D_{\rm A}(z) \right]^{2/3} \left[\frac{cz}{H(z)} \right]^{1/3}.
\ee
The BAO scale probed by the CMB anisotropy at high redshift and the galaxy distribution at low redshift 
should be the same. Moreover, the power spectrum with the acoustic features has been precisely determined for the CMB and interpreted using linear cosmological perturbation theory (see the left panel of Fig.~\ref{fig:bao}). 
Thus the detection of BAO in the galaxy distribution enables us to constrain $D_{\rm V}$, and hence the geometric quantities such as $H_0$, $w$, and $\Omega_{\rm DE}$ in equation (\ref{eq:hubblez}) \citep{Eisenstein:2005}.

\subsection{Anisotropy of BAO and Alcock-Paczynski}\label{sec:2d_bao}
The method presented above does not use all of the information encoded in BAO. 
To maximally extract the cosmological information, 
we need to measure the correlation function as functions of separations of galaxy pairs 
perpendicular ($s_\perp$) and parallel ($s_\parallel$) to the line of sight, $\xi(s_\perp,s_\parallel)$, where $s=\sqrt{s_\perp^2+s_\parallel^2}$.
In this way, in principle we can constrain $D_{\rm A}(z)$ and $H(z)$ using the transverse and radial BAO measurements, respectively. Given the cosmological dependence of angular and radial distances (equations \ref{eq:hubblez} and \ref{eq:angular_diameter_z}), the shape of the BAO peak is distorted if the wrong cosmology is assumed. This effect was first pointed out by \citet[][AP]{Alcock:1979}.

In fact, the AP test using the anisotropy of BAO has additional advantages. Since the BAO scale should correspond to the sound horizon at recombination, it should be a constant. Thus, we can determine the geometric quantities by requiring that the radial BAO scale equals to the angular BAO scale. We no longer need to know the exact value of $r_{\rm d}$ nor need to rely on the CMB experiment \cite[see e.g.,][]{Seo:2003, Matsubara:2004}.
This is particularly important in the context of measuring $H_0$ given the tension in $H_0$ between the local measurements and the inference by the Planck team in flat $\Lambda$CDM (see Section \ref{sec:gl:dist}).

In galaxy surveys, the distance to each galaxy is measured through redshift, thus it gives the sum of the true distance and the contribution from the peculiar motion of the galaxies and it produces an anisotropy in galaxy distribution along the line of sight, which are called redshift-space distortions (RSD).
Since the velocity field of a galaxy is caused by gravity, the anisotropy contains additional cosmological information. 
This effect is called the Kaiser effect, named after \citet{Kaiser:1987} who proposed RSD as a cosmological probe in the linear perturbation theory limit. 
Since the velocity field is related to the density field through the continuity equation, the anisotropy constrains the quantity $f$, defined as the logarithmic derivative of the density perturbation, $f={\rm d}\ln{\delta}/{\rm d}\ln{a}$.
See 
\cite{Okumura:2016} 
for the constraint on $f$ from RSD as a function of $z$ including the high-$z$ measurement. 

\begin{figure}[tb]
\begin{center}
\includegraphics[width=55mm]{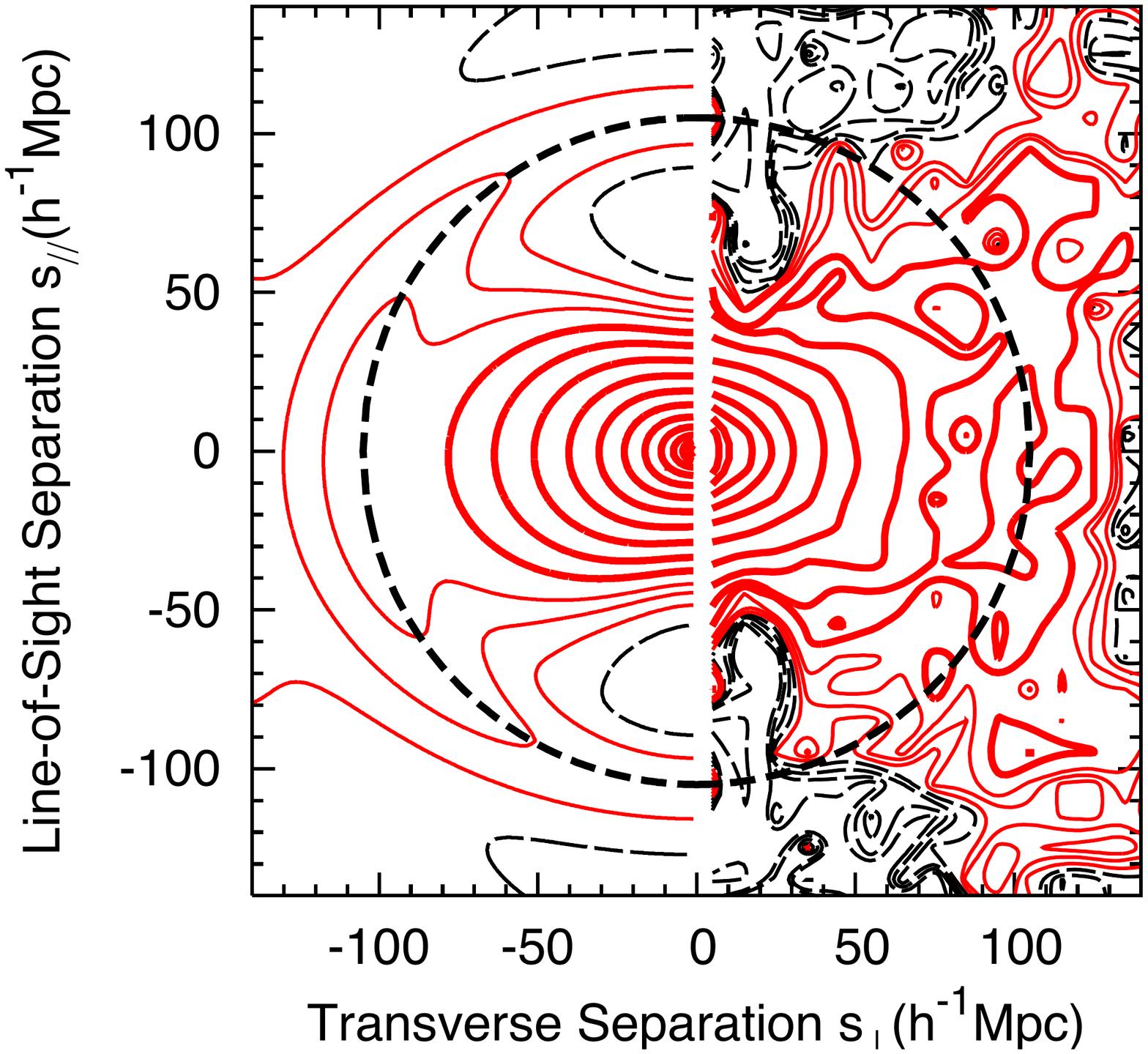}
\includegraphics[width=61mm]{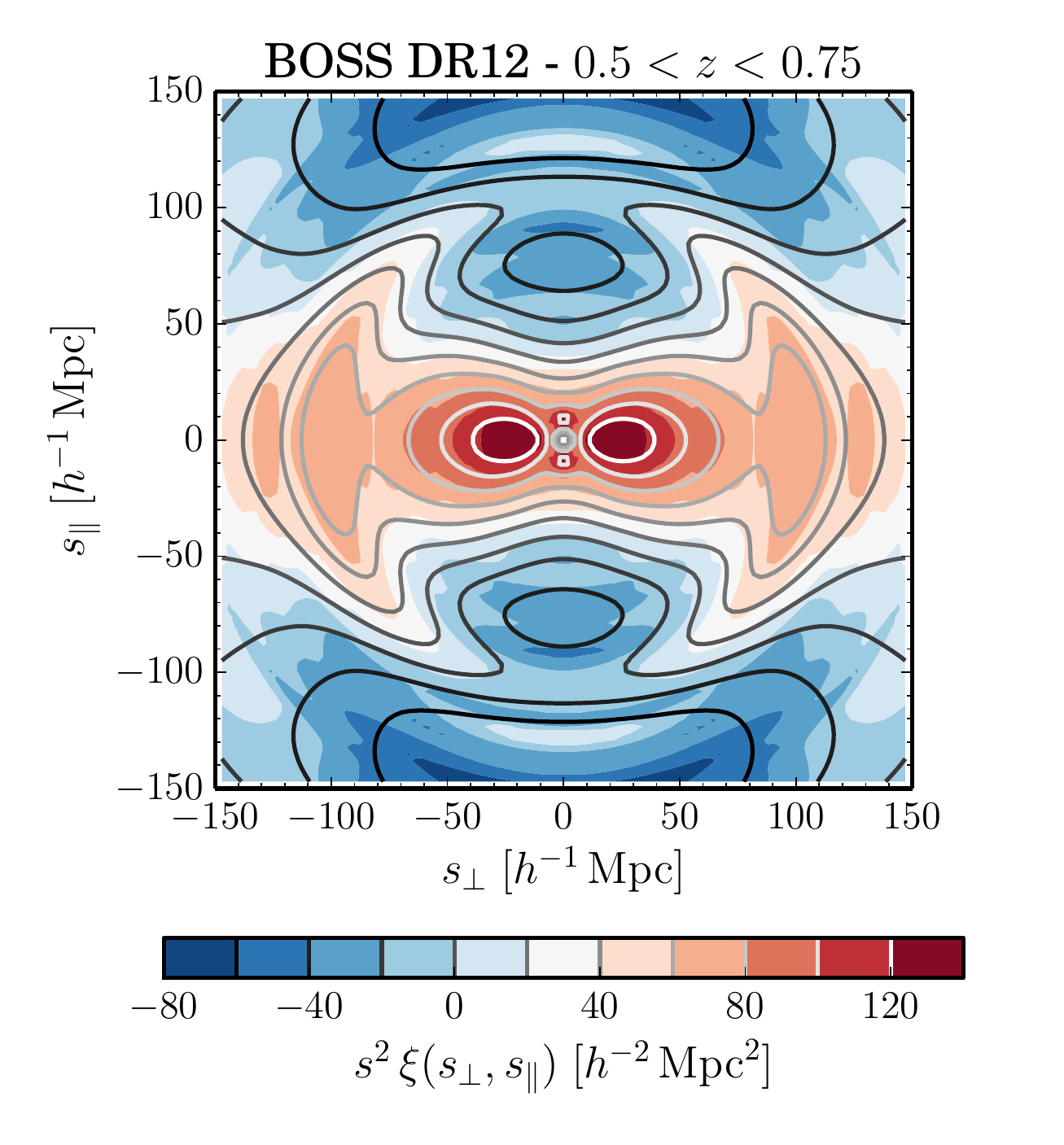}
 \caption{({\it Left}) Contour plots of the correlation function measured from the SDSS galaxy sample as functions of transverse ($s_\perp$) 
 and line-of-sight ($s_\parallel$) separations ({\it right half}) and the corresponding theoretical prediction ({\it left half}) \citep[reproduced with permission from][]{Okumura:2008}. 
The dashed thin black lines show $\xi<-0.01$ increasing logarithmically with 0.25 and $-0.01\leq \xi <0$ linearly with 0.0025. 
The solid thin lines colored red show $0\leq \xi <0.01$ increasing linearly with 0.0025 and the solid thick ones colored red are $\xi \geq 0.01$ logarithmically with 0.25. 
The baryonic feature slightly appears as ridge structures around the scale 
$s=(s_\perp^2 + s_\parallel^2)^{1/2} \simeq 100\himpc $ ( $= 150$ Mpc), 
and the dashed circle traces the peaks of the baryon ridges. 
({\it Right}) Similar to the left panel, but the correlation function from currently the largest galaxy sample from BOSS survey \citep[reproduced with permission from][]{Alam:2016a}. 
The correlation function is multiplied by the square of the distance, $s^2\xi$, in order to emphasize the BAO feature.
 \label{fig:2d_bao}}
\end{center}
\end{figure}

By simultaneously measuring the BAO and RSD and combining them with the CMB anisotropy power spectrum, we can obtain a further constraint on additional geometric quantities, such as the time evolution of $w$, the neutrino mass $m_\nu$, etc. 
The cosmological results, presented below, correspond to this case.

The correlation function in 2D space including the BAO scales has been measured by \cite{Okumura:2008} for the first time using the same galaxy sample as \cite{Eisenstein:2005}. 
The right-hand side of the left panel in Fig.~\ref{fig:2d_bao} shows the measured correlation function, while the left-hand side shows the best fitting model based on linear perturbation theory \citep{Matsubara:2004}.
The circle shown at the scale $s=(s_\perp^2 + s_\parallel^2)^{1/2}\simeq 105\himpc$ again corresponds to the sound horizon at recombination. The distorted, anisotropic contours shown at the smaller scales are the RSD effect caused by the velocity field.

The right panel of Fig.~\ref{fig:2d_bao} is the latest measurement of the correlation function using the SDSS-III Baryon Oscillation Spectroscopic Survey (BOSS) DR12 sample \citep{Alam:2016a}. The anisotropic feature of BAO is more clearly detected due to the improvement of the data, both in the number of galaxies and the survey volume:  the data of the DR12 sample used in \citet{Alam:2016a} comprised 1.2 million galaxies over the volume of $18.7 {\rm Gpc}^3 $, whose numbers are respectively 25 times and 9 times larger than those of the DR3 sample used in \citet{Okumura:2008}.

\begin{figure}[tb]
\begin{center}
\includegraphics[width=65mm]{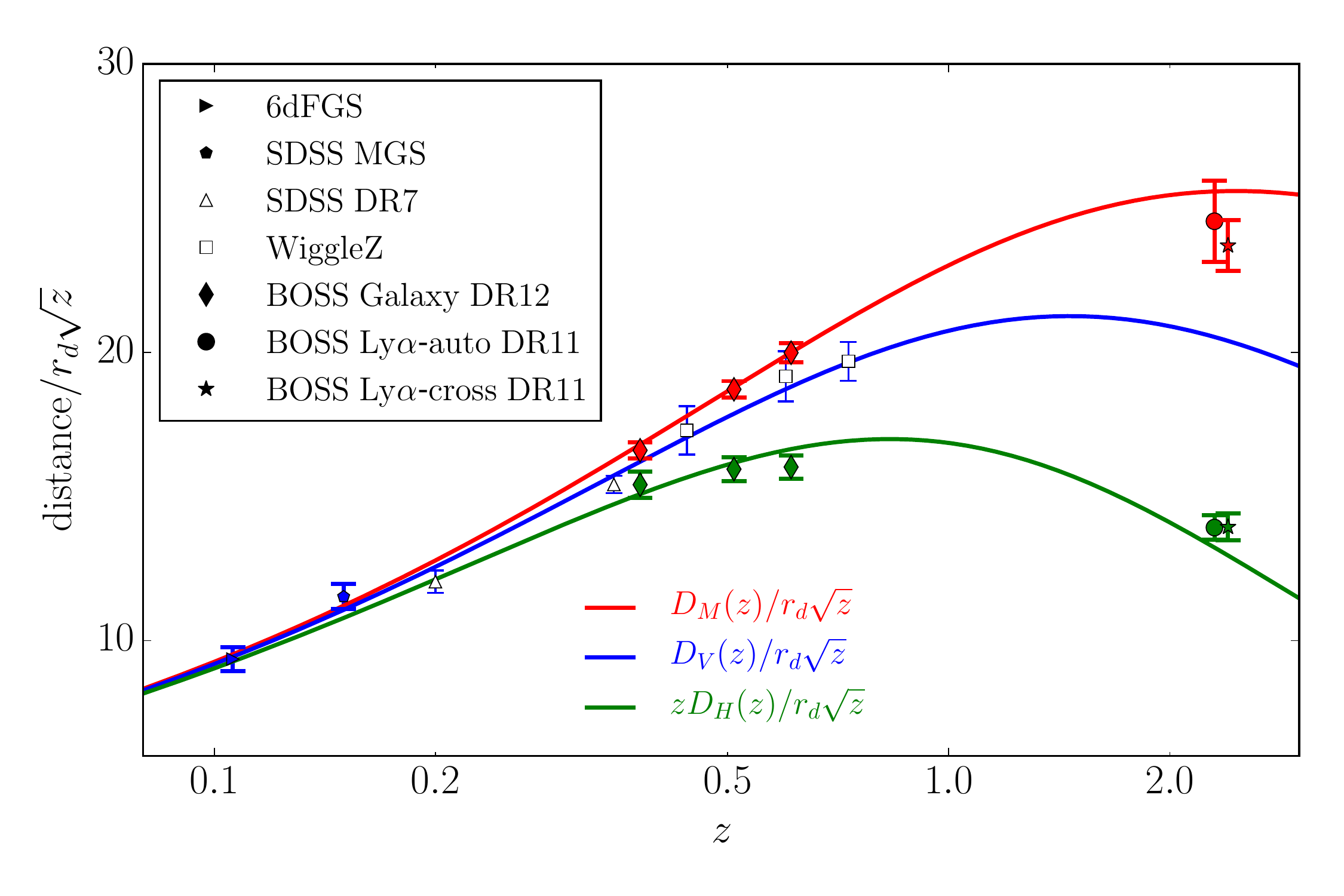}
\includegraphics[width=51mm]{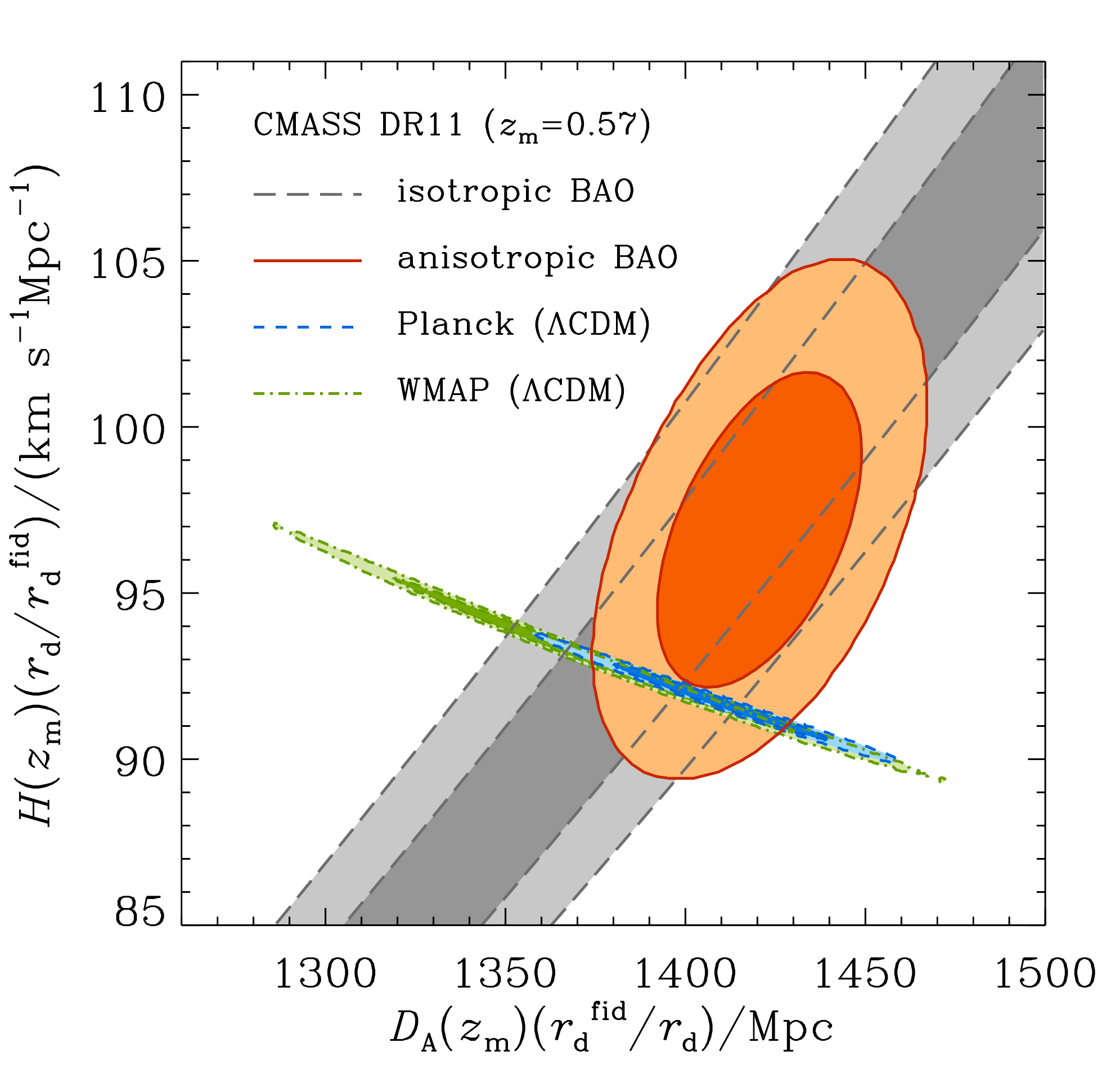}
 \caption{({\it Left}) Three distance measures obtained by BAO in various galaxy surveys \citep[reproduced with permission from][]{Aubourg:2015}. 
The $y$-axis is the ratio of each distance to $r_{\rm d}$, divided by $\sqrt{z}$. The top-red, middle-blue and bottom-green curves correspond to the distances $D_{\rm M}(z)$, $D_{\rm V}(z)$ and $zD_{\rm H}(z)$, respectively.
({\it Right}) Joint constraints on the angular diameter distance $D_{\rm A}(z)$ and the Hubble parameter $H(z)$ obtained from the 
correlation analysis of the BOSS galaxy sample at $z=0.5$ \citep[reproduced with permission from][]{Anderson:2014}. 
The inner and outer contours correspond to the 68\% and 95\% confidence levels, respectively.
The gray and orange contours are the constraints from 1D and 2D BAO analyses, respectively, while 
the blue and green contours are from CMB experiments (Planck and WMAP). 
 \label{fig:distance_acacia}}
\end{center}
\end{figure}


\subsection{Constraints on BAO distance scales and $H_0$}

The left panel of Fig.~\ref{fig:distance_acacia} shows the summary of the three distance measures obtained by BAO in various galaxy surveys \citep{Aubourg:2015}.  
The $y$-axis is the ratio of each distance and $r_{\rm d}$, divided by $\sqrt{z}$.
The blue points are the measurement of $D_{\rm V}$ from the angularly-averaged BAO, while the red and green points are respectively the measurements of $D_{\rm M}$ and $D_{\rm H}$ obtained from the anisotropy of BAO, where $D_{\rm H}$ is the radial distance defined as $D_{\rm H}(z)=c/H(z)$. The three lines with the same color as the points are the corresponding predictions of the $\Lambda$CDM model obtained by Planck \citep{Planck-Collaboration:2016a}.
Nice agreement between BAO measurements from galaxy surveys and Planck cosmology can be seen. 
However, the agreement with the WMAP cosmology is equivalently good in terms of distance measures
\citep{Anderson:2014}.

The right panel of Fig.~\ref{fig:distance_acacia} focuses on currently the largest survey, the BOSS survey at $z=0.5$ \citep{Anderson:2014}. Here the joint constraints on $H(z)$ and $D_{\rm A}(z)$ are shown.
The gray contours are obtained from the 1D BAO analysis (see section \ref{sec:1d_bao}). 
Because the 1D BAO constrains $D_{\rm V}\propto D_{\rm A}^{2/3}H^{-1/3}$, there is a perfect degeneracy between $D_{\rm A}$ and $H$. 
On the other hand, the solid orange contours are from the 2D BAO analysis where the degeneracy is broken to some extent (Section \ref{sec:2d_bao}). The obtained constraints on the distance scales are as tight as the flat $\lambda$CDM constraints from the CMB experiments, Planck (dashed blue) and WMAP (dot-dashed green). 
Future galaxy surveys will enable us to measure BAO more accurately and determine the cosmic distance scales with higher precision (see Section \ref{sec:conclusion}). 

\begin{figure*}[tb]
\begin{center}
\includegraphics[width=60mm]{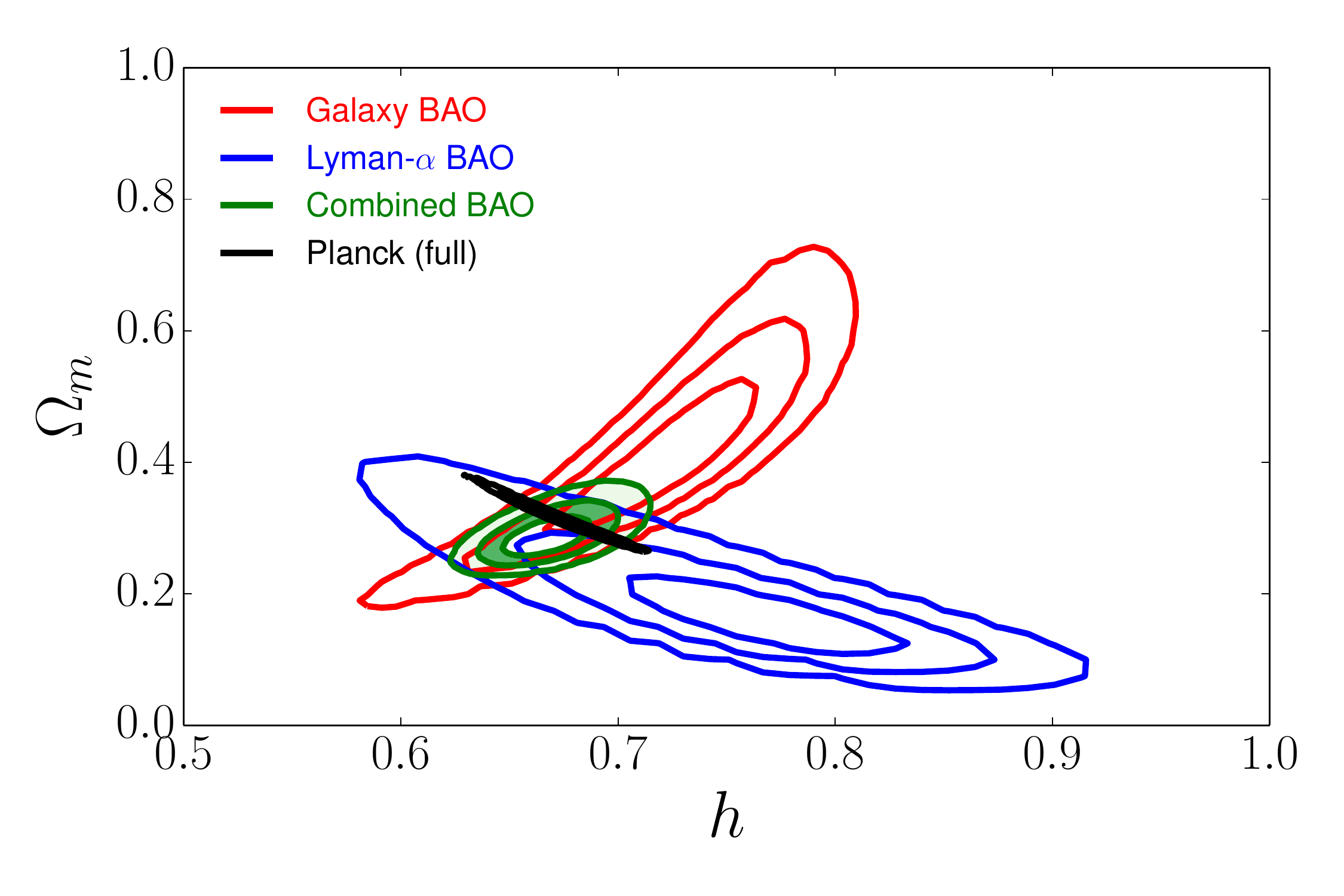}
\includegraphics[width=56mm]{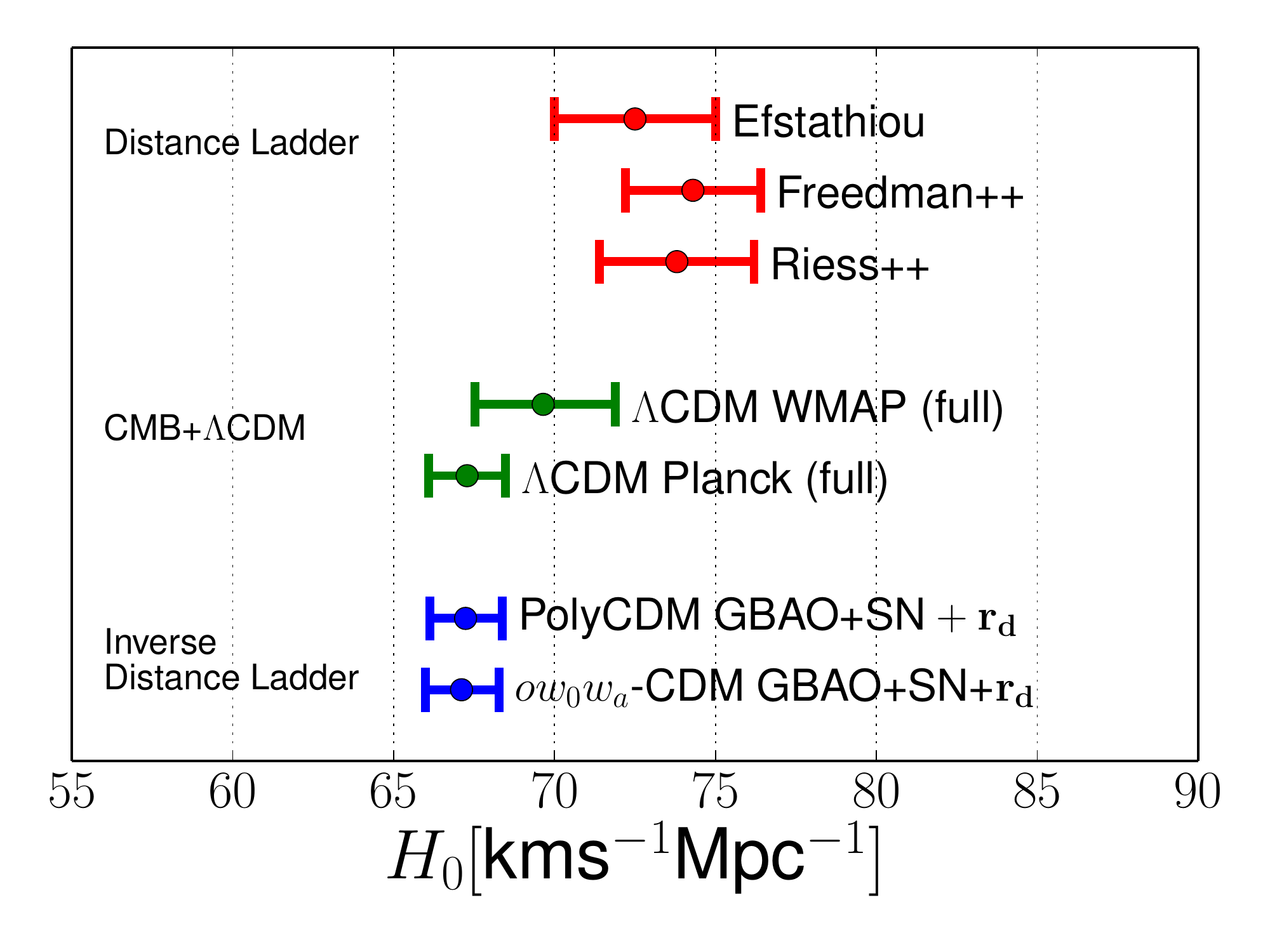}
 \caption{
({\it Left}) Joint constraints on the matter density parameter $\Omega_{\rm m}$ and the Hubble constant $h$ in a flat cosmology
 \citep[reproduced with permission from][]{Aubourg:2015}.
 Each contour shows the 68\%, 95\% and 99\% confidence levels from inward.
Galaxy BAO constraints (red) show strong correlations between $\Omega_{\rm m}$ and $h$, whereas that of Ly-$\alpha$ BAO (blue) show strong anti-correlations.  The combination of the two (``Combined BAO'' in green) thus breaks the degeneracies, resulting in constraints located at the intersection of the two.  Planck CMB constraints (black) show also anti-correlation between $\Omega_{\rm m}$ and $h$, but are substantially narrower than that of Combined BAO.
 ({\it Right}) Comparison of the constraints on $H_0$ \citep[reproduced with permission from][]{Aubourg:2015} from the distance ladder probes (local measurements, red), the CMB anisotropies (green), and the inverse distance ladder analysis (combination of BAO and supernovae; blue).
 \label{fig:hubble}}
\end{center}
\end{figure*}

Let us move onto the constraints on cosmological models using BAO observations. 
The left panel of Fig.~\ref{fig:hubble} shows the joint constraints on the matter density parameter $\Omega_{\rm m}$ and the Hubble constant $h=H_0/(100\,{\rm km\,s^{-1}\,Mpc^{-1}})$ 
obtained from the measurements of BAO anisotropy \citep{Aubourg:2015}. 
The red and blue contours are the constraints from the BAO measured from the various galaxy samples at $z<1$ and from the Ly$\alpha$ forest at $z\sim 2$, respectively, as shown in the left panel of Fig.~\ref{fig:distance_acacia}. Since these constraints are not very tight, the constrained $H_0$ from either galaxy BAO or Ly$\alpha$ BAO is consistent with other probes including the local measurements. The combination of these two BAO probes largely tightens the constraint on $H_0$ and causes a slight, $2\sigma$ tension as we will see below. 
With $\Omega_{\rm m}$ being marginalized over, the Hubble constant is constrained to $h=0.67\pm 0.013$ ($1\sigma$ C. L.) \citep{Aubourg:2015}. 

\begin{figure*}[tb]
\begin{center}
\includegraphics[width=100mm]{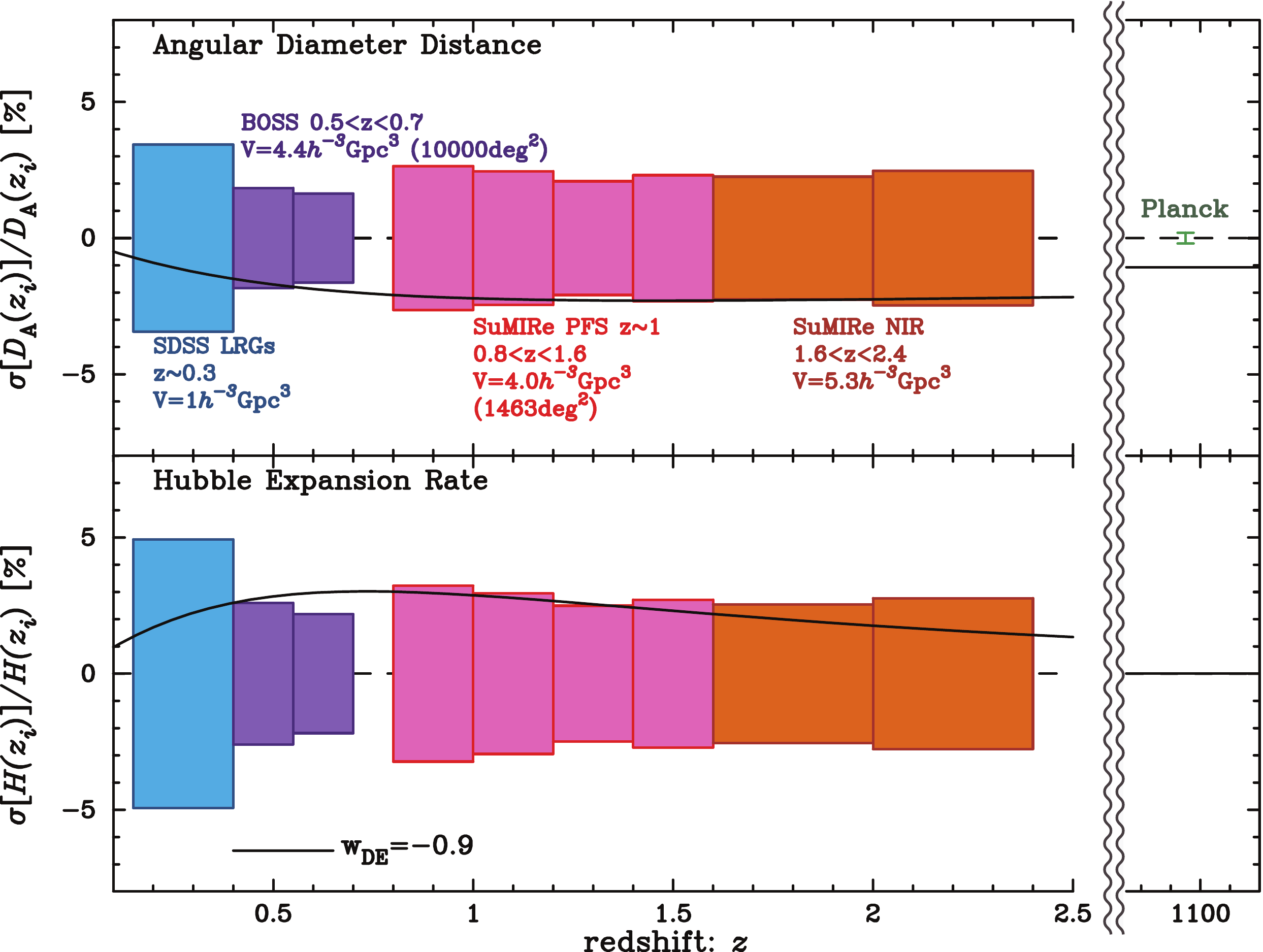}
 \caption{Fractional errors in the angular diameter distance and the expansion rate through the measurements of BAO \citep[reproduced with permission from][]{Takada:2014}. 
 The expected accuracies are compared to the existing SDSS and BOSS surveys at $z<0.7$. 
Each panel assumes $w=-1$ as the fiducial model, and
when the model is changed to $w=-0.9$, the baseline of the fractional errors is systematically shifted from the dashed line to the solid curve.
 \label{fig:da-h_future}}
\end{center}
\end{figure*}

The right panel of Fig.~\ref{fig:hubble} summarizes the comparison of $H_0$ constraints from the BAO measurement and other probes.
Augmenting Fig.~\ref{Fig:stateofart}, the top three points are obtained from the distance ladder analysis, showing constraints from the local universe from three independent teams \citep{Riess:2016, FreedmanEtal12, Efstathiou14}.
The green, two middle points are the constraints from the two CMB satellite probes, WMAP and Planck. 
The bottom two points are from the inverse distance ladder analysis, namely the combination of BAO and SN distance measures. 
As seen from the figure, those from the Planck and the inverse ladder measurements have a mild but non-negligible tension with the distance-ladder measurements, about $\sim 2\sigma$. 
The discrepancy could be due to just systematics, or a hint of new physics. 
We will need further observational constraints to resolve these discrepancies. 

\subsection{Future BAO Surveys}\label{sec:conclusion}
BAO are considered as a probe least affected by systematic biases to measure distance scales, even beyond the local universe ($z> 0$),
and thus are a promising tool to reveal the expansion history of the universe and constrain the cosmological model. 
To improve the precision of the distance measurement, a dominant source of error on BAO observations is the cosmic variance. 
There are larger, ongoing and planned galaxy redshift surveys, such as extended BOSS (eBOSS) \citep{Dawson:2016}, 
Subaru Prime Focus Spectrograph (PFS) \citep{Takada:2014}, and Dark Energy Spectroscopic Instrument (DESI) \citep{DESI-Collaboration:2016}. 
With the larger survey volumes, these surveys will measure distance scales using BAO with percent-level precisions.
These surveys are also deep and can reveal fainter sources, and hence enable us to extend the distance scales to more distant parts of the universe. 

As an example, Fig.~\ref{fig:da-h_future} presents the accuracies of constraining the angular diameter distance and Hubble expansion rate expected from the analysis of anisotropic BAO (see Section \ref{sec:2d_bao}) of the PFS survey at the Subaru Telescope.
The PFS will observe the universe at $0.8<z<2.4$ by using [OII] emitters as a tracer of the large-scale structure. 
The survey volume of each redshift slice is on the order of $[h^{-1}{\rm Gpc}]^3$ and the number density is larger than $ 10^{-4} [h^{-1}{\rm Mpc}]^{-3}$, which are comparable to the existing SDSS and BOSS surveys at $z<0.7$. 
Hence, one will be able to achieve a few percent constraints on $D_{\rm A}$ and $H$ at high redshifts, almost the same as those obtained from the low-$z$ surveys. Deep galaxy surveys such as the PFS allow for constraining not only the expansion history of the universe but also dark energy (see the solid line of Fig.~\ref{fig:da-h_future}).

Ultimately, we would like to survey the galaxies over the whole sky, which can be achieved by satellite missions,
such as Euclid \citep{Amendola:2013} and WFIRST \citep{Spergel:2013}.
These surveys will measure the cosmic distances with an unprecedentedly high precision. 





%
\section{Intensity Mapping}
\label{sec:im}


\subsection{21cm Intensity Mapping BAO}
\label{sec:im:subsectitle}

As we described in Section \ref{sec:bao}, current BAO measurements are enabled by large galaxy spectroscopic surveys, and the resulting constraining power on cosmological parameters generally scales as the effective survey comoving volume. 
Specifically, the precision of cosmological parameter constraints scales as
$\propto1/\sqrt{N}$, where $N$ is the number of modes, or in the case
of 3D map $\propto1/\sqrt{k_{\rm max}^3V}$ where $V$ is the comoving
volume and $k_{\rm max}$ the maximum useful comoving wavenumber. This scale is often given by the non-linearity scale, $k_{\rm
  nl}(z=0)\sim0.2\,h/{\rm Mpc}$, where the complexity of baryonic astrophysics on galaxy and galaxy-cluster scales 
limits our ability to extract cosmological parameters. Improving 
parametric precision will therefore require larger volumes, which
requires mapping higher redshift volumes that have the added benefit
of increasing $k_{\rm nl}(z)$.  The emerging technique of 21~cm Intensity Mapping appears to be a very promising way to reach this goal.
 
Galaxy redshifts can be measured at radio wavelengths using the 21~cm
hyperfine emission of atomic neutral hydrogen (HI). The 21~cm line is
unique in cosmology because for $\lambda>21\,$cm it is the dominant
astronomical line emission, i.e., for all positive redshifts and all
cosmological emission. So to a good approximation the wavelength of a
spectral feature can be converted to a Doppler redshift without the uncertainty and ambiguity of having 
to first identify the atomic transition. The direct determination of
redshifts using 21~cm data can be compared to the corresponding
optical technique, which requires identifying a suitable subset of
target galaxies (photometry), then obtaining an optical spectrum, and
finally finding some unique combination of emission and absorption
lines that allow an unambiguous determination of the redshift for that
galaxy (spectroscopy). 

The 21~cm signal has been used to conduct galaxy redshift surveys
in the local Universe around $z\sim 0.1$
\citep{2010ApJ...723.1359M,Zwaan2001} and out to $z\sim 0.4$ \citep{2016ApJ...824L...1F}.  Beyond this
redshift, current radio telescopes do not have sufficient collecting
area or sensitivity to make 21~cm surveys using individual galaxies. 
 
A radically different technique, HI
intensity mapping (HIM), has been proposed \citep{Chang08, WL08}. It uses maps of 21~cm emission where individual galaxies are
not resolved. Instead, it detects the combined emission from the many
galaxies that occupy large ($1000~{\rm Mpc}^3$) voxels. The technique
allows 100~m class telescopes such as the Green Bank Telescope (GBT), which only
have angular resolution of several arc-minutes, to rapidly survey
enormous comoving volumes at $z\sim1$
\citep{2005MNRAS.360...27A,2006astro.ph..6104P, 2007arXiv0709.2955W,
  Chang08, WL08, 2009PhRvD..79h3530T, 2010PhRvD..82j3501T, 2010ApJ...721..164S}.
A number of authors \citep{2010ApJ...721..164S,Xu:2014bya,Bull:2014rha} have studied the overall promise of the intensity mapping technique.
\citet{Chang2010, Masui2013} and \citet{Switzer13} have reported the first detections in cross-correlations and upper limits to the 21cm auto-power spectrum using the Green Bank Telescope (GBT).

\subsubsection{Challenges}
One of the major challenges for 21~cm intensity mapping is the mitigation of radio foregrounds, which are predominantly Galatic and extragalactic synchrotron emissions, and are at least $\sim10^4$ times brighter in intensity than the 21~cm emission.  The two can be distinguished because the former have smooth spectra and the latter trace the underlying large-scale structure and have spectral structures.  The brightness temperature of the synchrotron foreground typically has a spectral dependence of $\nu^{-2.6}$, or $(1+z)^{2.6}$, and is thus more severe at higher redshifts.  Note it has not yet been demonstrated whether the synchrotron radiation is indeed spectrally smooth down to one part of  $10^4$ or higher and therefore can in principle be suppressed by this factor to reveal the 21cm fluctuation signals.  However, since the BAO wiggles have very specific structures,
we can potentially select Fourier modes that are observationally accessible in scale and redshifts \citep{Chang08, 2010ApJ...721..164S}.

The 21~cm features we are most interested in are the relatively non-smooth BAO `wiggles'. Unfortunately radio telescopes are diffraction-limited and the beam patterns depend on frequency, which mixes angular and frequency dependence. Since the foregrounds are not smooth in position across the sky it is a nontrivial task to identify and subtract the smooth frequency foregrounds with sufficient accuracy so as to reveal the 21~cm emission. It is easier if 
we go to very small radial scales, but to get the most cosmological information out of the data 
we would need to remove the foregrounds over the largest range of scales possible.  \citet{Shaw:2013wza,Shaw:2014khi} have demonstrated that this is achievable in principle.

To achieve the foreground subtraction goal and to make accurate 3D
maps 
we need a very accurate model of the beam patterns and 
characterization of the mapping between the observed and true skies. \cite{Liao16}
have recently demonstrated accurate measurements of the polarized GBT
beam to sub-percent level, which is critical for
polarized foreground mitigation. Developing and demonstrating the
efficacy of methods to model and calibrate large dataset is also
necessary to achieve the main objective.   

\subsubsection{Future Prospects}

As discussed in Section \ref{sec:bao}, baryon acoustic oscillations provide a convenient standard ruler in the cosmological large scale structure (LSS) allowing a precise measurement of the distance-redshift relation over cosmic time.  This distance redshift relation is measured, whether by BAO, SNe-Ia surveys, or weak lensing, to characterize the dark energy; it is augmented by the growth rate of inhomogeneities 
as well as redshift-space distortions.  All three of these quantities can be measured in HI surveys even though to-date, only optical instruments have detected BAO features in the power spectrum. 

Going forward requires the most cost-effective way to map the largest cosmological volume, and this may be radio spectroscopy through intensity mapping.  One unique advantage of 21cm Intensity Mapping is the fact that the 21~cm signal is in principle observable from $z=0$ up to a redshift of $\sim100$, when its spin temperature decouples from the Cosmic Microwave Background radiation.  The vast majority of the cosmic volume is only visible during the dark ages via the 21~cm radiation from neutral hydrogen, before the onset of galaxy formation.  21~cm Intensity Mapping thus provides a unique access to measuring LSS during this period \citep{2009PhRvD..79h3530T, 2010PhRvD..82j3501T}. Besides, 21~cm intensity mapping has a set of observational systematics that should be largely uncorrelated with the systematic effects in optical
surveys. 

\begin{figure*}
\centering
\includegraphics[width=3.in]{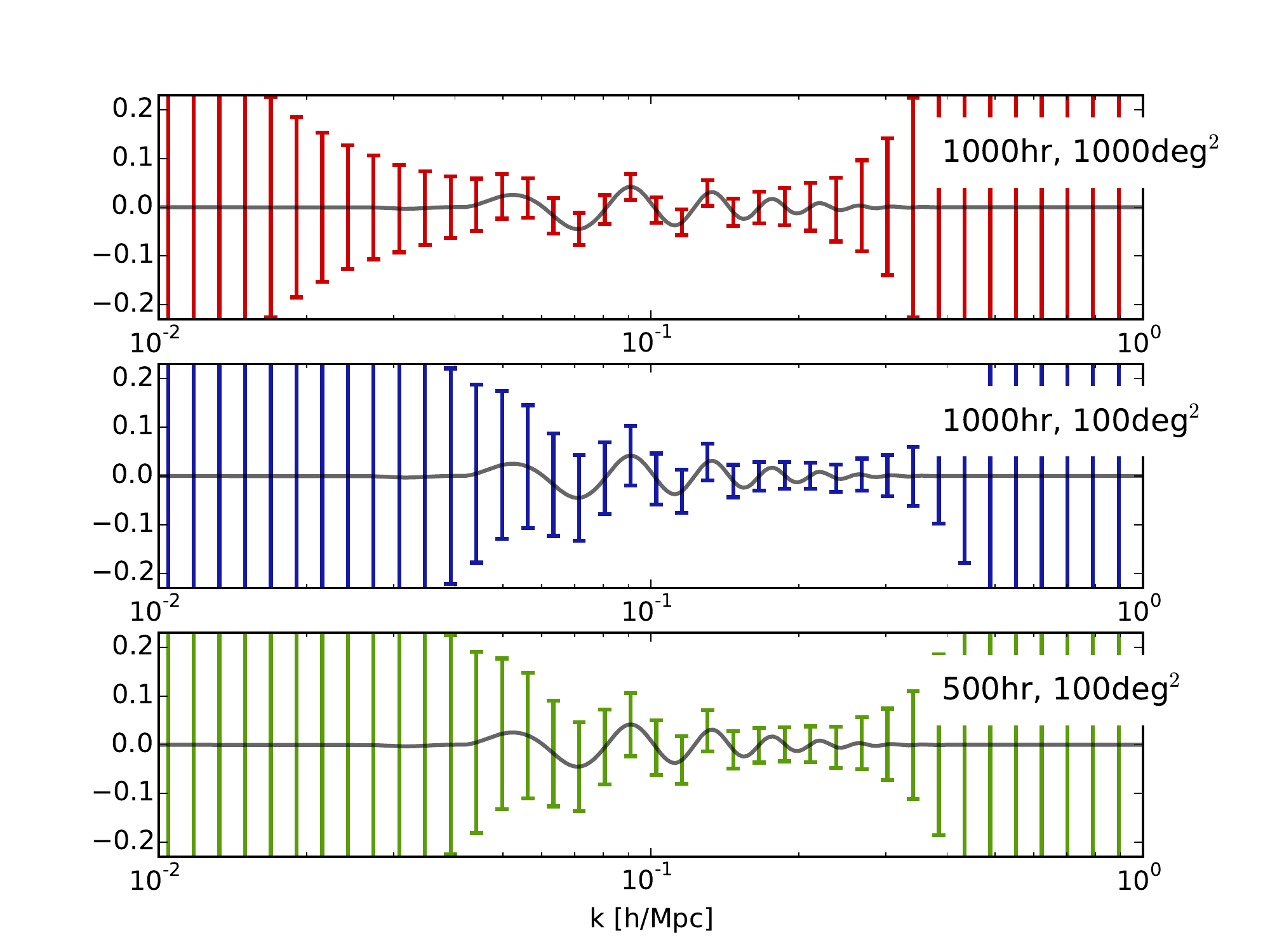}
\includegraphics[width=3.in]{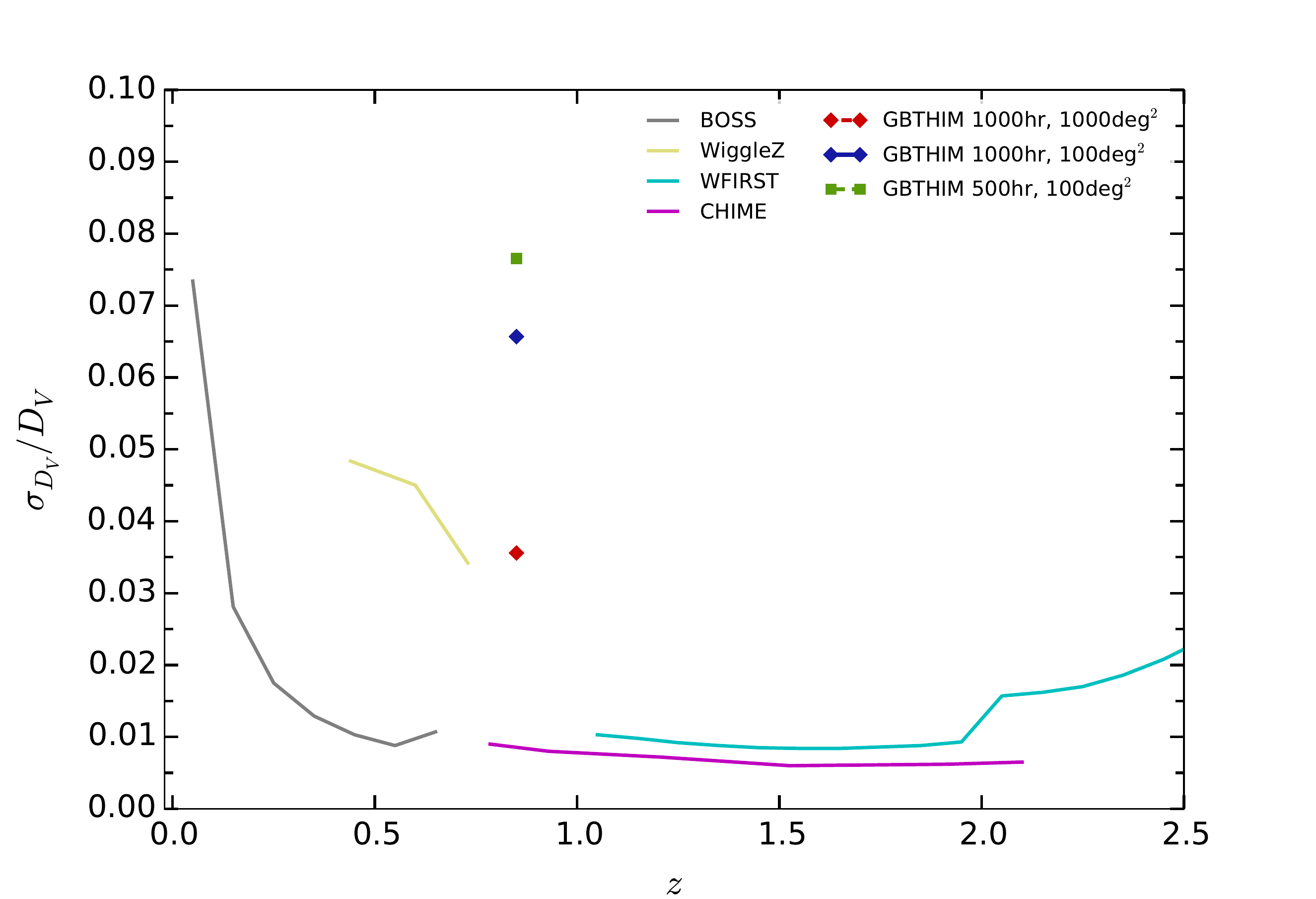}
\caption{\label{f:bao} 
{\it Top:} Expected errors on the BAO signature at $z\sim0.8$ of the GBT-HIM array, assuming different observing depth and sky coverage: (1000 hrs, 1000 deg$^2$), (1000 hrs, 100 deg$^2$), and (500 hrs, 100 deg$^2$). BAO signatures can be detected in all three. {\it Bottom:} The expected fractional error on the BAO distance scale of the three scenarios. We anticipate a 3.5\% error on the BAO distance at $z\sim0.8$ with 1000 hours of GBT observing time with the array. A detection of BAO will validate HI intensity mapping as a viable tool for large-scale structure and cosmology, and serve as a systematic check and alternative approach to the optical spectroscopic redshift surveys.} 
\end{figure*}

The on-going low-z GBT-HIM survey is a step along the way to a dark ages radio survey. We have made BAO forecasts based on the expected performance of the array. We assume the seven-beam array has a 700-850 MHz frequency coverage with a total system temperature of 33K.  The BAO forecasts are consistent with predictions in \cite{Masui2010} and are in reasonable agreement with those of \cite{Bull:2014rha}.  We consider three scenarios with different observing depth and sky coverage: 500 or 1000 hours of on-sky GBT observations, covering 100 or 1000 deg$^2$ of sky areas. The expected errors on the BAO wiggles and the fractional distance constraints are shown in Fig.~\ref{f:bao}.  We anticipate to yield a 3.5\% error on the BAO distance at $z\sim0.8$ with 1000 hours of GBT observing time. The bottom panel of Fig.~\ref{f:bao} also shows recent constraints from WiggleZ \citep{WiggleZ} and the BOSS surveys \citep{Anderson:2014} at lower redshifts, and forecasts for CHIME \citep{CHIME} and WFIRST \citep{wfirst}.  With the demonstrated results and good understanding of systematic effects at the GBT, and with very different astrophysical and measurement systematics from optical/IR spectroscopic redshift surveys, we anticipate the GBT-HIM array can make a firm detection of the BAO signature at z$\sim$0.8 with the HI intensity mapping technique, and contribute to the future of large-scale structure surveys and the field of 21-cm cosmology. Other on-going experiments such as CHIME \citep{CHIME} and HIRAX \citep{HIRAX} will reach $z=0.8-2.5$ and probe even larger cosmological volume.

%
\begin{acknowledgement}
Open access funding provided by the Max Planck Society.  We thank ISSI-BJ for the kind hospitality and an engaging workshop.  We would like to acknowledge Vivien Bonvin for providing some of the figures in this review.  T.O.~thanks Anze Slosar, Shun Saito and Atsushi Taruya for useful discussions on interpretations of cosmological constraints from the BOSS survey.
S.H.S.~thanks the Max Planck Society for support through the Max Planck Research Group.  
F.C.~acknowledges support from the Swiss National Science Foundation (SNSF).
T.O.~acknowledges support from the Ministry of Science and Technology of Taiwan under the grant MOST 106-2119-M-001-031-MY3.

\end{acknowledgement}
%


\bibliography{cosmodist}

\begin{thebibliography}{189}
\providecommand{\natexlab}[1]{#1}
\providecommand{\url}[1]{{#1}}
\providecommand{\urlprefix}{URL }
\expandafter\ifx\csname urlstyle\endcsname\relax
  \providecommand{\doi}[1]{DOI~\discretionary{}{}{}#1}\else
  \providecommand{\doi}{DOI~\discretionary{}{}{}\begingroup
  \urlstyle{rm}\Url}\fi
\providecommand{\eprint}[2][]{\url{#2}}

\bibitem[{{Abdalla} and {Rawlings}(2005)}]{2005MNRAS.360...27A}
{Abdalla} FB, {Rawlings} S (2005) {Probing dark energy with baryonic
  oscillations and future radio surveys of neutral hydrogen}. MNRAS 360:27--40,
  \doi{10.1111/j.1365-2966.2005.08650.x}, \eprint{arXiv:astro-ph/0411342}

\bibitem[{{Agnello} et~al(2017){Agnello}, {Lin}, {Buckley-Geer}, {Treu},
  {Bonvin}, {Courbin}, {Lemon}, {Morishita}, {Amara}, {Auger}, {Birrer},
  {Chan}, {Collett}, {More}, {Fassnacht}, {Frieman}, {Marshall}, {McMahon},
  {Meylan}, {Suyu}, {Castander}, {Finley}, {Howell}, {Kochanek}, {Makler},
  {Martini}, {Morgan}, {Nord}, {Ostrovski}, {Schechter}, {Tucker}, {Wechsler},
  {Abbott}, {Abdalla}, {Allam}, {Benoit-L{\'e}vy}, {Bertin}, {Brooks}, {Burke},
  {Rosell}, {Kind}, {Carretero}, {Crocce}, {Cunha}, {D'Andrea}, {da Costa},
  {Desai}, {Dietrich}, {Eifler}, {Flaugher}, {Fosalba},
  {Garc{\'{\i}}a-Bellido}, {Gaztanaga}, {Gill}, {Goldstein}, {Gruen},
  {Gruendl}, {Gschwend}, {Gutierrez}, {Honscheid}, {James}, {Kuehn},
  {Kuropatkin}, {Li}, {Lima}, {Maia}, {March}, {Marshall}, {Melchior},
  {Menanteau}, {Miquel}, {Ogando}, {Plazas}, {Romer}, {Sanchez}, {Schindler},
  {Schubnell}, {Sevilla-Noarbe}, {Smith}, {Smith}, {Sobreira}, {Suchyta},
  {Swanson}, {Tarle}, {Thomas}, and {Walker}}]{Agnello2017}
{Agnello} A, {Lin} H, {Buckley-Geer} L, {Treu} T, {Bonvin} V, {Courbin} F,
  {Lemon} C, {Morishita} T, {Amara} A, {Auger} MW, {Birrer} S, {Chan} J,
  {Collett} T, {More} A, {Fassnacht} CD, {Frieman} J, {Marshall} PJ, {McMahon}
  RG, {Meylan} G, {Suyu} SH, {Castander} F, {Finley} D, {Howell} A, {Kochanek}
  C, {Makler} M, {Martini} P, {Morgan} N, {Nord} B, {Ostrovski} F, {Schechter}
  P, {Tucker} D, {Wechsler} R, {Abbott} TMC, {Abdalla} FB, {Allam} S,
  {Benoit-L{\'e}vy} A, {Bertin} E, {Brooks} D, {Burke} DL, {Rosell} AC, {Kind}
  MC, {Carretero} J, {Crocce} M, {Cunha} CE, {D'Andrea} CB, {da Costa} LN,
  {Desai} S, {Dietrich} JP, {Eifler} TF, {Flaugher} B, {Fosalba} P,
  {Garc{\'{\i}}a-Bellido} J, {Gaztanaga} E, {Gill} MS, {Goldstein} DA, {Gruen}
  D, {Gruendl} RA, {Gschwend} J, {Gutierrez} G, {Honscheid} K, {James} DJ,
  {Kuehn} K, {Kuropatkin} N, {Li} TS, {Lima} M, {Maia} MAG, {March} M,
  {Marshall} JL, {Melchior} P, {Menanteau} F, {Miquel} R, {Ogando} RLC,
  {Plazas} AA, {Romer} AK, {Sanchez} E, {Schindler} R, {Schubnell} M,
  {Sevilla-Noarbe} I, {Smith} M, {Smith} RC, {Sobreira} F, {Suchyta} E,
  {Swanson} MEC, {Tarle} G, {Thomas} D, {Walker} AR (2017) {Models of the
  strongly lensed quasar DES J0408-5354}. \mnras 472:4038--4050,
  \doi{10.1093/mnras/stx2242}, \eprint{1702.00406}

\bibitem[{{Alam} et~al(2016){Alam}, {Ata}, {Bailey}, {Beutler}, {Bizyaev},
  {Blazek}, {Bolton}, {Brownstein}, {Burden}, {Chuang}, {Comparat}, {Cuesta},
  {Dawson}, {Eisenstein}, {Escoffier}, {Gil-Mar{\'{\i}}n}, {Grieb}, {Hand},
  {Ho}, {Kinemuchi}, {Kirkby}, {Kitaura}, {Malanushenko}, {Malanushenko},
  {Maraston}, {McBride}, {Nichol}, {Olmstead}, {Oravetz}, {Padmanabhan},
  {Palanque-Delabrouille}, {Pan}, {Pellejero-Ibanez}, {Percival}, {Petitjean},
  {Prada}, {Price-Whelan}, {Reid}, {Rodr{\'{\i}}guez-Torres}, {Roe}, {Ross},
  {Ross}, {Rossi}, {Rubi{\~n}o-Mart{\'{\i}}n}, {S{\'a}nchez}, {Saito},
  {Salazar-Albornoz}, {Samushia}, {Satpathy}, {Sc{\'o}ccola}, {Schlegel},
  {Schneider}, {Seo}, {Simmons}, {Slosar}, {Strauss}, {Swanson}, {Thomas},
  {Tinker}, {Tojeiro}, {Vargas Maga{\~n}a}, {Vazquez}, {Verde}, {Wake}, {Wang},
  {Weinberg}, {White}, {Wood-Vasey}, {Y{\`e}che}, {Zehavi}, {Zhai}, and
  {Zhao}}]{Alam:2016a}
{Alam} S, {Ata} M, {Bailey} S, {Beutler} F, {Bizyaev} D, {Blazek} JA, {Bolton}
  AS, {Brownstein} JR, {Burden} A, {Chuang} CH, {Comparat} J, {Cuesta} AJ,
  {Dawson} KS, {Eisenstein} DJ, {Escoffier} S, {Gil-Mar{\'{\i}}n} H, {Grieb}
  JN, {Hand} N, {Ho} S, {Kinemuchi} K, {Kirkby} D, {Kitaura} F, {Malanushenko}
  E, {Malanushenko} V, {Maraston} C, {McBride} CK, {Nichol} RC, {Olmstead} MD,
  {Oravetz} D, {Padmanabhan} N, {Palanque-Delabrouille} N, {Pan} K,
  {Pellejero-Ibanez} M, {Percival} WJ, {Petitjean} P, {Prada} F, {Price-Whelan}
  AM, {Reid} BA, {Rodr{\'{\i}}guez-Torres} SA, {Roe} NA, {Ross} AJ, {Ross} NP,
  {Rossi} G, {Rubi{\~n}o-Mart{\'{\i}}n} JA, {S{\'a}nchez} AG, {Saito} S,
  {Salazar-Albornoz} S, {Samushia} L, {Satpathy} S, {Sc{\'o}ccola} CG,
  {Schlegel} DJ, {Schneider} DP, {Seo} HJ, {Simmons} A, {Slosar} A, {Strauss}
  MA, {Swanson} MEC, {Thomas} D, {Tinker} JL, {Tojeiro} R, {Vargas Maga{\~n}a}
  M, {Vazquez} JA, {Verde} L, {Wake} DA, {Wang} Y, {Weinberg} DH, {White} M,
  {Wood-Vasey} WM, {Y{\`e}che} C, {Zehavi} I, {Zhai} Z, {Zhao} GB (2016) {The
  clustering of galaxies in the completed SDSS-III Baryon Oscillation
  Spectroscopic Survey: cosmological analysis of the DR12 galaxy sample}. ArXiv
  e-prints \eprint{1607.03155}

\bibitem[{{Alcock} and {Paczynski}(1979)}]{Alcock:1979}
{Alcock} C, {Paczynski} B (1979) {An evolution free test for non-zero
  cosmological constant}. \nat 281:358, \doi{10.1038/281358a0}

\bibitem[{{Amendola} et~al(2013){Amendola}, {Appleby}, {Bacon}, {Baker},
  {Baldi}, {Bartolo}, {Blanchard}, {Bonvin}, {Borgani}, {Branchini}, {Burrage},
  {Camera}, {Carbone}, {Casarini}, {Cropper}, {de Rham}, {Di Porto}, {Ealet},
  {Ferreira}, {Finelli}, {Garc{\'{\i}}a-Bellido}, {Giannantonio}, {Guzzo},
  {Heavens}, {Heisenberg}, {Heymans}, {Hoekstra}, {Hollenstein}, {Holmes},
  {Horst}, {Jahnke}, {Kitching}, {Koivisto}, {Kunz}, {La Vacca}, {March},
  {Majerotto}, {Markovic}, {Marsh}, {Marulli}, {Massey}, {Mellier}, {Mota},
  {Nunes}, {Percival}, {Pettorino}, {Porciani}, {Quercellini}, {Read},
  {Rinaldi}, {Sapone}, {Scaramella}, {Skordis}, {Simpson}, {Taylor}, {Thomas},
  {Trotta}, {Verde}, {Vernizzi}, {Vollmer}, {Wang}, {Weller}, and
  {Zlosnik}}]{Amendola:2013}
{Amendola} L, {Appleby} S, {Bacon} D, {Baker} T, {Baldi} M, {Bartolo} N,
  {Blanchard} A, {Bonvin} C, {Borgani} S, {Branchini} E, {Burrage} C, {Camera}
  S, {Carbone} C, {Casarini} L, {Cropper} M, {de Rham} C, {Di Porto} C, {Ealet}
  A, {Ferreira} PG, {Finelli} F, {Garc{\'{\i}}a-Bellido} J, {Giannantonio} T,
  {Guzzo} L, {Heavens} A, {Heisenberg} L, {Heymans} C, {Hoekstra} H,
  {Hollenstein} L, {Holmes} R, {Horst} O, {Jahnke} K, {Kitching} TD, {Koivisto}
  T, {Kunz} M, {La Vacca} G, {March} M, {Majerotto} E, {Markovic} K, {Marsh} D,
  {Marulli} F, {Massey} R, {Mellier} Y, {Mota} DF, {Nunes} NJ, {Percival} W,
  {Pettorino} V, {Porciani} C, {Quercellini} C, {Read} J, {Rinaldi} M, {Sapone}
  D, {Scaramella} R, {Skordis} C, {Simpson} F, {Taylor} A, {Thomas} S, {Trotta}
  R, {Verde} L, {Vernizzi} F, {Vollmer} A, {Wang} Y, {Weller} J, {Zlosnik} T
  (2013) {Cosmology and Fundamental Physics with the Euclid Satellite}. Living
  Reviews in Relativity 16:6, \doi{10.12942/lrr-2013-6}, \eprint{1206.1225}

\bibitem[{{Anderson} et~al(2014){Anderson}, {Aubourg}, {Bailey}, {Beutler},
  {Bhardwaj}, {Blanton}, {Bolton}, {Brinkmann}, {Brownstein}, {Burden},
  {Chuang}, {Cuesta}, {Dawson}, {Eisenstein}, {Escoffier}, {Gunn}, {Guo}, {Ho},
  {Honscheid}, {Howlett}, {Kirkby}, {Lupton}, {Manera}, {Maraston}, {McBride},
  {Mena}, {Montesano}, {Nichol}, {Nuza}, {Olmstead}, {Padmanabhan},
  {Palanque-Delabrouille}, {Parejko}, {Percival}, {Petitjean}, {Prada},
  {Price-Whelan}, {Reid}, {Roe}, {Ross}, {Ross}, {Sabiu}, {Saito}, {Samushia},
  {S{\'a}nchez}, {Schlegel}, {Schneider}, {Scoccola}, {Seo}, {Skibba},
  {Strauss}, {Swanson}, {Thomas}, {Tinker}, {Tojeiro}, {Maga{\~n}a}, {Verde},
  {Wake}, {Weaver}, {Weinberg}, {White}, {Xu}, {Y{\`e}che}, {Zehavi}, and
  {Zhao}}]{Anderson:2014}
{Anderson} L, {Aubourg} {\'E}, {Bailey} S, {Beutler} F, {Bhardwaj} V, {Blanton}
  M, {Bolton} AS, {Brinkmann} J, {Brownstein} JR, {Burden} A, {Chuang} CH,
  {Cuesta} AJ, {Dawson} KS, {Eisenstein} DJ, {Escoffier} S, {Gunn} JE, {Guo} H,
  {Ho} S, {Honscheid} K, {Howlett} C, {Kirkby} D, {Lupton} RH, {Manera} M,
  {Maraston} C, {McBride} CK, {Mena} O, {Montesano} F, {Nichol} RC, {Nuza} SE,
  {Olmstead} MD, {Padmanabhan} N, {Palanque-Delabrouille} N, {Parejko} J,
  {Percival} WJ, {Petitjean} P, {Prada} F, {Price-Whelan} AM, {Reid} B, {Roe}
  NA, {Ross} AJ, {Ross} NP, {Sabiu} CG, {Saito} S, {Samushia} L, {S{\'a}nchez}
  AG, {Schlegel} DJ, {Schneider} DP, {Scoccola} CG, {Seo} HJ, {Skibba} RA,
  {Strauss} MA, {Swanson} MEC, {Thomas} D, {Tinker} JL, {Tojeiro} R,
  {Maga{\~n}a} MV, {Verde} L, {Wake} DA, {Weaver} BA, {Weinberg} DH, {White} M,
  {Xu} X, {Y{\`e}che} C, {Zehavi} I, {Zhao} GB (2014) {The clustering of
  galaxies in the SDSS-III Baryon Oscillation Spectroscopic Survey: baryon
  acoustic oscillations in the Data Releases 10 and 11 Galaxy samples}. \mnras
  441:24--62, \doi{10.1093/mnras/stu523}, \eprint{1312.4877}

\bibitem[{{Aubourg} et~al(2015){Aubourg}, {Bailey}, {Bautista}, {Beutler},
  {Bhardwaj}, {Bizyaev}, {Blanton}, {Blomqvist}, {Bolton}, {Bovy},
  {Brewington}, {Brinkmann}, {Brownstein}, {Burden}, {Busca}, {Carithers},
  {Chuang}, {Comparat}, {Croft}, {Cuesta}, {Dawson}, {Delubac}, {Eisenstein},
  {Font-Ribera}, {Ge}, {Le Goff}, {Gontcho}, {Gott}, {Gunn}, {Guo}, {Guy},
  {Hamilton}, {Ho}, {Honscheid}, {Howlett}, {Kirkby}, {Kitaura}, {Kneib},
  {Lee}, {Long}, {Lupton}, {Maga{\~n}a}, {Malanushenko}, {Malanushenko},
  {Manera}, {Maraston}, {Margala}, {McBride}, {Miralda-Escud{\'e}}, {Myers},
  {Nichol}, {Noterdaeme}, {Nuza}, {Olmstead}, {Oravetz}, {P{\^a}ris},
  {Padmanabhan}, {Palanque-Delabrouille}, {Pan}, {Pellejero-Ibanez},
  {Percival}, {Petitjean}, {Pieri}, {Prada}, {Reid}, {Rich}, {Roe}, {Ross},
  {Ross}, {Rossi}, {Rubi{\~n}o-Mart{\'{\i}}n}, {S{\'a}nchez}, {Samushia},
  {Santos}, {Sc{\'o}ccola}, {Schlegel}, {Schneider}, {Seo}, {Sheldon},
  {Simmons}, {Skibba}, {Slosar}, {Strauss}, {Thomas}, {Tinker}, {Tojeiro},
  {Vazquez}, {Viel}, {Wake}, {Weaver}, {Weinberg}, {Wood-Vasey}, {Y{\`e}che},
  {Zehavi}, {Zhao}, and {BOSS Collaboration}}]{Aubourg:2015}
{Aubourg} {\'E}, {Bailey} S, {Bautista} JE, {Beutler} F, {Bhardwaj} V,
  {Bizyaev} D, {Blanton} M, {Blomqvist} M, {Bolton} AS, {Bovy} J, {Brewington}
  H, {Brinkmann} J, {Brownstein} JR, {Burden} A, {Busca} NG, {Carithers} W,
  {Chuang} CH, {Comparat} J, {Croft} RAC, {Cuesta} AJ, {Dawson} KS, {Delubac}
  T, {Eisenstein} DJ, {Font-Ribera} A, {Ge} J, {Le Goff} JM, {Gontcho} SGA,
  {Gott} JR, {Gunn} JE, {Guo} H, {Guy} J, {Hamilton} JC, {Ho} S, {Honscheid} K,
  {Howlett} C, {Kirkby} D, {Kitaura} FS, {Kneib} JP, {Lee} KG, {Long} D,
  {Lupton} RH, {Maga{\~n}a} MV, {Malanushenko} V, {Malanushenko} E, {Manera} M,
  {Maraston} C, {Margala} D, {McBride} CK, {Miralda-Escud{\'e}} J, {Myers} AD,
  {Nichol} RC, {Noterdaeme} P, {Nuza} SE, {Olmstead} MD, {Oravetz} D,
  {P{\^a}ris} I, {Padmanabhan} N, {Palanque-Delabrouille} N, {Pan} K,
  {Pellejero-Ibanez} M, {Percival} WJ, {Petitjean} P, {Pieri} MM, {Prada} F,
  {Reid} B, {Rich} J, {Roe} NA, {Ross} AJ, {Ross} NP, {Rossi} G,
  {Rubi{\~n}o-Mart{\'{\i}}n} JA, {S{\'a}nchez} AG, {Samushia} L, {Santos} RTG,
  {Sc{\'o}ccola} CG, {Schlegel} DJ, {Schneider} DP, {Seo} HJ, {Sheldon} E,
  {Simmons} A, {Skibba} RA, {Slosar} A, {Strauss} MA, {Thomas} D, {Tinker} JL,
  {Tojeiro} R, {Vazquez} JA, {Viel} M, {Wake} DA, {Weaver} BA, {Weinberg} DH,
  {Wood-Vasey} WM, {Y{\`e}che} C, {Zehavi} I, {Zhao} GB, {BOSS Collaboration}
  (2015) {Cosmological implications of baryon acoustic oscillation
  measurements}. \prd 92(12):123516, \doi{10.1103/PhysRevD.92.123516},
  \eprint{1411.1074}

\bibitem[{{Bandura} et~al(2014){Bandura}, {Addison}, {Amiri}, {Bond},
  {Campbell-Wilson}, {Connor}, {Cliche}, {Davis}, {Deng}, {Denman}, {Dobbs},
  {Fandino}, {Gibbs}, {Gilbert}, {Halpern}, {Hanna}, {Hincks}, {Hinshaw},
  {H{\"o}fer}, {Klages}, {Landecker}, {Masui}, {Mena Parra}, {Newburgh}, {Pen},
  {Peterson}, {Recnik}, {Shaw}, {Sigurdson}, {Sitwell}, {Smecher}, {Smegal},
  {Vanderlinde}, and {Wiebe}}]{CHIME}
{Bandura} K, {Addison} GE, {Amiri} M, {Bond} JR, {Campbell-Wilson} D, {Connor}
  L, {Cliche} JF, {Davis} G, {Deng} M, {Denman} N, {Dobbs} M, {Fandino} M,
  {Gibbs} K, {Gilbert} A, {Halpern} M, {Hanna} D, {Hincks} AD, {Hinshaw} G,
  {H{\"o}fer} C, {Klages} P, {Landecker} TL, {Masui} K, {Mena Parra} J,
  {Newburgh} LB, {Pen} Ul, {Peterson} JB, {Recnik} A, {Shaw} JR, {Sigurdson} K,
  {Sitwell} M, {Smecher} G, {Smegal} R, {Vanderlinde} K, {Wiebe} D (2014)
  {Canadian Hydrogen Intensity Mapping Experiment (CHIME) pathfinder}. In:
  Ground-based and Airborne Telescopes V, \procspie, vol 9145, p 914522,
  \doi{10.1117/12.2054950}, \eprint{1406.2288}

\bibitem[{{Barkana}(1998)}]{Barkana98}
{Barkana} R (1998) {Fast Calculation of a Family of Elliptical Mass
  Gravitational Lens Models}. \apj 502:531, \doi{10.1086/305950},
  \eprint{arXiv:astro-ph/9802002}

\bibitem[{{Barnab{\`e}} et~al(2009){Barnab{\`e}}, {Czoske}, {Koopmans}, {Treu},
  {Bolton}, and {Gavazzi}}]{BarnabeEtal09}
{Barnab{\`e}} M, {Czoske} O, {Koopmans} LVE, {Treu} T, {Bolton} AS, {Gavazzi} R
  (2009) {Two-dimensional kinematics of SLACS lenses - II. Combined lensing and
  dynamics analysis of early-type galaxies at z = 0.08-0.33}. \mnras
  399:21--36, \doi{10.1111/j.1365-2966.2009.14941.x}, \eprint{0904.3861}

\bibitem[{{Barnab{\`e}} et~al(2011){Barnab{\`e}}, {Czoske}, {Koopmans}, {Treu},
  and {Bolton}}]{BarnabeEtal11}
{Barnab{\`e}} M, {Czoske} O, {Koopmans} LVE, {Treu} T, {Bolton} AS (2011)
  {Two-dimensional kinematics of SLACS lenses - III. Mass structure and
  dynamics of early-type lens galaxies beyond z {$\sim$} 0.1}. \mnras
  415:2215--2232, \doi{10.1111/j.1365-2966.2011.18842.x}, \eprint{1102.2261}

\bibitem[{{Bassett} and {Hlozek}(2010)}]{Bassett:2010}
{Bassett} B, {Hlozek} R (2010) {Baryon acoustic oscillations}, p 246

\bibitem[{{Bennett} et~al(2013){Bennett}, {Larson}, {Weiland}, {Jarosik},
  {Hinshaw}, {Odegard}, {Smith}, {Hill}, {Gold}, {Halpern}, {Komatsu}, {Nolta},
  {Page}, {Spergel}, {Wollack}, {Dunkley}, {Kogut}, {Limon}, {Meyer}, {Tucker},
  and {Wright}}]{Bennett2013}
{Bennett} CL, {Larson} D, {Weiland} JL, {Jarosik} N, {Hinshaw} G, {Odegard} N,
  {Smith} KM, {Hill} RS, {Gold} B, {Halpern} M, {Komatsu} E, {Nolta} MR, {Page}
  L, {Spergel} DN, {Wollack} E, {Dunkley} J, {Kogut} A, {Limon} M, {Meyer} SS,
  {Tucker} GS, {Wright} EL (2013) {Nine-year Wilkinson Microwave Anisotropy
  Probe (WMAP) Observations: Final Maps and Results}. \apjs 208:20,
  \doi{10.1088/0067-0049/208/2/20}, \eprint{1212.5225}

\bibitem[{{Birrer} et~al(2015{\natexlab{a}}){Birrer}, {Amara}, and
  {Refregier}}]{BirrerEtal15}
{Birrer} S, {Amara} A, {Refregier} A (2015{\natexlab{a}}) {Gravitational Lens
  Modeling with Basis Sets}. \apj 813:102, \doi{10.1088/0004-637X/813/2/102},
  \eprint{1504.07629}

\bibitem[{{Birrer} et~al(2015{\natexlab{b}}){Birrer}, {Amara}, and
  {Refregier}}]{BirrerEtal16}
{Birrer} S, {Amara} A, {Refregier} A (2015{\natexlab{b}}) {The mass-sheet
  degeneracy and time-delay cosmography: Analysis of the strong lens
  RXJ1131-1231}. ArXiv e-prints (151103662) \eprint{1511.03662}

\bibitem[{{Blandford} and {Narayan}(1986)}]{BlandfordNarayan86}
{Blandford} R, {Narayan} R (1986) {Fermat's principle, caustics, and the
  classification of gravitatio nal lens images}. \apj 310:568--582,
  \doi{10.1086/164709}

\bibitem[{{Blandford} et~al(2001){Blandford}, {Surpi}, and
  {Kundi{\'c}}}]{BlandfordEtal01}
{Blandford} R, {Surpi} G, {Kundi{\'c}} T (2001) {Modeling Galaxy Lenses}. In:
  {Brainerd} TG, {Kochanek} CS (eds) {ASP Conf. Ser. 237: Gravitational
  Lensing: Recent Progress and Future Goals}, San Francisco: Astron. Soc. Pac.,
  p~65

\bibitem[{{Bonvin} et~al(2016){Bonvin}, {Tewes}, {Courbin}, {Kuntzer}, {Sluse},
  and {Meylan}}]{Bonvin2016}
{Bonvin} V, {Tewes} M, {Courbin} F, {Kuntzer} T, {Sluse} D, {Meylan} G (2016)
  {COSMOGRAIL: the COSmological MOnitoring of GRAvItational Lenses. XV.
  Assessing the achievability and precision of time-delay measurements}. \aap
  585:A88, \doi{10.1051/0004-6361/201526704}, \eprint{1506.07524}

\bibitem[{{Bonvin} et~al(2017){Bonvin}, {Courbin}, {Suyu}, {Marshall}, {Rusu},
  {Sluse}, {Tewes}, {Wong}, {Collett}, {Fassnacht}, {Treu}, {Auger}, {Hilbert},
  {Koopmans}, {Meylan}, {Rumbaugh}, {Sonnenfeld}, and {Spiniello}}]{Bonvin2017}
{Bonvin} V, {Courbin} F, {Suyu} SH, {Marshall} PJ, {Rusu} CE, {Sluse} D,
  {Tewes} M, {Wong} KC, {Collett} T, {Fassnacht} CD, {Treu} T, {Auger} MW,
  {Hilbert} S, {Koopmans} LVE, {Meylan} G, {Rumbaugh} N, {Sonnenfeld} A,
  {Spiniello} C (2017) {H0LiCOW - V. New COSMOGRAIL time delays of HE
  0435-1223: H$_{0}$ to 3.8 per cent precision from strong lensing in a flat
  {$\Lambda$}CDM model}. \mnras 465:4914--4930, \doi{10.1093/mnras/stw3006},
  \eprint{1607.01790}

\bibitem[{{Bonvin} et~al(2018){Bonvin}, {Tihhonova}, {Millon}, {Chan},
  {Savary}, {Huber}, and {Courbin}}]{Bonvin2018}
{Bonvin} V, {Tihhonova} O, {Millon} M, {Chan} JHH, {Savary} E, {Huber} S,
  {Courbin} F (2018) {Impact of the 3D source geometry on time-delay
  measurements of lensed type-Ia Supernovae}. ArXiv e-prints
  \eprint{1805.04525}

\bibitem[{{Brewer} and {Lewis}(2008)}]{BrewerLewis08}
{Brewer} BJ, {Lewis} GF (2008) {Unlensing HST observations of the Einstein ring
  1RXS J1131-1231: a Bayesian analysis}. \mnras 390:39--48,
  \doi{10.1111/j.1365-2966.2008.13715.x}, \eprint{0807.2145}

\bibitem[{Bull et~al(2015)Bull, Ferreira, Patel, and Santos}]{Bull:2014rha}
Bull P, Ferreira PG, Patel P, Santos MG (2015) {Late-time cosmology with 21cm
  intensity mapping experiments}. Astrophys J 803(1):21,
  \doi{10.1088/0004-637X/803/1/21}, \eprint{1405.1452}

\bibitem[{{Burud} et~al(2000){Burud}, {Hjorth}, {Jaunsen}, {Andersen},
  {Korhonen}, {Clasen}, {Pelt}, {Pijpers}, {Magain}, and
  {{\O}stensen}}]{Burud2000}
{Burud} I, {Hjorth} J, {Jaunsen} AO, {Andersen} MI, {Korhonen} H, {Clasen} JW,
  {Pelt} J, {Pijpers} FP, {Magain} P, {{\O}stensen} R (2000) {An Optical Time
  Delay Estimate for the Double Gravitational Lens System B1600+434}. \apj
  544:117--122, \doi{10.1086/317213}, \eprint{astro-ph/0007136}

\bibitem[{{Burud} et~al(2002{\natexlab{a}}){Burud}, {Courbin}, {Magain},
  {Lidman}, {Hutsem{\'e}kers}, {Kneib}, {Hjorth}, {Brewer}, {Pompei},
  {Germany}, {Pritchard}, {Jaunsen}, {Letawe}, and {Meylan}}]{Burud2002b}
{Burud} I, {Courbin} F, {Magain} P, {Lidman} C, {Hutsem{\'e}kers} D, {Kneib}
  JP, {Hjorth} J, {Brewer} J, {Pompei} E, {Germany} L, {Pritchard} J, {Jaunsen}
  AO, {Letawe} G, {Meylan} G (2002{\natexlab{a}}) {An optical time-delay for
  the lensed BAL quasar HE 2149-2745}. \aap 383:71--81,
  \doi{10.1051/0004-6361:20011731}, \eprint{astro-ph/0112225}

\bibitem[{{Burud} et~al(2002{\natexlab{b}}){Burud}, {Hjorth}, {Courbin},
  {Cohen}, {Magain}, {Jaunsen}, {Kaas}, {Faure}, and {Letawe}}]{Burud2002}
{Burud} I, {Hjorth} J, {Courbin} F, {Cohen} JG, {Magain} P, {Jaunsen} AO,
  {Kaas} AA, {Faure} C, {Letawe} G (2002{\natexlab{b}}) {Time delay and lens
  redshift for the doubly imaged BAL quasar SBS 1520+530}. \aap 391:481--486,
  \doi{10.1051/0004-6361:20020856}, \eprint{astro-ph/0206084}

\bibitem[{{Cantale} et~al(2016){Cantale}, {Courbin}, {Tewes}, {Jablonka}, and
  {Meylan}}]{Cantale2016}
{Cantale} N, {Courbin} F, {Tewes} M, {Jablonka} P, {Meylan} G (2016) {Firedec:
  a two-channel finite-resolution image deconvolution algorithm}. \aap 589:A81,
  \doi{10.1051/0004-6361/201424003}, \eprint{1602.02167}

\bibitem[{{Cappellari} et~al(2015){Cappellari}, {Romanowsky}, {Brodie},
  {Forbes}, {Strader}, {Foster}, {Kartha}, {Pastorello}, {Pota}, {Spitler},
  {Usher}, and {Arnold}}]{CappellariEtal15}
{Cappellari} M, {Romanowsky} AJ, {Brodie} JP, {Forbes} DA, {Strader} J,
  {Foster} C, {Kartha} SS, {Pastorello} N, {Pota} V, {Spitler} LR, {Usher} C,
  {Arnold} JA (2015) {Small Scatter and Nearly Isothermal Mass Profiles to Four
  Half-light Radii from Two-dimensional Stellar Dynamics of Early-type
  Galaxies}. \apjl 804:L21, \doi{10.1088/2041-8205/804/1/L21},
  \eprint{1504.00075}

\bibitem[{{Chang} et~al(2008){Chang}, {Pen}, {Peterson}, and
  {McDonald}}]{Chang08}
{Chang} TC, {Pen} UL, {Peterson} JB, {McDonald} P (2008) {Baryon Acoustic
  Oscillation Intensity Mapping of Dark Energy}. Phys Rev Lett
  100(9):091,303--+, \doi{10.1103/PhysRevLett.100.091303}, \eprint{0709.3672}

\bibitem[{{Chang} et~al(2010){Chang}, {Pen}, {Bandura}, and
  {Peterson}}]{Chang2010}
{Chang} TC, {Pen} UL, {Bandura} K, {Peterson} JB (2010) {An intensity map of
  hydrogen 21-cm emission at redshift z\~{}0.8}. \nat 466:463--465,
  \doi{10.1038/nature09187}

\bibitem[{{Chen} et~al(2016){Chen}, {Suyu}, {Wong}, {Fassnacht}, {Chiueh},
  {Halkola}, {Hu}, {Auger}, {Koopmans}, {Lagattuta}, {McKean}, and
  {Vegetti}}]{ChenEtal16}
{Chen} GCF, {Suyu} SH, {Wong} KC, {Fassnacht} CD, {Chiueh} T, {Halkola} A, {Hu}
  IS, {Auger} MW, {Koopmans} LVE, {Lagattuta} DJ, {McKean} JP, {Vegetti} S
  (2016) {SHARP - III. First use of adaptive-optics imaging to constrain
  cosmology with gravitational lens time delays}. \mnras 462:3457--3475,
  \doi{10.1093/mnras/stw991}, \eprint{1601.01321}

\bibitem[{{Chen} et~al(2018){Chen}, {Fassnacht}, {Chan}, {Bonvin}, {Rojas},
  {Millon}, {Courbin}, {Suyu}, {Wong}, {Sluse}, {Treu}, {Shajib}, {Hsueh},
  {Lagattuta}, and {McKean}}]{Chen2018}
{Chen} GCF, {Fassnacht} CD, {Chan} JHH, {Bonvin} V, {Rojas} K, {Millon} M,
  {Courbin} F, {Suyu} SH, {Wong} KC, {Sluse} D, {Treu} T, {Shajib} AJ, {Hsueh}
  JW, {Lagattuta} DJ, {McKean} JP (2018) {Constraining the microlensing effect
  on time delays with new time-delay prediction model in $H\_{0}$
  measurements}. ArXiv180409390 \eprint{1804.09390}

\bibitem[{{Cole} et~al(2005){Cole}, {Percival}, {Peacock}, {Norberg}, {Baugh},
  {Frenk}, {Baldry}, {Bland-Hawthorn}, {Bridges}, {Cannon}, {Colless},
  {Collins}, {Couch}, {Cross}, {Dalton}, {Eke}, {De Propris}, {Driver},
  {Efstathiou}, {Ellis}, {Glazebrook}, {Jackson}, {Jenkins}, {Lahav}, {Lewis},
  {Lumsden}, {Maddox}, {Madgwick}, {Peterson}, {Sutherland}, and
  {Taylor}}]{Cole:2005}
{Cole} S, {Percival} WJ, {Peacock} JA, {Norberg} P, {Baugh} CM, {Frenk} CS,
  {Baldry} I, {Bland-Hawthorn} J, {Bridges} T, {Cannon} R, {Colless} M,
  {Collins} C, {Couch} W, {Cross} NJG, {Dalton} G, {Eke} VR, {De Propris} R,
  {Driver} SP, {Efstathiou} G, {Ellis} RS, {Glazebrook} K, {Jackson} C,
  {Jenkins} A, {Lahav} O, {Lewis} I, {Lumsden} S, {Maddox} S, {Madgwick} D,
  {Peterson} BA, {Sutherland} W, {Taylor} K (2005) {The 2dF Galaxy Redshift
  Survey: power-spectrum analysis of the final data set and cosmological
  implications}. \mnras 362:505--534, \doi{10.1111/j.1365-2966.2005.09318.x},
  \eprint{arXiv:astro-ph/0501174}

\bibitem[{{Collett} et~al(2013){Collett}, {Marshall}, {Auger}, {Hilbert},
  {Suyu}, {Greene}, {Treu}, {Fassnacht}, {Koopmans}, {Brada{\v c}}, and
  {Blandford}}]{CollettEtal13}
{Collett} TE, {Marshall} PJ, {Auger} MW, {Hilbert} S, {Suyu} SH, {Greene} Z,
  {Treu} T, {Fassnacht} CD, {Koopmans} LVE, {Brada{\v c}} M, {Blandford} RD
  (2013) {Reconstructing the lensing mass in the Universe from photometric
  catalogue data}. \mnras 432:679--692, \doi{10.1093/mnras/stt504},
  \eprint{1303.6564}

\bibitem[{{Colley} et~al(2003){Colley}, {Schild}, {Abajas}, {Alcalde}, {Aslan},
  {Bikmaev}, {Chavushyan}, {Chinarro}, {Cournoyer}, {Crowe}, {Dudinov},
  {Evans}, {Jeon}, {Goicoechea}, {Golbasi}, {Khamitov}, {Kjernsmo}, {Lee},
  {Lee}, {Lee}, {Lee}, {Lopez-Cruz}, {Mediavilla}, {Moffat}, {Mujica}, {Ullan},
  {Mu{\~n}oz}, {Oscoz}, {Park}, {Purves}, {Saanum}, {Sakhibullin},
  {Serra-Ricart}, {Sinelnikov}, {Stabell}, {Stockton}, {Teuber}, {Thompson},
  {Woo}, and {Zheleznyak}}]{Colley2003}
{Colley} WN, {Schild} RE, {Abajas} C, {Alcalde} D, {Aslan} Z, {Bikmaev} I,
  {Chavushyan} V, {Chinarro} L, {Cournoyer} JP, {Crowe} R, {Dudinov} V, {Evans}
  AKD, {Jeon} YB, {Goicoechea} LJ, {Golbasi} O, {Khamitov} I, {Kjernsmo} K,
  {Lee} HJ, {Lee} J, {Lee} KW, {Lee} MG, {Lopez-Cruz} O, {Mediavilla} E,
  {Moffat} AFJ, {Mujica} R, {Ullan} A, {Mu{\~n}oz} J, {Oscoz} A, {Park} MG,
  {Purves} N, {Saanum} O, {Sakhibullin} N, {Serra-Ricart} M, {Sinelnikov} I,
  {Stabell} R, {Stockton} A, {Teuber} J, {Thompson} R, {Woo} HS, {Zheleznyak} A
  (2003) {Around-the-Clock Observations of the Q0957+561A,B Gravitationally
  Lensed Quasar. II. Results for the Second Observing Season}. \apj 587:71--79,
  \doi{10.1086/368076}, \eprint{astro-ph/0210400}

\bibitem[{{Courbin} et~al(2005){Courbin}, {Eigenbrod}, {Vuissoz}, {Meylan}, and
  {Magain}}]{Courbin2005}
{Courbin} F, {Eigenbrod} A, {Vuissoz} C, {Meylan} G, {Magain} P (2005)
  {COSMOGRAIL: the COSmological MOnitoring of GRAvItational Lenses}. In:
  {Mellier} Y, {Meylan} G (eds) Gravitational Lensing Impact on Cosmology, IAU
  Symposium, vol 225, pp 297--303, \doi{10.1017/S1743921305002097}

\bibitem[{{Courbin} et~al(2011{\natexlab{a}}){Courbin}, {Chantry}, {Revaz},
  {Sluse}, {Faure}, {Tewes}, {Eulaers}, {Koleva}, {Asfandiyarov}, {Dye},
  {Magain}, {van Winckel}, {Coles}, {Saha}, {Ibrahimov}, and
  {Meylan}}]{Courbin2011}
{Courbin} F, {Chantry} V, {Revaz} Y, {Sluse} D, {Faure} C, {Tewes} M, {Eulaers}
  E, {Koleva} M, {Asfandiyarov} I, {Dye} S, {Magain} P, {van Winckel} H,
  {Coles} J, {Saha} P, {Ibrahimov} M, {Meylan} G (2011{\natexlab{a}})
  {COSMOGRAIL: the COSmological MOnitoring of GRAvItational Lenses. IX. Time
  delays, lens dynamics and baryonic fraction in HE 0435-1223}. \aap 536:A53,
  \doi{10.1051/0004-6361/201015709}, \eprint{1009.1473}

\bibitem[{{Courbin} et~al(2011{\natexlab{b}}){Courbin}, {Chantry}, {Revaz},
  {Sluse}, {Faure}, {Tewes}, {Eulaers}, {Koleva}, and {et al.}}]{CourbinEtal11}
{Courbin} F, {Chantry} V, {Revaz} Y, {Sluse} D, {Faure} C, {Tewes} M, {Eulaers}
  E, {Koleva} M, {et al} (2011{\natexlab{b}}) {COSMOGRAIL: the COSmological
  MOnitoring of GRAvItational Lenses. IX. Time delays, lens dynamics and
  baryonic fraction in HE 0435-1223}. \aap 536:A53,
  \doi{10.1051/0004-6361/201015709}, \eprint{1009.1473}

\bibitem[{{Courbin} et~al(2017){Courbin}, {Bonvin}, {Buckley-Geer},
  {Fassnacht}, {Frieman}, {Lin}, {Marshall}, {Suyu}, {Treu}, {Anguita},
  {Motta}, {Meylan}, {Paic}, {Tewes}, {Agnello}, {Chao}, {Chijani}, {Gilman},
  {Rojas}, {Williams}, {Hempel}, {Kim}, {Lachaume}, {Rabus}, {Abbott}, {Allam},
  {Annis}, {Banerji}, {Bechtol}, {Benoit-L{\'e}vy}, {Brooks}, {Burke}, {Carnero
  Rosell}, {Carrasco Kind}, {Carretero}, {D'Andrea}, {da Costa}, {Davis},
  {DePoy}, {Desai}, {Flaugher}, {Fosalba}, {Garcia-Bellido}, {Gaztanaga},
  {Goldstein}, {Gruen}, {Gruendl}, {Gschwend}, {Gutierrez}, {Honscheid},
  {James}, {Kuehn}, {Kuhlmann}, {Kuropatkin}, {Lahav}, {Lima}, {Maia}, {March},
  {Marshall}, {McMahon}, {Menanteau}, {Miquel}, {Nord}, {Plazas}, {Sanchez},
  {Scarpine}, {Schindler}, {Schubnell}, {Sevilla-Noarbe}, {Smith},
  {Soares-Santos}, {Sobreira}, {Suchyta}, {Tarle}, {Tucker}, {Walker}, and
  {Wester}}]{Courbin2017}
{Courbin} F, {Bonvin} V, {Buckley-Geer} E, {Fassnacht} CD, {Frieman} J, {Lin}
  H, {Marshall} PJ, {Suyu} SH, {Treu} T, {Anguita} T, {Motta} V, {Meylan} G,
  {Paic} E, {Tewes} M, {Agnello} A, {Chao} DCY, {Chijani} M, {Gilman} D,
  {Rojas} K, {Williams} P, {Hempel} A, {Kim} S, {Lachaume} R, {Rabus} M,
  {Abbott} TMC, {Allam} S, {Annis} J, {Banerji} M, {Bechtol} K,
  {Benoit-L{\'e}vy} A, {Brooks} D, {Burke} DL, {Carnero Rosell} A, {Carrasco
  Kind} M, {Carretero} J, {D'Andrea} CB, {da Costa} LN, {Davis} C, {DePoy} DL,
  {Desai} S, {Flaugher} B, {Fosalba} P, {Garcia-Bellido} J, {Gaztanaga} E,
  {Goldstein} DA, {Gruen} D, {Gruendl} RA, {Gschwend} J, {Gutierrez} G,
  {Honscheid} K, {James} DJ, {Kuehn} K, {Kuhlmann} S, {Kuropatkin} N, {Lahav}
  O, {Lima} M, {Maia} MAG, {March} M, {Marshall} JL, {McMahon} RG, {Menanteau}
  F, {Miquel} R, {Nord} B, {Plazas} AA, {Sanchez} E, {Scarpine} V, {Schindler}
  R, {Schubnell} M, {Sevilla-Noarbe} I, {Smith} M, {Soares-Santos} M,
  {Sobreira} F, {Suchyta} E, {Tarle} G, {Tucker} DL, {Walker} AR, {Wester} W
  (2017) {COSMOGRAIL XVI: Time delays for the quadruply imaged quasar DES
  J0408-5354 with high-cadence photometric monitoring}. ArXiv e-prints
  \eprint{1706.09424}

\bibitem[{{Dawson} et~al(2016){Dawson}, {Kneib}, {Percival}, {Alam},
  {Albareti}, {Anderson}, {Armengaud}, {Aubourg}, {Bailey}, {Bautista},
  {Berlind}, {Bershady}, {Beutler}, {Bizyaev}, {Blanton}, {Blomqvist},
  {Bolton}, {Bovy}, {Brandt}, {Brinkmann}, {Brownstein}, {Burtin}, {Busca},
  {Cai}, {Chuang}, {Clerc}, {Comparat}, {Cope}, {Croft}, {Cruz-Gonzalez}, {da
  Costa}, {Cousinou}, {Darling}, {de la Macorra}, {de la Torre}, {Delubac}, {du
  Mas des Bourboux}, {Dwelly}, {Ealet}, {Eisenstein}, {Eracleous}, {Escoffier},
  {Fan}, {Finoguenov}, {Font-Ribera}, {Frinchaboy}, {Gaulme}, {Georgakakis},
  {Green}, {Guo}, {Guy}, {Ho}, {Holder}, {Huehnerhoff}, {Hutchinson}, {Jing},
  {Jullo}, {Kamble}, {Kinemuchi}, {Kirkby}, {Kitaura}, {Klaene}, {Laher},
  {Lang}, {Laurent}, {Le Goff}, {Li}, {Liang}, {Lima}, {Lin}, {Lin}, {Lin},
  {Long}, {Lundgren}, {MacDonald}, {Geimba Maia}, {Malanushenko},
  {Malanushenko}, {Mariappan}, {McBride}, {McGreer}, {M{\'e}nard}, {Merloni},
  {Meza}, {Montero-Dorta}, {Muna}, {Myers}, {Nandra}, {Naugle}, {Newman},
  {Noterdaeme}, {Nugent}, {Ogando}, {Olmstead}, {Oravetz}, {Oravetz},
  {Padmanabhan}, {Palanque-Delabrouille}, {Pan}, {Parejko}, {P{\^a}ris},
  {Peacock}, {Petitjean}, {Pieri}, {Pisani}, {Prada}, {Prakash}, {Raichoor},
  {Reid}, {Rich}, {Ridl}, {Rodriguez-Torres}, {Carnero Rosell}, {Ross},
  {Rossi}, {Ruan}, {Salvato}, {Sayres}, {Schneider}, {Schlegel}, {Seljak},
  {Seo}, {Sesar}, {Shandera}, {Shu}, {Slosar}, {Sobreira}, {Streblyanska},
  {Suzuki}, {Taylor}, {Tao}, {Tinker}, {Tojeiro}, {Vargas-Maga{\~n}a}, {Wang},
  {Weaver}, {Weinberg}, {White}, {Wood-Vasey}, {Yeche}, {Zhai}, {Zhao}, {Zhao},
  {Zheng}, {Ben Zhu}, and {Zou}}]{Dawson:2016}
{Dawson} KS, {Kneib} JP, {Percival} WJ, {Alam} S, {Albareti} FD, {Anderson} SF,
  {Armengaud} E, {Aubourg} {\'E}, {Bailey} S, {Bautista} JE, {Berlind} AA,
  {Bershady} MA, {Beutler} F, {Bizyaev} D, {Blanton} MR, {Blomqvist} M,
  {Bolton} AS, {Bovy} J, {Brandt} WN, {Brinkmann} J, {Brownstein} JR, {Burtin}
  E, {Busca} NG, {Cai} Z, {Chuang} CH, {Clerc} N, {Comparat} J, {Cope} F,
  {Croft} RAC, {Cruz-Gonzalez} I, {da Costa} LN, {Cousinou} MC, {Darling} J,
  {de la Macorra} A, {de la Torre} S, {Delubac} T, {du Mas des Bourboux} H,
  {Dwelly} T, {Ealet} A, {Eisenstein} DJ, {Eracleous} M, {Escoffier} S, {Fan}
  X, {Finoguenov} A, {Font-Ribera} A, {Frinchaboy} P, {Gaulme} P, {Georgakakis}
  A, {Green} P, {Guo} H, {Guy} J, {Ho} S, {Holder} D, {Huehnerhoff} J,
  {Hutchinson} T, {Jing} Y, {Jullo} E, {Kamble} V, {Kinemuchi} K, {Kirkby} D,
  {Kitaura} FS, {Klaene} MA, {Laher} RR, {Lang} D, {Laurent} P, {Le Goff} JM,
  {Li} C, {Liang} Y, {Lima} M, {Lin} Q, {Lin} W, {Lin} YT, {Long} DC,
  {Lundgren} B, {MacDonald} N, {Geimba Maia} MA, {Malanushenko} E,
  {Malanushenko} V, {Mariappan} V, {McBride} CK, {McGreer} ID, {M{\'e}nard} B,
  {Merloni} A, {Meza} A, {Montero-Dorta} AD, {Muna} D, {Myers} AD, {Nandra} K,
  {Naugle} T, {Newman} JA, {Noterdaeme} P, {Nugent} P, {Ogando} R, {Olmstead}
  MD, {Oravetz} A, {Oravetz} DJ, {Padmanabhan} N, {Palanque-Delabrouille} N,
  {Pan} K, {Parejko} JK, {P{\^a}ris} I, {Peacock} JA, {Petitjean} P, {Pieri}
  MM, {Pisani} A, {Prada} F, {Prakash} A, {Raichoor} A, {Reid} B, {Rich} J,
  {Ridl} J, {Rodriguez-Torres} S, {Carnero Rosell} A, {Ross} AJ, {Rossi} G,
  {Ruan} J, {Salvato} M, {Sayres} C, {Schneider} DP, {Schlegel} DJ, {Seljak} U,
  {Seo} HJ, {Sesar} B, {Shandera} S, {Shu} Y, {Slosar} A, {Sobreira} F,
  {Streblyanska} A, {Suzuki} N, {Taylor} D, {Tao} C, {Tinker} JL, {Tojeiro} R,
  {Vargas-Maga{\~n}a} M, {Wang} Y, {Weaver} BA, {Weinberg} DH, {White} M,
  {Wood-Vasey} WM, {Yeche} C, {Zhai} Z, {Zhao} C, {Zhao} Gb, {Zheng} Z, {Ben
  Zhu} G, {Zou} H (2016) {The SDSS-IV Extended Baryon Oscillation Spectroscopic
  Survey: Overview and Early Data}. \aj 151:44,
  \doi{10.3847/0004-6256/151/2/44}, \eprint{1508.04473}

\bibitem[{{DESI Collaboration} et~al(2016){DESI Collaboration}, {Aghamousa},
  {Aguilar}, {Ahlen}, {Alam}, {Allen}, {Allende Prieto}, {Annis}, {Bailey},
  {Balland}, and et~al.}]{DESI-Collaboration:2016}
{DESI Collaboration}, {Aghamousa} A, {Aguilar} J, {Ahlen} S, {Alam} S, {Allen}
  LE, {Allende Prieto} C, {Annis} J, {Bailey} S, {Balland} C, et~al (2016) {The
  DESI Experiment Part I: Science,Targeting, and Survey Design}. ArXiv e-prints
  \eprint{1611.00036}

\bibitem[{{Dobler} and {Keeton}(2006)}]{DoblerKeeton06}
{Dobler} G, {Keeton} CR (2006) {Microlensing of Lensed Supernovae}. \apj
  653:1391--1399, \doi{10.1086/508769}, \eprint{astro-ph/0608391}

\bibitem[{{Dobler} et~al(2015){Dobler}, {Fassnacht}, {Treu}, {Marshall},
  {Liao}, {Hojjati}, {Linder}, and {Rumbaugh}}]{Dobler2015}
{Dobler} G, {Fassnacht} CD, {Treu} T, {Marshall} P, {Liao} K, {Hojjati} A,
  {Linder} E, {Rumbaugh} N (2015) {Strong Lens Time Delay Challenge. I.
  Experimental Design}. \apj 799:168, \doi{10.1088/0004-637X/799/2/168}

\bibitem[{{Drinkwater} et~al(2010){Drinkwater}, {Jurek}, {Blake}, {Woods},
  {Pimbblet}, {Glazebrook}, {Sharp}, {Pracy}, {Brough}, {Colless}, {Couch},
  {Croom}, {Davis}, {Forbes}, {Forster}, {Gilbank}, {Gladders}, {Jelliffe},
  {Jones}, {Li}, {Madore}, {Martin}, {Poole}, {Small}, {Wisnioski}, {Wyder},
  and {Yee}}]{WiggleZ}
{Drinkwater} MJ, {Jurek} RJ, {Blake} C, {Woods} D, {Pimbblet} KA, {Glazebrook}
  K, {Sharp} R, {Pracy} MB, {Brough} S, {Colless} M, {Couch} WJ, {Croom} SM,
  {Davis} TM, {Forbes} D, {Forster} K, {Gilbank} DG, {Gladders} M, {Jelliffe}
  B, {Jones} N, {Li} IH, {Madore} B, {Martin} DC, {Poole} GB, {Small} T,
  {Wisnioski} E, {Wyder} T, {Yee} HKC (2010) {The WiggleZ Dark Energy Survey:
  survey design and first data release}. MNRAS 401:1429--1452,
  \doi{10.1111/j.1365-2966.2009.15754.x}, \eprint{0911.4246}

\bibitem[{{Dye} and {Warren}(2005)}]{DyeWarren05}
{Dye} S, {Warren} SJ (2005) {Decomposition of the Visible and Dark Matter in
  the Einstein Ring 0047-2808 by Semilinear Inversion}. \apj 623:31--41,
  \doi{10.1086/428340}

\bibitem[{{Efstathiou}(2014)}]{Efstathiou14}
{Efstathiou} G (2014) {H$_{0}$ revisited}. \mnras 440:1138--1152,
  \doi{10.1093/mnras/stu278}, \eprint{1311.3461}

\bibitem[{{Eigenbrod} et~al(2005){Eigenbrod}, {Courbin}, {Vuissoz}, {Meylan},
  {Saha}, and {Dye}}]{Eigenbrod2005}
{Eigenbrod} A, {Courbin} F, {Vuissoz} C, {Meylan} G, {Saha} P, {Dye} S (2005)
  {COSMOGRAIL: The COSmological MOnitoring of GRAvItational Lenses. I. How to
  sample the light curves of gravitationally lensed quasars to measure accurate
  time delays}. \aap 436:25--35, \doi{10.1051/0004-6361:20042422},
  \eprint{astro-ph/0503019}

\bibitem[{{Eisenstein} and {Hu}(1998)}]{Eisenstein:1998}
{Eisenstein} DJ, {Hu} W (1998) {Baryonic Features in the Matter Transfer
  Function}. \apj 496:605--+, \doi{10.1086/305424},
  \eprint{arXiv:astro-ph/9709112}

\bibitem[{{Eisenstein} et~al(2005){Eisenstein}, {Zehavi}, {Hogg},
  {Scoccimarro}, {Blanton}, {Nichol}, {Scranton}, {Seo}, {Tegmark}, {Zheng},
  {Anderson}, {Annis}, {Bahcall}, {Brinkmann}, {Burles}, {Castander},
  {Connolly}, {Csabai}, {Doi}, {Fukugita}, {Frieman}, {Glazebrook}, {Gunn},
  {Hendry}, {Hennessy}, {Ivezi{\'c}}, {Kent}, {Knapp}, {Lin}, {Loh}, {Lupton},
  {Margon}, {McKay}, {Meiksin}, {Munn}, {Pope}, {Richmond}, {Schlegel},
  {Schneider}, {Shimasaku}, {Stoughton}, {Strauss}, {SubbaRao}, {Szalay},
  {Szapudi}, {Tucker}, {Yanny}, and {York}}]{Eisenstein:2005}
{Eisenstein} DJ, {Zehavi} I, {Hogg} DW, {Scoccimarro} R, {Blanton} MR, {Nichol}
  RC, {Scranton} R, {Seo} H, {Tegmark} M, {Zheng} Z, {Anderson} SF, {Annis} J,
  {Bahcall} N, {Brinkmann} J, {Burles} S, {Castander} FJ, {Connolly} A,
  {Csabai} I, {Doi} M, {Fukugita} M, {Frieman} JA, {Glazebrook} K, {Gunn} JE,
  {Hendry} JS, {Hennessy} G, {Ivezi{\'c}} Z, {Kent} S, {Knapp} GR, {Lin} H,
  {Loh} Y, {Lupton} RH, {Margon} B, {McKay} TA, {Meiksin} A, {Munn} JA, {Pope}
  A, {Richmond} MW, {Schlegel} D, {Schneider} DP, {Shimasaku} K, {Stoughton} C,
  {Strauss} MA, {SubbaRao} M, {Szalay} AS, {Szapudi} I, {Tucker} DL, {Yanny} B,
  {York} DG (2005) {Detection of the Baryon Acoustic Peak in the Large-Scale
  Correlation Function of SDSS Luminous Red Galaxies}. \apj 633:560--574,
  \doi{10.1086/466512}, \eprint{arXiv:astro-ph/0501171}

\bibitem[{{Eulaers} et~al(2013){Eulaers}, {Tewes}, {Magain}, {Courbin},
  {Asfandiyarov}, {Ehgamberdiev}, {Rathna Kumar}, {Stalin}, {Prabhu}, {Meylan},
  and {Van Winckel}}]{Eulaers2013}
{Eulaers} E, {Tewes} M, {Magain} P, {Courbin} F, {Asfandiyarov} I,
  {Ehgamberdiev} S, {Rathna Kumar} S, {Stalin} CS, {Prabhu} TP, {Meylan} G,
  {Van Winckel} H (2013) {COSMOGRAIL: the COSmological MOnitoring of
  GRAvItational Lenses. XII. Time delays of the doubly lensed quasars SDSS
  J1206+4332 and HS 2209+1914}. \aap 553:A121,
  \doi{10.1051/0004-6361/201321140}, \eprint{1304.4474}

\bibitem[{{Fadely} et~al(2010){Fadely}, {Keeton}, {Nakajima}, and
  {Bernstein}}]{FadelyEtal10}
{Fadely} R, {Keeton} CR, {Nakajima} R, {Bernstein} GM (2010) {Improved
  Constraints on the Gravitational Lens Q0957+561. II. Strong Lensing}. \apj
  711:246--267, \doi{10.1088/0004-637X/711/1/246}, \eprint{0909.1807}

\bibitem[{{Falco} et~al(1985){Falco}, {Gorenstein}, and
  {Shapiro}}]{FalcoEtal85}
{Falco} EE, {Gorenstein} MV, {Shapiro} II (1985) {On model-dependent bounds on
  H(0) from gravitational images Application of Q0957 + 561A,B}. \apjl
  289:L1--L4, \doi{10.1086/184422}

\bibitem[{{Fassnacht} et~al(1999){Fassnacht}, {Pearson}, {Readhead}, {Browne},
  {Koopmans}, {Myers}, and {Wilkinson}}]{Fassnacht1999}
{Fassnacht} CD, {Pearson} TJ, {Readhead} ACS, {Browne} IWA, {Koopmans} LVE,
  {Myers} ST, {Wilkinson} PN (1999) {A Determination of H$_{0}$ with the CLASS
  Gravitational Lens B1608+656. I. Time Delay Measurements with the VLA}. \apj
  527:498--512, \doi{10.1086/308118}, \eprint{astro-ph/9907257}

\bibitem[{{Fassnacht} et~al(2002){Fassnacht}, {Xanthopoulos}, {Koopmans}, and
  {Rusin}}]{FassnachtEtal02}
{Fassnacht} CD, {Xanthopoulos} E, {Koopmans} LVE, {Rusin} D (2002) {A
  Determination of $H_{0}$ with the CLASS Gravitational Lens B1608+656. III. A
  Significant Improvement in the Precision of the Time Delay Measuremxents}.
  \apj 581:823--835, \doi{10.1086/344368}, \eprint{astro-ph/0208420}

\bibitem[{{Fassnacht} et~al(2006){Fassnacht}, {Gal}, {Lubin}, {McKean},
  {Squires}, and {Readhead}}]{FassnachtEtal06}
{Fassnacht} CD, {Gal} RR, {Lubin} LM, {McKean} JP, {Squires} GK, {Readhead} ACS
  (2006) {Mass along the Line of Sight to the Gravitational Lens B1608+656:
  Galaxy Groups and Implications for $H_{0}$ }. \apj 642:30--38,
  \doi{10.1086/500927}, \eprint{arXiv:astro-ph/0510728}

\bibitem[{{Fern{\'a}ndez} et~al(2016){Fern{\'a}ndez}, {Gim}, {van Gorkom},
  {Yun}, {Momjian}, {Popping}, {Chomiuk}, {Hess}, {Hunt}, {Kreckel}, {Lucero},
  {Maddox}, {Oosterloo}, {Pisano}, {Verheijen}, {Hales}, {Chung}, {Dodson},
  {Golap}, {Gross}, {Henning}, {Hibbard}, {Jaff{\'e}}, {Donovan Meyer},
  {Meyer}, {Sanchez-Barrantes}, {Schiminovich}, {Wicenec}, {Wilcots},
  {Bershady}, {Scoville}, {Strader}, {Tremou}, {Salinas}, and
  {Ch{\'a}vez}}]{2016ApJ...824L...1F}
{Fern{\'a}ndez} X, {Gim} HB, {van Gorkom} JH, {Yun} MS, {Momjian} E, {Popping}
  A, {Chomiuk} L, {Hess} KM, {Hunt} L, {Kreckel} K, {Lucero} D, {Maddox} N,
  {Oosterloo} T, {Pisano} DJ, {Verheijen} MAW, {Hales} CA, {Chung} A, {Dodson}
  R, {Golap} K, {Gross} J, {Henning} P, {Hibbard} J, {Jaff{\'e}} YL, {Donovan
  Meyer} J, {Meyer} M, {Sanchez-Barrantes} M, {Schiminovich} D, {Wicenec} A,
  {Wilcots} E, {Bershady} M, {Scoville} N, {Strader} J, {Tremou} E, {Salinas}
  R, {Ch{\'a}vez} R (2016) {Highest Redshift Image of Neutral Hydrogen in
  Emission: A CHILES Detection of a Starbursting Galaxy at z = 0.376}. \apjl
  824:L1, \doi{10.3847/2041-8205/824/1/L1}, \eprint{1606.00013}

\bibitem[{{Fohlmeister} et~al(2008){Fohlmeister}, {Kochanek}, {Falco},
  {Morgan}, and {Wambsganss}}]{Fohlmeister2008}
{Fohlmeister} J, {Kochanek} CS, {Falco} EE, {Morgan} CW, {Wambsganss} J (2008)
  {The Rewards of Patience: An 822 Day Time Delay in the Gravitational Lens
  SDSS J1004+4112}. \apj 676:761-766, \doi{10.1086/528789}, \eprint{0710.1634}

\bibitem[{{Fohlmeister} et~al(2013){Fohlmeister}, {Kochanek}, {Falco},
  {Wambsganss}, {Oguri}, and {Dai}}]{Fohlmeister2013}
{Fohlmeister} J, {Kochanek} CS, {Falco} EE, {Wambsganss} J, {Oguri} M, {Dai} X
  (2013) {A Two-year Time Delay for the Lensed Quasar SDSS J1029+2623}. \apj
  764:186, \doi{10.1088/0004-637X/764/2/186}, \eprint{1207.5776}

\bibitem[{{Foxley-Marrable} et~al(2018){Foxley-Marrable}, {Collett},
  {Vernardos}, {Goldstein}, and {Bacon}}]{FM18}
{Foxley-Marrable} M, {Collett} TE, {Vernardos} G, {Goldstein} DA, {Bacon} D
  (2018) {The Impact of Microlensing on the Standardisation of Strongly Lensed
  Type Ia Supernovae}. ArXiv e-prints \eprint{1802.07738}

\bibitem[{{Freedman} et~al(2012){Freedman}, {Madore}, {Scowcroft}, {Burns},
  {Monson}, {Persson}, {Seibert}, and {Rigby}}]{FreedmanEtal12}
{Freedman} WL, {Madore} BF, {Scowcroft} V, {Burns} C, {Monson} A, {Persson} SE,
  {Seibert} M, {Rigby} J (2012) {Carnegie Hubble Program: A Mid-infrared
  Calibration of the Hubble Constant}. \apj 758:24,
  \doi{10.1088/0004-637X/758/1/24}, \eprint{1208.3281}

\bibitem[{{Gavazzi} et~al(2008){Gavazzi}, {Treu}, {Koopmans}, {Bolton},
  {Moustakas}, {Burles}, and {Marshall}}]{GavazziEtal08}
{Gavazzi} R, {Treu} T, {Koopmans} LVE, {Bolton} AS, {Moustakas} LA, {Burles} S,
  {Marshall} PJ (2008) {The Sloan Lens ACS Survey. VI. Discovery and Analysis
  of a Double Einstein Ring}. \apj 677:1046-1059, \doi{10.1086/529541},
  \eprint{0801.1555}

\bibitem[{{Giannini} et~al(2017){Giannini}, {Schmidt}, {Wambsganss}, {Alsubai},
  {Andersen}, {Anguita}, {Bozza}, {Bramich}, {Browne}, {Calchi Novati},
  {Damerdji}, {Diehl}, {Dodds}, {Dominik}, {Elyiv}, {Fang}, {Figuera Jaimes},
  {Finet}, {Gerner}, {Gu}, {Hardis}, {Harps{\o}e}, {Hinse}, {Hornstrup},
  {Hundertmark}, {Jessen-Hansen}, {J{\o}rgensen}, {Juncher}, {Kains}, {Kerins},
  {Korhonen}, {Liebig}, {Lund}, {Lundkvist}, {Maier}, {Mancini}, {Masi},
  {Mathiasen}, {Penny}, {Proft}, {Rabus}, {Rahvar}, {Ricci}, {Scarpetta},
  {Sahu}, {Sch{\"a}fer}, {Sch{\"o}nebeck}, {Skottfelt}, {Snodgrass},
  {Southworth}, {Surdej}, {Tregloan-Reed}, {Vilela}, {Wertz}, and
  {Zimmer}}]{Giannini2017}
{Giannini} E, {Schmidt} RW, {Wambsganss} J, {Alsubai} K, {Andersen} JM,
  {Anguita} T, {Bozza} V, {Bramich} DM, {Browne} P, {Calchi Novati} S,
  {Damerdji} Y, {Diehl} C, {Dodds} P, {Dominik} M, {Elyiv} A, {Fang} X,
  {Figuera Jaimes} R, {Finet} F, {Gerner} T, {Gu} S, {Hardis} S, {Harps{\o}e}
  K, {Hinse} TC, {Hornstrup} A, {Hundertmark} M, {Jessen-Hansen} J,
  {J{\o}rgensen} UG, {Juncher} D, {Kains} N, {Kerins} E, {Korhonen} H, {Liebig}
  C, {Lund} MN, {Lundkvist} MS, {Maier} G, {Mancini} L, {Masi} G, {Mathiasen}
  M, {Penny} M, {Proft} S, {Rabus} M, {Rahvar} S, {Ricci} D, {Scarpetta} G,
  {Sahu} K, {Sch{\"a}fer} S, {Sch{\"o}nebeck} F, {Skottfelt} J, {Snodgrass} C,
  {Southworth} J, {Surdej} J, {Tregloan-Reed} J, {Vilela} C, {Wertz} O,
  {Zimmer} F (2017) {MiNDSTEp differential photometry of the gravitationally
  lensed quasars WFI 2033-4723 and HE 0047-1756: microlensing and a new time
  delay}. \aap 597:A49, \doi{10.1051/0004-6361/201527422}, \eprint{1610.03732}

\bibitem[{{Goldstein} and {Nugent}(2017)}]{Goldstein2017}
{Goldstein} DA, {Nugent} PE (2017) {How to Find Gravitationally Lensed Type Ia
  Supernovae}. \apjl 834:L5, \doi{10.3847/2041-8213/834/1/L5},
  \eprint{1611.09459}

\bibitem[{{Goldstein} et~al(2018){Goldstein}, {Nugent}, {Kasen}, and
  {Collett}}]{GoldsteinEtal18}
{Goldstein} DA, {Nugent} PE, {Kasen} DN, {Collett} TE (2018) {Precise Time
  Delays from Strongly Gravitationally Lensed Type Ia Supernovae with
  Chromatically Microlensed Images}. \apj 855:22,
  \doi{10.3847/1538-4357/aaa975}, \eprint{1708.00003}

\bibitem[{{Golse} and {Kneib}(2002)}]{GolseKneib02}
{Golse} G, {Kneib} JP (2002) {Pseudo elliptical lensing mass model: Application
  to the NFW mass distribution}. \aap 390:821--827,
  \doi{10.1051/0004-6361:20020639}, \eprint{arXiv:astro-ph/0112138}

\bibitem[{{Goobar}(2017)}]{Goobar2017}
{Goobar} A (2017) {iPTF16geu: A multiply imaged, gravitationally lensed type Ia
  supernova}. Science 356:291--295, \doi{10.1126/science.aal2729},
  \eprint{1611.00014}

\bibitem[{{Greene} et~al(2013){Greene}, {Suyu}, {Treu}, {Hilbert}, {Auger},
  {Collett}, {Marshall}, {Fassnacht}, {Blandford}, {Brada{\v c}}, and
  {Koopmans}}]{GreeneEtal13}
{Greene} ZS, {Suyu} SH, {Treu} T, {Hilbert} S, {Auger} MW, {Collett} TE,
  {Marshall} PJ, {Fassnacht} CD, {Blandford} RD, {Brada{\v c}} M, {Koopmans}
  LVE (2013) {Improving the Precision of Time-delay Cosmography with
  Observations of Galaxies along the Line of Sight}. \apj 768:39,
  \doi{10.1088/0004-637X/768/1/39}, \eprint{1303.3588}

\bibitem[{{Grillo} et~al(2016){Grillo}, {Karman}, {Suyu}, {Rosati}, {Balestra},
  {Mercurio}, {Lombardi}, {Treu}, {Caminha}, {Halkola}, {Rodney}, {Gavazzi},
  and {Caputi}}]{GrilloEtal16}
{Grillo} C, {Karman} W, {Suyu} SH, {Rosati} P, {Balestra} I, {Mercurio} A,
  {Lombardi} M, {Treu} T, {Caminha} GB, {Halkola} A, {Rodney} SA, {Gavazzi} R,
  {Caputi} KI (2016) {The Story of Supernova Refsdal Told by Muse}. \apj
  822:78, \doi{10.3847/0004-637X/822/2/78}, \eprint{1511.04093}

\bibitem[{{Grillo} et~al(2018){Grillo}, {Rosati}, {Suyu}, {Balestra},
  {Caminha}, {Halkola}, {Kelly}, {Lombardi}, {Mercurio}, {Rodney}, and
  {Treu}}]{GrilloEtal18}
{Grillo} C, {Rosati} P, {Suyu} SH, {Balestra} I, {Caminha} GB, {Halkola} A,
  {Kelly} PL, {Lombardi} M, {Mercurio} A, {Rodney} SA, {Treu} T (2018)
  {Measuring the Value of the Hubble Constant ``{\`a} la Refsdal"}. \apj
  860:94, \doi{10.3847/1538-4357/aac2c9}, \eprint{1802.01584}

\bibitem[{{Grogin} and {Narayan}(1996)}]{GroginNarayan96a}
{Grogin} NA, {Narayan} R (1996) {A New Model of the Gravitational Lens 0957+561
  and a Limit on the Hubble Constant}. \apj 464:92, \doi{10.1086/177302}

\bibitem[{{Hainline} et~al(2013){Hainline}, {Morgan}, {MacLeod}, {Landaal},
  {Kochanek}, {Harris}, {Tilleman}, {Goicoechea}, {Shalyapin}, and
  {Falco}}]{Hainline2013}
{Hainline} LJ, {Morgan} CW, {MacLeod} CL, {Landaal} ZD, {Kochanek} CS, {Harris}
  HC, {Tilleman} T, {Goicoechea} LJ, {Shalyapin} VN, {Falco} EE (2013) {Time
  Delay and Accretion Disk Size Measurements in the Lensed Quasar SBS 0909+532
  from Multiwavelength Microlensing Analysis}. \apj 774:69,
  \doi{10.1088/0004-637X/774/1/69}, \eprint{1307.3254}

\bibitem[{{Hazard} et~al(1963){Hazard}, {Mackey}, and {Shimmins}}]{Hazard1963}
{Hazard} C, {Mackey} MB, {Shimmins} AJ (1963) {Investigation of the Radio
  Source 3C 273 By The Method of Lunar Occultations}. \nat 197:1037--1039,
  \doi{10.1038/1971037a0}

\bibitem[{{Hilbert} et~al(2007){Hilbert}, {White}, {Hartlap}, and
  {Schneider}}]{HilbertEtal07}
{Hilbert} S, {White} SDM, {Hartlap} J, {Schneider} P (2007) {Strong lensing
  optical depths in a {$\Lambda$}CDM universe}. \mnras 382:121--132,
  \doi{10.1111/j.1365-2966.2007.12391.x}, \eprint{arXiv:astro-ph/0703803}

\bibitem[{{Hjorth} et~al(2002){Hjorth}, {Burud}, {Jaunsen}, {Schechter},
  {Kneib}, {Andersen}, {Korhonen}, {Clasen}, {Kaas}, {{\O}stensen}, {Pelt}, and
  {Pijpers}}]{Hjorth2002}
{Hjorth} J, {Burud} I, {Jaunsen} AO, {Schechter} PL, {Kneib} JP, {Andersen} MI,
  {Korhonen} H, {Clasen} JW, {Kaas} AA, {{\O}stensen} R, {Pelt} J, {Pijpers} FP
  (2002) {The Time Delay of the Quadruple Quasar RX J0911.4+0551}. \apjl
  572:L11--L14, \doi{10.1086/341603}, \eprint{astro-ph/0205124}

\bibitem[{{Hojjati} and {Linder}(2014)}]{Hojjati2014}
{Hojjati} A, {Linder} EV (2014) {Next generation strong lensing time delay
  estimation with Gaussian processes}. \prd 90(12):123501,
  \doi{10.1103/PhysRevD.90.123501}, \eprint{1408.5143}

\bibitem[{{Hojjati} et~al(2013){Hojjati}, {Kim}, and {Linder}}]{Hojjati2013}
{Hojjati} A, {Kim} AG, {Linder} EV (2013) {Robust strong lensing time delay
  estimation}. \prd 87(12):123512, \doi{10.1103/PhysRevD.87.123512},
  \eprint{1304.0309}

\bibitem[{{Hu} and {Sugiyama}(1996)}]{Hu:1996}
{Hu} W, {Sugiyama} N (1996) {Small-Scale Cosmological Perturbations: an
  Analytic Approach}. \apj 471:542, \doi{10.1086/177989},
  \eprint{astro-ph/9510117}

\bibitem[{{Huchra} et~al(1985){Huchra}, {Gorenstein}, {Kent}, {Shapiro},
  {Smith}, {Horine}, and {Perley}}]{Huchra1985}
{Huchra} J, {Gorenstein} M, {Kent} S, {Shapiro} I, {Smith} G, {Horine} E,
  {Perley} R (1985) {2237 + 0305: A new and unusual gravitational lens}. \aj
  90:691--696, \doi{10.1086/113777}

\bibitem[{{Jakobsson} et~al(2005){Jakobsson}, {Hjorth}, {Burud}, {Letawe},
  {Lidman}, and {Courbin}}]{Jakobsson2005}
{Jakobsson} P, {Hjorth} J, {Burud} I, {Letawe} G, {Lidman} C, {Courbin} F
  (2005) {An optical time delay for the double gravitational lens system FBQ
  0951+2635}. \aap 431:103--109, \doi{10.1051/0004-6361:20041432},
  \eprint{astro-ph/0409444}

\bibitem[{{Jauzac} et~al(2016){Jauzac}, {Richard}, {Limousin}, {Knowles},
  {Mahler}, {Smith}, {Kneib}, {Jullo}, {Natarajan}, {Ebeling}, {Atek},
  {Cl{\'e}ment}, {Eckert}, {Egami}, {Massey}, and {Rexroth}}]{JauzacEtal16}
{Jauzac} M, {Richard} J, {Limousin} M, {Knowles} K, {Mahler} G, {Smith} GP,
  {Kneib} JP, {Jullo} E, {Natarajan} P, {Ebeling} H, {Atek} H, {Cl{\'e}ment} B,
  {Eckert} D, {Egami} E, {Massey} R, {Rexroth} M (2016) {Hubble Frontier
  Fields: predictions for the return of SN Refsdal with the MUSE and GMOS
  spectrographs}. \mnras 457:2029--2042, \doi{10.1093/mnras/stw069},
  \eprint{1509.08914}

\bibitem[{{Jee} et~al(2015){Jee}, {Komatsu}, and {Suyu}}]{JeeEtal15}
{Jee} I, {Komatsu} E, {Suyu} SH (2015) {Measuring angular diameter distances of
  strong gravitational lenses}. \jcap 11:033,
  \doi{10.1088/1475-7516/2015/11/033}, \eprint{1410.7770}

\bibitem[{{Jee} et~al(2016){Jee}, {Komatsu}, {Suyu}, and {Huterer}}]{JeeEtal16}
{Jee} I, {Komatsu} E, {Suyu} SH, {Huterer} D (2016) {Time-delay cosmography:
  increased leverage with angular diameter distances}. \jcap 4:031,
  \doi{10.1088/1475-7516/2016/04/031}, \eprint{1509.03310}

\bibitem[{{Jullo} et~al(2007){Jullo}, {Kneib}, {Limousin},
  {El{\'{\i}}asd{\'o}ttir}, {Marshall}, and {Verdugo}}]{JulloEtal2007}
{Jullo} E, {Kneib} J, {Limousin} M, {El{\'{\i}}asd{\'o}ttir} {\'A}, {Marshall}
  PJ, {Verdugo} T (2007) {A Bayesian approach to strong lensing modelling of
  galaxy clusters}. New Journal of Physics 9:447,
  \doi{10.1088/1367-2630/9/12/447}, \eprint{0706.0048}

\bibitem[{{Kaiser}(1987)}]{Kaiser:1987}
{Kaiser} N (1987) {Clustering in real space and in redshift space}. \mnras
  227:1--21

\bibitem[{{Kawamata} et~al(2016){Kawamata}, {Oguri}, {Ishigaki}, {Shimasaku},
  and {Ouchi}}]{KawamataEtal16}
{Kawamata} R, {Oguri} M, {Ishigaki} M, {Shimasaku} K, {Ouchi} M (2016) {Precise
  Strong Lensing Mass Modeling of Four Hubble Frontier Field Clusters and a
  Sample of Magnified High-redshift Galaxies}. \apj 819:114,
  \doi{10.3847/0004-637X/819/2/114}, \eprint{1510.06400}

\bibitem[{{Keeton}(2001)}]{Keeton01}
{Keeton} CR (2001) {Computational Methods for Gravitational Lensing}. e-prints
  (astro-ph/0102340) \eprint{astro-ph/0102340}

\bibitem[{{Kelly} et~al(2015){Kelly}, {Rodney}, {Treu}, {Foley}, {Brammer},
  {Schmidt}, {Zitrin}, {Sonnenfeld}, {Strolger}, {Graur}, {Filippenko}, {Jha},
  {Riess}, {Bradac}, {Weiner}, {Scolnic}, {Malkan}, {von der Linden}, {Trenti},
  {Hjorth}, {Gavazzi}, {Fontana}, {Merten}, {McCully}, {Jones}, {Postman},
  {Dressler}, {Patel}, {Cenko}, {Graham}, and {Tucker}}]{Kelly2015}
{Kelly} PL, {Rodney} SA, {Treu} T, {Foley} RJ, {Brammer} G, {Schmidt} KB,
  {Zitrin} A, {Sonnenfeld} A, {Strolger} LG, {Graur} O, {Filippenko} AV, {Jha}
  SW, {Riess} AG, {Bradac} M, {Weiner} BJ, {Scolnic} D, {Malkan} MA, {von der
  Linden} A, {Trenti} M, {Hjorth} J, {Gavazzi} R, {Fontana} A, {Merten} JC,
  {McCully} C, {Jones} T, {Postman} M, {Dressler} A, {Patel} B, {Cenko} SB,
  {Graham} ML, {Tucker} BE (2015) {Multiple images of a highly magnified
  supernova formed by an early-type cluster galaxy lens}. Science
  347:1123--1126, \doi{10.1126/science.aaa3350}, \eprint{1411.6009}

\bibitem[{{Kelly} et~al(2016{\natexlab{a}}){Kelly}, {Rodney}, {Treu},
  {Strolger}, {Foley}, {Jha}, {Selsing}, {Brammer}, {Brada{\v c}}, {Cenko},
  {Graur}, {Filippenko}, {Hjorth}, {McCully}, {Molino}, {Nonino}, {Riess},
  {Schmidt}, {Tucker}, {von der Linden}, {Weiner}, and {Zitrin}}]{Kelly2016}
{Kelly} PL, {Rodney} SA, {Treu} T, {Strolger} LG, {Foley} RJ, {Jha} SW,
  {Selsing} J, {Brammer} G, {Brada{\v c}} M, {Cenko} SB, {Graur} O,
  {Filippenko} AV, {Hjorth} J, {McCully} C, {Molino} A, {Nonino} M, {Riess} AG,
  {Schmidt} KB, {Tucker} B, {von der Linden} A, {Weiner} BJ, {Zitrin} A
  (2016{\natexlab{a}}) {Deja Vu All Over Again: The Reappearance of Supernova
  Refsdal}. \apjl 819:L8, \doi{10.3847/2041-8205/819/1/L8}, \eprint{1512.04654}

\bibitem[{{Kelly} et~al(2016{\natexlab{b}}){Kelly}, {Rodney}, {Treu},
  {Strolger}, {Foley}, {Jha}, {Selsing}, {Brammer}, {Brada{\v c}}, {Cenko},
  {Graur}, {Filippenko}, {Hjorth}, {McCully}, {Molino}, {Nonino}, {Riess},
  {Schmidt}, {Tucker}, {von der Linden}, {Weiner}, and {Zitrin}}]{KellyEtal16}
{Kelly} PL, {Rodney} SA, {Treu} T, {Strolger} LG, {Foley} RJ, {Jha} SW,
  {Selsing} J, {Brammer} G, {Brada{\v c}} M, {Cenko} SB, {Graur} O,
  {Filippenko} AV, {Hjorth} J, {McCully} C, {Molino} A, {Nonino} M, {Riess} AG,
  {Schmidt} KB, {Tucker} B, {von der Linden} A, {Weiner} BJ, {Zitrin} A
  (2016{\natexlab{b}}) {Deja Vu All Over Again: The Reappearance of Supernova
  Refsdal}. \apjl 819:L8, \doi{10.3847/2041-8205/819/1/L8}, \eprint{1512.04654}

\bibitem[{{Kochanek}(2002)}]{Kochanek02}
{Kochanek} CS (2002) {What Do Gravitational Lens Time Delays Measure?} \apj
  578:25--32, \doi{10.1086/342476}, \eprint{arXiv:astro-ph/0205319}

\bibitem[{{Kochanek} et~al(2006){Kochanek}, {Morgan}, {Falco}, {McLeod},
  {Winn}, {Dembicky}, and {Ketzeback}}]{Kochanek2006}
{Kochanek} CS, {Morgan} ND, {Falco} EE, {McLeod} BA, {Winn} JN, {Dembicky} J,
  {Ketzeback} B (2006) {The Time Delays of Gravitational Lens HE 0435-1223: An
  Early-Type Galaxy with a Rising Rotation Curve}. \apj 640:47--61,
  \doi{10.1086/499766}, \eprint{astro-ph/0508070}

\bibitem[{{Koopmans}(2005)}]{Koopmans05}
{Koopmans} LVE (2005) {Gravitational imaging of cold dark matter
  substructures}. \mnras 363:1136--1144, \doi{10.1111/j.1365-2966.2005.09523.x}

\bibitem[{{Koopmans} and {Treu}(2002)}]{KoopmansTreu02}
{Koopmans} LVE, {Treu} T (2002) {The Stellar Velocity Dispersion of the Lens
  Galaxy in MG 2016+112 at z=1.004}. \apjl 568:L5--L8, \doi{10.1086/340143},
  \eprint{arXiv:astro-ph/0201017}

\bibitem[{{Koopmans} et~al(2003){Koopmans}, {Treu}, {Fassnacht}, {Blandford},
  and {Surpi}}]{KoopmansEtal03}
{Koopmans} LVE, {Treu} T, {Fassnacht} CD, {Blandford} RD, {Surpi} G (2003) {The
  Hubble Constant from the Gravitational Lens B1608+656}. \apj 599:70--85,
  \doi{10.1086/379226}, \eprint{astro-ph/0306216}

\bibitem[{{Koopmans} et~al(2006){Koopmans}, {Treu}, {Bolton}, {Burles}, and
  {Moustakas}}]{KoopmansEtal06}
{Koopmans} LVE, {Treu} T, {Bolton} AS, {Burles} S, {Moustakas} LA (2006) {The
  Sloan Lens ACS Survey. III. The Structure and Formation of Early-Type
  Galaxies and Their Evolution since z \~{} 1}. \apj 649:599--615,
  \doi{10.1086/505696}, \eprint{arXiv:astro-ph/0601628}

\bibitem[{{Kormann} et~al(1994){Kormann}, {Schneider}, and
  {Bartelmann}}]{KormannEtal94}
{Kormann} R, {Schneider} P, {Bartelmann} M (1994) {Isothermal elliptical
  gravitational lens models}. \aap 284:285--299

\bibitem[{{Kundi{\'c}} et~al(1997){Kundi{\'c}}, {Turner}, {Colley}, {Gott},
  {Rhoads}, {Wang}, {Bergeron}, {Gloria}, {Long}, {Malhotra}, and
  {Wambsganss}}]{Kundic1997}
{Kundi{\'c}} T, {Turner} EL, {Colley} WN, {Gott} JR III, {Rhoads} JE, {Wang} Y,
  {Bergeron} LE, {Gloria} KA, {Long} DC, {Malhotra} S, {Wambsganss} J (1997) {A
  Robust Determination of the Time Delay in 0957+561A, B and a Measurement of
  the Global Value of Hubble's Constant}. \apj 482:75--82,
  \doi{10.1086/304147}, \eprint{astro-ph/9610162}

\bibitem[{{Lehar} et~al(1992){Lehar}, {Hewitt}, {Burke}, and
  {Roberts}}]{Lehar1992}
{Lehar} J, {Hewitt} JN, {Burke} BF, {Roberts} DH (1992) {The radio time delay
  in the double quasar 0957 + 561}. \apj 384:453--466, \doi{10.1086/170887}

\bibitem[{{Liao} et~al(2015){Liao}, {Treu}, {Marshall}, {Fassnacht},
  {Rumbaugh}, {Dobler}, {Aghamousa}, {Bonvin}, {Courbin}, {Hojjati}, {Jackson},
  {Kashyap}, {Rathna Kumar}, {Linder}, {Mandel}, {Meng}, {Meylan}, {Moustakas},
  {Prabhu}, {Romero-Wolf}, {Shafieloo}, {Siemiginowska}, {Stalin}, {Tak},
  {Tewes}, and {van Dyk}}]{Liao2015}
{Liao} K, {Treu} T, {Marshall} P, {Fassnacht} CD, {Rumbaugh} N, {Dobler} G,
  {Aghamousa} A, {Bonvin} V, {Courbin} F, {Hojjati} A, {Jackson} N, {Kashyap}
  V, {Rathna Kumar} S, {Linder} E, {Mandel} K, {Meng} XL, {Meylan} G,
  {Moustakas} LA, {Prabhu} TP, {Romero-Wolf} A, {Shafieloo} A, {Siemiginowska}
  A, {Stalin} CS, {Tak} H, {Tewes} M, {van Dyk} D (2015) {Strong Lens Time
  Delay Challenge. II. Results of TDC1}. \apj 800:11,
  \doi{10.1088/0004-637X/800/1/11}, \eprint{1409.1254}

\bibitem[{{Liao} et~al(2016){Liao}, {Chang}, {Kuo}, {Masui}, {Oppermann},
  {Pen}, and {Peterson}}]{Liao16}
{Liao} YW, {Chang} TC, {Kuo} CY, {Masui} KW, {Oppermann} N, {Pen} UL,
  {Peterson} JB (2016) {Accurate Polarization Calibration at 800 MHz with the
  Green Bank Telescope}. \apj 833:289, \doi{10.3847/1538-4357/833/2/289},
  \eprint{1610.04365}

\bibitem[{{Lin} et~al(2017){Lin}, {Buckley-Geer}, {Agnello}, {Ostrovski},
  {McMahon}, {Nord}, {Kuropatkin}, {Tucker}, {Treu}, {Chan}, {Suyu}, {Diehl},
  {Collett}, {Gill}, {More}, {Amara}, {Auger}, {Courbin}, {Fassnacht},
  {Frieman}, {Marshall}, {Meylan}, {Rusu}, {Abbott}, {Abdalla}, {Allam},
  {Banerji}, {Bechtol}, {Benoit-L{\'e}vy}, {Bertin}, {Brooks}, {Burke},
  {Carnero Rosell}, {Carrasco Kind}, {Carretero}, {Castander}, {Crocce},
  {D'Andrea}, {da Costa}, {Desai}, {Dietrich}, {Eifler}, {Finley}, {Flaugher},
  {Fosalba}, {Garc{\'{\i}}a-Bellido}, {Gaztanaga}, {Gerdes}, {Goldstein},
  {Gruen}, {Gruendl}, {Gschwend}, {Gutierrez}, {Honscheid}, {James}, {Kuehn},
  {Lahav}, {Li}, {Lima}, {Maia}, {March}, {Marshall}, {Martini}, {Melchior},
  {Menanteau}, {Miquel}, {Ogando}, {Plazas}, {Romer}, {Sanchez}, {Schindler},
  {Schubnell}, {Sevilla-Noarbe}, {Smith}, {Smith}, {Sobreira}, {Suchyta},
  {Swanson}, {Tarle}, {Thomas}, {Walker}, and {DES Collaboration}}]{Lin2017}
{Lin} H, {Buckley-Geer} E, {Agnello} A, {Ostrovski} F, {McMahon} RG, {Nord} B,
  {Kuropatkin} N, {Tucker} DL, {Treu} T, {Chan} JHH, {Suyu} SH, {Diehl} HT,
  {Collett} T, {Gill} MSS, {More} A, {Amara} A, {Auger} MW, {Courbin} F,
  {Fassnacht} CD, {Frieman} J, {Marshall} PJ, {Meylan} G, {Rusu} CE, {Abbott}
  TMC, {Abdalla} FB, {Allam} S, {Banerji} M, {Bechtol} K, {Benoit-L{\'e}vy} A,
  {Bertin} E, {Brooks} D, {Burke} DL, {Carnero Rosell} A, {Carrasco Kind} M,
  {Carretero} J, {Castander} FJ, {Crocce} M, {D'Andrea} CB, {da Costa} LN,
  {Desai} S, {Dietrich} JP, {Eifler} TF, {Finley} DA, {Flaugher} B, {Fosalba}
  P, {Garc{\'{\i}}a-Bellido} J, {Gaztanaga} E, {Gerdes} DW, {Goldstein} DA,
  {Gruen} D, {Gruendl} RA, {Gschwend} J, {Gutierrez} G, {Honscheid} K, {James}
  DJ, {Kuehn} K, {Lahav} O, {Li} TS, {Lima} M, {Maia} MAG, {March} M,
  {Marshall} JL, {Martini} P, {Melchior} P, {Menanteau} F, {Miquel} R, {Ogando}
  RLC, {Plazas} AA, {Romer} AK, {Sanchez} E, {Schindler} R, {Schubnell} M,
  {Sevilla-Noarbe} I, {Smith} M, {Smith} RC, {Sobreira} F, {Suchyta} E,
  {Swanson} MEC, {Tarle} G, {Thomas} D, {Walker} AR, {DES Collaboration} (2017)
  {Discovery of the Lensed Quasar System DES J0408-5354}. \apjl 838:L15,
  \doi{10.3847/2041-8213/aa624e}, \eprint{1702.00072}

\bibitem[{{Magain} et~al(1988){Magain}, {Surdej}, {Swings}, {Borgeest}, and
  {Kayser}}]{Magain1988}
{Magain} P, {Surdej} J, {Swings} JP, {Borgeest} U, {Kayser} R (1988) {Discovery
  of a quadruply lensed quasar - The 'clover leaf' H1413 + 117}. \nat
  334:325--327, \doi{10.1038/334325a0}

\bibitem[{{Magain} et~al(1998){Magain}, {Courbin}, and {Sohy}}]{Magain1998}
{Magain} P, {Courbin} F, {Sohy} S (1998) {Deconvolution with Correct Sampling}.
  \apj 494:472--477, \doi{10.1086/305187}, \eprint{astro-ph/9704059}

\bibitem[{{Marshall} et~al(2007){Marshall}, {Treu}, {Melbourne}, {Gavazzi},
  {Bundy}, {Ammons}, {Bolton}, {Burles}, {Larkin}, {Le Mignant}, {Koo},
  {Koopmans}, {Max}, {Moustakas}, {Steinbring}, and {Wright}}]{MarshallEtal07}
{Marshall} PJ, {Treu} T, {Melbourne} J, {Gavazzi} R, {Bundy} K, {Ammons} SM,
  {Bolton} AS, {Burles} S, {Larkin} JE, {Le Mignant} D, {Koo} DC, {Koopmans}
  LVE, {Max} CE, {Moustakas} LA, {Steinbring} E, {Wright} SA (2007)
  {Superresolving Distant Galaxies with Gravitational Telescopes: Keck Laser
  Guide Star Adaptive Optics and Hubble Space Telescope Imaging of the Lens
  System SDSS J0737+3216}. \apj 671:1196--1211, \doi{10.1086/523091},
  \eprint{0710.0637}

\bibitem[{{Martin} et~al(2010){Martin}, {Papastergis}, {Giovanelli}, {Haynes},
  {Springob}, and {Stierwalt}}]{2010ApJ...723.1359M}
{Martin} AM, {Papastergis} E, {Giovanelli} R, {Haynes} MP, {Springob} CM,
  {Stierwalt} S (2010) {The Arecibo Legacy Fast ALFA Survey. X. The H I Mass
  Function and {$\Omega$}\_H I from the 40\% ALFALFA Survey}. \apj
  723:1359--1374, \doi{10.1088/0004-637X/723/2/1359}, \eprint{1008.5107}

\bibitem[{{Masui} et~al(2010){Masui}, {McDonald}, and {Pen}}]{Masui2010}
{Masui} KW, {McDonald} P, {Pen} UL (2010) {Near-term measurements with 21 cm
  intensity mapping: Neutral hydrogen fraction and BAO at z$<$2}. \prd
  81(10):103527, \doi{10.1103/PhysRevD.81.103527}, \eprint{1001.4811}

\bibitem[{{Masui} et~al(2013){Masui}, {Switzer}, {Banavar}, {Bandura}, {Blake},
  {Calin}, {Chang}, {Chen}, {Li}, {Liao}, {Natarajan}, {Pen}, {Peterson},
  {Shaw}, and {Voytek}}]{Masui2013}
{Masui} KW, {Switzer} ER, {Banavar} N, {Bandura} K, {Blake} C, {Calin} LM,
  {Chang} TC, {Chen} X, {Li} YC, {Liao} YW, {Natarajan} A, {Pen} UL, {Peterson}
  JB, {Shaw} JR, {Voytek} TC (2013) {Measurement of 21 cm Brightness
  Fluctuations at z \~{} 0.8 in Cross-correlation}. \apjl 763:L20,
  \doi{10.1088/2041-8205/763/1/L20}, \eprint{1208.0331}

\bibitem[{{Matsubara}(2004)}]{Matsubara:2004}
{Matsubara} T (2004) {Correlation Function in Deep Redshift Space as a
  Cosmological Probe}. \apj 615:573--585, \doi{10.1086/424561},
  \eprint{arXiv:astro-ph/w349}

\bibitem[{{McCully} et~al(2017){McCully}, {Keeton}, {Wong}, and
  {Zabludoff}}]{McCullyEtal17}
{McCully} C, {Keeton} CR, {Wong} KC, {Zabludoff} AI (2017) {Quantifying
  Environmental and Line-of-sight Effects in Models of Strong Gravitational
  Lens Systems}. \apj 836:141, \doi{10.3847/1538-4357/836/1/141},
  \eprint{1601.05417}

\bibitem[{{Momcheva} et~al(2006){Momcheva}, {Williams}, {Keeton}, and
  {Zabludoff}}]{MomchevaEtal06}
{Momcheva} I, {Williams} K, {Keeton} C, {Zabludoff} A (2006) {A Spectroscopic
  Study of the Environments of Gravitational Lens Galaxies}. \apj 641:169--189,
  \doi{10.1086/500382}, \eprint{arXiv:astro-ph/0511594}

\bibitem[{{More} et~al(2017){More}, {Suyu}, {Oguri}, {More}, and
  {Lee}}]{More2017}
{More} A, {Suyu} SH, {Oguri} M, {More} S, {Lee} CH (2017) {Interpreting the
  Strongly Lensed Supernova iPTF16geu: Time Delay Predictions, Microlensing,
  and Lensing Rates}. \apjl 835:L25, \doi{10.3847/2041-8213/835/2/L25},
  \eprint{1611.04866}

\bibitem[{{Morgan} et~al(2012){Morgan}, {Hainline}, {Chen}, {Tewes},
  {Kochanek}, {Dai}, {Kozlowski}, {Blackburne}, {Mosquera}, {Chartas},
  {Courbin}, and {Meylan}}]{Morgan2012}
{Morgan} CW, {Hainline} LJ, {Chen} B, {Tewes} M, {Kochanek} CS, {Dai} X,
  {Kozlowski} S, {Blackburne} JA, {Mosquera} AM, {Chartas} G, {Courbin} F,
  {Meylan} G (2012) {Further Evidence that Quasar X-Ray Emitting Regions are
  Compact: X-Ray and Optical Microlensing in the Lensed Quasar Q J0158-4325}.
  \apj 756:52, \doi{10.1088/0004-637X/756/1/52}, \eprint{1205.4727}

\bibitem[{{Newburgh} et~al(2016){Newburgh}, {Bandura}, {Bucher}, {Chang},
  {Chiang}, {Cliche}, {Dav{\'e}}, {Dobbs}, {Clarkson}, {Ganga}, {Gogo},
  {Gumba}, {Gupta}, {Hilton}, {Johnstone}, {Karastergiou}, {Kunz}, {Lokhorst},
  {Maartens}, {Macpherson}, {Mdlalose}, {Moodley}, {Ngwenya}, {Parra},
  {Peterson}, {Recnik}, {Saliwanchik}, {Santos}, {Sievers}, {Smirnov},
  {Stronkhorst}, {Taylor}, {Vanderlinde}, {Van Vuuren}, {Weltman}, and
  {Witzemann}}]{HIRAX}
{Newburgh} LB, {Bandura} K, {Bucher} MA, {Chang} TC, {Chiang} HC, {Cliche} JF,
  {Dav{\'e}} R, {Dobbs} M, {Clarkson} C, {Ganga} KM, {Gogo} T, {Gumba} A,
  {Gupta} N, {Hilton} M, {Johnstone} B, {Karastergiou} A, {Kunz} M, {Lokhorst}
  D, {Maartens} R, {Macpherson} S, {Mdlalose} M, {Moodley} K, {Ngwenya} L,
  {Parra} JM, {Peterson} J, {Recnik} O, {Saliwanchik} B, {Santos} MG, {Sievers}
  JL, {Smirnov} O, {Stronkhorst} P, {Taylor} R, {Vanderlinde} K, {Van Vuuren}
  G, {Weltman} A, {Witzemann} A (2016) {HIRAX: a probe of dark energy and radio
  transients}. In: Ground-based and Airborne Telescopes VI, \procspie, vol
  9906, p 99065X, \doi{10.1117/12.2234286}, \eprint{1607.02059}

\bibitem[{{Nightingale} and {Dye}(2015)}]{NightingaleDye15}
{Nightingale} JW, {Dye} S (2015) {Adaptive semi-linear inversion of strong
  gravitational lens imaging}. \mnras 452:2940--2959,
  \doi{10.1093/mnras/stv1455}, \eprint{1412.7436}

\bibitem[{{Ofek} and {Maoz}(2003)}]{Ofek2003}
{Ofek} EO, {Maoz} D (2003) {Time-Delay Measurement of the Lensed Quasar HE
  1104-1805}. \apj 594:101--106, \doi{10.1086/376903},
  \eprint{astro-ph/0305200}

\bibitem[{{Oguri}(2007)}]{Oguri07}
{Oguri} M (2007) {Gravitational Lens Time Delays: A Statistical Assessment of
  Lens Model Dependences and Implications for the Global Hubble Constant}. \apj
  660:1--15, \doi{10.1086/513093}, \eprint{arXiv:astro-ph/0609694}

\bibitem[{{Oguri}(2010)}]{Oguri10}
{Oguri} M (2010) {The Mass Distribution of SDSS J1004+4112 Revisited}. \pasj
  62:1017--1024, \doi{10.1093/pasj/62.4.1017}, \eprint{1005.3103}

\bibitem[{{Oguri} and {Marshall}(2010)}]{OguriMarshall10}
{Oguri} M, {Marshall} PJ (2010) {Gravitationally lensed quasars and supernovae
  in future wide-field optical imaging surveys}. \mnras 405:2579--2593,
  \doi{10.1111/j.1365-2966.2010.16639.x}, \eprint{1001.2037}

\bibitem[{{Okumura} et~al(2008){Okumura}, {Matsubara}, {Eisenstein}, {Kayo},
  {Hikage}, {Szalay}, and {Schneider}}]{Okumura:2008}
{Okumura} T, {Matsubara} T, {Eisenstein} DJ, {Kayo} I, {Hikage} C, {Szalay} AS,
  {Schneider} DP (2008) {Large-Scale Anisotropic Correlation Function of SDSS
  Luminous Red Galaxies}. \apj 676:889-898, \doi{10.1086/528951},
  \eprint{0711.3640}

\bibitem[{{Okumura} et~al(2016){Okumura}, {Hikage}, {Totani}, {Tonegawa},
  {Okada}, {Glazebrook}, {Blake}, {Ferreira}, {More}, {Taruya}, {Tsujikawa},
  {Akiyama}, {Dalton}, {Goto}, {Ishikawa}, {Iwamuro}, {Matsubara},
  {Nishimichi}, {Ohta}, {Shimizu}, {Takahashi}, {Takato}, {Tamura}, {Yabe}, and
  {Yoshida}}]{Okumura:2016}
{Okumura} T, {Hikage} C, {Totani} T, {Tonegawa} M, {Okada} H, {Glazebrook} K,
  {Blake} C, {Ferreira} PG, {More} S, {Taruya} A, {Tsujikawa} S, {Akiyama} M,
  {Dalton} G, {Goto} T, {Ishikawa} T, {Iwamuro} F, {Matsubara} T, {Nishimichi}
  T, {Ohta} K, {Shimizu} I, {Takahashi} R, {Takato} N, {Tamura} N, {Yabe} K,
  {Yoshida} N (2016) {The Subaru FMOS galaxy redshift survey (FastSound). IV.
  New constraint on gravity theory from redshift space distortions at z 1.4}.
  \pasj 68:38, \doi{10.1093/pasj/psw029}, \eprint{1511.08083}

\bibitem[{{Oldham} et~al(2017){Oldham}, {Auger}, {Fassnacht}, {Treu}, {Brewer},
  {Koopmans}, {Lagattuta}, {Marshall}, {McKean}, and {Vegetti}}]{OldhamEtal17}
{Oldham} L, {Auger} MW, {Fassnacht} CD, {Treu} T, {Brewer} BJ, {Koopmans} LVE,
  {Lagattuta} D, {Marshall} P, {McKean} J, {Vegetti} S (2017) {Red nuggets grow
  inside-out: evidence from gravitational lensing}. \mnras 465:3185--3202,
  \doi{10.1093/mnras/stw2832}, \eprint{1611.00008}

\bibitem[{{Oscoz} et~al(1997){Oscoz}, {Mediavilla}, {Goicoechea},
  {Serra-Ricart}, and {Buitrago}}]{Oscoz1997}
{Oscoz} A, {Mediavilla} E, {Goicoechea} LJ, {Serra-Ricart} M, {Buitrago} J
  (1997) {Time Delay of QSO 0957+561 and Cosmological Implications}. \apjl
  479:L89--L92, \doi{10.1086/310599}

\bibitem[{{Paraficz} and {Hjorth}(2009)}]{ParaficzHjorth09}
{Paraficz} D, {Hjorth} J (2009) {Gravitational lenses as cosmic rulers:
  {$\Omega$}$_{m}$, {$\Omega$}$_{Λ}$ from time delays and velocity
  dispersions}. \aap 507:L49--L52, \doi{10.1051/0004-6361/200913307},
  \eprint{0910.5823}

\bibitem[{{Peebles} and {Yu}(1970)}]{Peebles:1970}
{Peebles} PJE, {Yu} JT (1970) {Primeval Adiabatic Perturbation in an Expanding
  Universe}. \apj 162:815, \doi{10.1086/150713}

\bibitem[{{Peterson} et~al(2006){Peterson}, {Bandura}, and
  {Pen}}]{2006astro.ph..6104P}
{Peterson} JB, {Bandura} K, {Pen} UL (2006) {The Hubble Sphere Hydrogen
  Survey}. ArXiv Astrophysics e-prints \eprint{arXiv:astro-ph/0606104}

\bibitem[{{Planck Collaboration} et~al(2016{\natexlab{a}}){Planck
  Collaboration}, {Adam}, {Ade}, {Aghanim}, {Akrami}, {Alves}, {Arg{\"u}eso},
  {Arnaud}, {Arroja}, {Ashdown}, and et~al.}]{Planck-Collaboration:2016a}
{Planck Collaboration}, {Adam} R, {Ade} PAR, {Aghanim} N, {Akrami} Y, {Alves}
  MIR, {Arg{\"u}eso} F, {Arnaud} M, {Arroja} F, {Ashdown} M, et~al
  (2016{\natexlab{a}}) {Planck 2015 results. I. Overview of products and
  scientific results}. \aap 594:A1, \doi{10.1051/0004-6361/201527101},
  \eprint{1502.01582}

\bibitem[{{Planck Collaboration} et~al(2016{\natexlab{b}}){Planck
  Collaboration}, {Ade}, {Aghanim}, {Arnaud}, {Ashdown}, {Aumont},
  {Baccigalupi}, {Banday}, {Barreiro}, {Bartlett}, and et~al.}]{Planck2016}
{Planck Collaboration}, {Ade} PAR, {Aghanim} N, {Arnaud} M, {Ashdown} M,
  {Aumont} J, {Baccigalupi} C, {Banday} AJ, {Barreiro} RB, {Bartlett} JG, et~al
  (2016{\natexlab{b}}) {Planck 2015 results. XIII. Cosmological parameters}.
  \aap 594:A13, \doi{10.1051/0004-6361/201525830}, \eprint{1502.01589}

\bibitem[{{Rathna Kumar} et~al(2013){Rathna Kumar}, {Tewes}, {Stalin},
  {Courbin}, {Asfandiyarov}, {Meylan}, {Eulaers}, {Prabhu}, {Magain}, {Van
  Winckel}, and {Ehgamberdiev}}]{Kumar2013}
{Rathna Kumar} S, {Tewes} M, {Stalin} CS, {Courbin} F, {Asfandiyarov} I,
  {Meylan} G, {Eulaers} E, {Prabhu} TP, {Magain} P, {Van Winckel} H,
  {Ehgamberdiev} S (2013) {COSMOGRAIL: the COSmological MOnitoring of
  GRAvItational Lenses. XIV. Time delay of the doubly lensed quasar SDSS
  J1001+5027}. \aap 557:A44, \doi{10.1051/0004-6361/201322116},
  \eprint{1306.5105}

\bibitem[{{Rathna Kumar} et~al(2015){Rathna Kumar}, {Stalin}, and
  {Prabhu}}]{Kumar2015}
{Rathna Kumar} S, {Stalin} CS, {Prabhu} TP (2015) {H$_{0}$ from ten
  well-measured time delay lenses}. \aap 580:A38,
  \doi{10.1051/0004-6361/201423977}, \eprint{1404.2920}

\bibitem[{{Refsdal}(1964)}]{Refsdal64}
{Refsdal} S (1964) {On the possibility of determining Hubble's parameter and
  the masses of galaxies from the gravitational lens effect}. \mnras 128:307

\bibitem[{{Riess} et~al(2016{\natexlab{a}}){Riess}, {Macri}, {Hoffmann},
  {Scolnic}, {Casertano}, {Filippenko}, {Tucker}, {Reid}, {Jones}, {Silverman},
  {Chornock}, {Challis}, {Yuan}, {Brown}, and {Foley}}]{Riess:2016}
{Riess} AG, {Macri} LM, {Hoffmann} SL, {Scolnic} D, {Casertano} S, {Filippenko}
  AV, {Tucker} BE, {Reid} MJ, {Jones} DO, {Silverman} JM, {Chornock} R,
  {Challis} P, {Yuan} W, {Brown} PJ, {Foley} RJ (2016{\natexlab{a}}) {A 2.4\%
  Determination of the Local Value of the Hubble Constant}. \apj 826:56,
  \doi{10.3847/0004-637X/826/1/56}, \eprint{1604.01424}

\bibitem[{{Riess} et~al(2016{\natexlab{b}}){Riess}, {Macri}, {Hoffmann},
  {Scolnic}, {Casertano}, {Filippenko}, {Tucker}, {Reid}, {Jones}, {Silverman},
  {Chornock}, {Challis}, {Yuan}, and {Foley}}]{RiessEtal16}
{Riess} AG, {Macri} LM, {Hoffmann} SL, {Scolnic} D, {Casertano} S, {Filippenko}
  AV, {Tucker} BE, {Reid} MJ, {Jones} DO, {Silverman} JM, {Chornock} R,
  {Challis} P, {Yuan} W, {Foley} RJ (2016{\natexlab{b}}) {A 2.4\% Determination
  of the Local Value of the Hubble Constant}. ArXiv e-prints (160401424)
  \eprint{1604.01424}

\bibitem[{{Riess} et~al(2018){Riess}, {Casertano}, {Yuan}, {Macri}, {Anderson},
  {Mackenty}, {Bowers}, {Clubb}, {Filippenko}, {Jones}, and
  {Tucker}}]{RiessEtal18}
{Riess} AG, {Casertano} S, {Yuan} W, {Macri} L, {Anderson} J, {Mackenty} JW,
  {Bowers} JB, {Clubb} KI, {Filippenko} AV, {Jones} DO, {Tucker} BE (2018) {New
  Parallaxes of Galactic Cepheids from Spatially Scanning the Hubble Space
  Telescope: Implications for the Hubble Constant}. ArXiv e-prints (180101120)
  \eprint{1801.01120}

\bibitem[{{Rodney} et~al(2016){Rodney}, {Strolger}, {Kelly}, {Brada{\v c}},
  {Brammer}, {Filippenko}, {Foley}, {Graur}, {Hjorth}, {Jha}, {McCully},
  {Molino}, {Riess}, {Schmidt}, {Selsing}, {Sharon}, {Treu}, {Weiner}, and
  {Zitrin}}]{Rodney2016}
{Rodney} SA, {Strolger} LG, {Kelly} PL, {Brada{\v c}} M, {Brammer} G,
  {Filippenko} AV, {Foley} RJ, {Graur} O, {Hjorth} J, {Jha} SW, {McCully} C,
  {Molino} A, {Riess} AG, {Schmidt} KB, {Selsing} J, {Sharon} K, {Treu} T,
  {Weiner} BJ, {Zitrin} A (2016) {SN Refsdal: Photometry and Time Delay
  Measurements of the First Einstein Cross Supernova}. \apj 820:50,
  \doi{10.3847/0004-637X/820/1/50}, \eprint{1512.05734}

\bibitem[{{Rusu} et~al(2017){Rusu}, {Fassnacht}, {Sluse}, {Hilbert}, {Wong},
  {Huang}, {Suyu}, {Collett}, {Marshall}, {Treu}, and {Koopmans}}]{RusuEtal17}
{Rusu} CE, {Fassnacht} CD, {Sluse} D, {Hilbert} S, {Wong} KC, {Huang} KH,
  {Suyu} SH, {Collett} TE, {Marshall} PJ, {Treu} T, {Koopmans} LVE (2017)
  {H0LiCOW - III. Quantifying the effect of mass along the line of sight to the
  gravitational lens HE 0435-1223 through weighted galaxy counts}. \mnras
  467:4220--4242, \doi{10.1093/mnras/stx285}, \eprint{1607.01047}

\bibitem[{{Saha} et~al(2006){Saha}, {Coles}, {Macci{\`o}}, and
  {Williams}}]{SahaEtal06}
{Saha} P, {Coles} J, {Macci{\`o}} AV, {Williams} LLR (2006) {The Hubble Time
  Inferred from 10 Time Delay Lenses}. \apjl 650:L17--L20,
  \doi{10.1086/507583}, \eprint{astro-ph/0607240}

\bibitem[{{Schechter} et~al(1997){Schechter}, {Bailyn}, {Barr}, {Barvainis},
  {Becker}, {Bernstein}, {Blakeslee}, {Bus}, {Dressler}, {Falco}, {Fesen},
  {Fischer}, {Gebhardt}, {Harmer}, {Hewitt}, {Hjorth}, {Hurt}, {Jaunsen},
  {Mateo}, {Mehlert}, {Richstone}, {Sparke}, {Thorstensen}, {Tonry}, {Wegner},
  {Willmarth}, and {Worthey}}]{Schechter1997}
{Schechter} PL, {Bailyn} CD, {Barr} R, {Barvainis} R, {Becker} CM, {Bernstein}
  GM, {Blakeslee} JP, {Bus} SJ, {Dressler} A, {Falco} EE, {Fesen} RA, {Fischer}
  P, {Gebhardt} K, {Harmer} D, {Hewitt} JN, {Hjorth} J, {Hurt} T, {Jaunsen} AO,
  {Mateo} M, {Mehlert} D, {Richstone} DO, {Sparke} LS, {Thorstensen} JR,
  {Tonry} JL, {Wegner} G, {Willmarth} DW, {Worthey} G (1997) {The Quadruple
  Gravitational Lens PG 1115+080: Time Delays and Models}. \apjl 475:L85--L88,
  \doi{10.1086/310478}, \eprint{astro-ph/9611051}

\bibitem[{{Schmidt}(1963)}]{Schmidt1963}
{Schmidt} M (1963) {3C 273 : A Star-Like Object with Large Red-Shift}. \nat
  197:1040, \doi{10.1038/1971040a0}

\bibitem[{{Schneider}(2014)}]{Schneider14}
{Schneider} P (2014) {Generalized multi-plane gravitational lensing: time
  delays, recursive lens equation, and the mass-sheet transformation}. ArXiv
  e-prints (14090015) \eprint{1409.0015}

\bibitem[{{Schneider} and {Sluse}(2013)}]{SchneiderSluse13}
{Schneider} P, {Sluse} D (2013) {Mass-sheet degeneracy, power-law models and
  external convergence: Impact on the determination of the Hubble constant from
  gravitational lensing}. \aap 559:A37, \doi{10.1051/0004-6361/201321882},
  \eprint{1306.0901}

\bibitem[{{Schneider} and {Sluse}(2014)}]{SchneiderSluse14}
{Schneider} P, {Sluse} D (2014) {Source-position transformation: an approximate
  invariance in strong gravitational lensing}. \aap 564:A103,
  \doi{10.1051/0004-6361/201322106}, \eprint{1306.4675}

\bibitem[{{Schneider} et~al(1992){Schneider}, {Ehlers}, and
  {Falco}}]{SchneiderEtal92}
{Schneider} P, {Ehlers} J, {Falco} EE (1992) {Gravitational Lenses}.
  Gravitational Lenses, XIV, 560 pp~112 figs~Springer-Verlag Berlin Heidelberg
  New York

\bibitem[{{Schneider} et~al(2006){Schneider}, {Kochanek}, and
  {Wambsganss}}]{SchneiderEtal06}
{Schneider} P, {Kochanek} CS, {Wambsganss} J (2006) {Gravitational Lensing:
  Strong, Weak and Micro (Springer)}. \doi{10.1007/978-3-540-30310-7}

\bibitem[{{Seo} and {Eisenstein}(2003)}]{Seo:2003}
{Seo} HJ, {Eisenstein} DJ (2003) {Probing Dark Energy with Baryonic Acoustic
  Oscillations from Future Large Galaxy Redshift Surveys}. \apj 598:720--740,
  \doi{10.1086/379122}, \eprint{astro-ph/0307460}

\bibitem[{{Seo} et~al(2010){Seo}, {Dodelson}, {Marriner}, {Mcginnis},
  {Stebbins}, {Stoughton}, and {Vallinotto}}]{2010ApJ...721..164S}
{Seo} HJ, {Dodelson} S, {Marriner} J, {Mcginnis} D, {Stebbins} A, {Stoughton}
  C, {Vallinotto} A (2010) {A Ground-based 21 cm Baryon Acoustic Oscillation
  Survey}. Astrophys J 721:164--173, \doi{10.1088/0004-637X/721/1/164},
  \eprint{0910.5007}

\bibitem[{{Sereno} and {Paraficz}(2014)}]{SerenoParaficz14}
{Sereno} M, {Paraficz} D (2014) {Hubble constant and dark energy inferred from
  free-form determined time delay distances}. \mnras 437:600--605,
  \doi{10.1093/mnras/stt1938}, \eprint{1310.2251}

\bibitem[{{Shajib} et~al(2017){Shajib}, {Treu}, and {Agnello}}]{ShajibEtal17}
{Shajib} AJ, {Treu} T, {Agnello} A (2017) {Improving time-delay cosmography
  with spatially resolved kinematics}. ArXiv e-prints \eprint{1709.01517}

\bibitem[{{Shalyapin} and {Goicoechea}(2017)}]{Shalyapin2017}
{Shalyapin} VN, {Goicoechea} LJ (2017) {Doubly Imaged Quasar SDSS J1515+1511:
  Time Delay and Lensing Galaxies}. \apj 836:14,
  \doi{10.3847/1538-4357/836/1/14}, \eprint{1701.04272}

\bibitem[{Shaw et~al(2013)Shaw, Sigurdson, Pen, Stebbins, and
  Sitwell}]{Shaw:2013wza}
Shaw JR, Sigurdson K, Pen UL, Stebbins A, Sitwell M (2013) {All-Sky
  Interferometry with Spherical Harmonic Transit Telescopes}. Astrophys J
  \doi{10.1088/0004-637X/781/2/57}, \eprint{1302.0327}

\bibitem[{Shaw et~al(2015)Shaw, Sigurdson, Sitwell, Stebbins, and
  Pen}]{Shaw:2014khi}
Shaw JR, Sigurdson K, Sitwell M, Stebbins A, Pen UL (2015) {Coaxing cosmic 21
  cm fluctuations from the polarized sky using m-mode analysis}. Phys Rev
  D91(8):083,514, \doi{10.1103/PhysRevD.91.083514}, \eprint{1401.2095}

\bibitem[{{Sluse} et~al(2017){Sluse}, {Sonnenfeld}, {Rumbaugh}, {Rusu},
  {Fassnacht}, {Treu}, {Suyu}, {Wong}, {Auger}, {Bonvin}, {Collett}, {Courbin},
  {Hilbert}, {Koopmans}, {Marshall}, {Meylan}, {Spiniello}, and
  {Tewes}}]{SluseEtal17}
{Sluse} D, {Sonnenfeld} A, {Rumbaugh} N, {Rusu} CE, {Fassnacht} CD, {Treu} T,
  {Suyu} SH, {Wong} KC, {Auger} MW, {Bonvin} V, {Collett} T, {Courbin} F,
  {Hilbert} S, {Koopmans} LVE, {Marshall} PJ, {Meylan} G, {Spiniello} C,
  {Tewes} M (2017) {H0LiCOW - II. Spectroscopic survey and galaxy-group
  identification of the strong gravitational lens system HE 0435-1223}. \mnras
  470:4838--4857, \doi{10.1093/mnras/stx1484}, \eprint{1607.00382}

\bibitem[{{Spergel} et~al(2013){Spergel}, {Gehrels}, {Breckinridge}, {Donahue},
  {Dressler}, {Gaudi}, {Greene}, {Guyon}, {Hirata}, {Kalirai}, {Kasdin},
  {Moos}, {Perlmutter}, {Postman}, {Rauscher}, {Rhodes}, {Wang}, {Weinberg},
  {Centrella}, {Traub}, {Baltay}, {Colbert}, {Bennett}, {Kiessling},
  {Macintosh}, {Merten}, {Mortonson}, {Penny}, {Rozo}, {Savransky},
  {Stapelfeldt}, {Zu}, {Baker}, {Cheng}, {Content}, {Dooley}, {Foote},
  {Goullioud}, {Grady}, {Jackson}, {Kruk}, {Levine}, {Melton}, {Peddie},
  {Ruffa}, and {Shaklan}}]{Spergel:2013}
{Spergel} D, {Gehrels} N, {Breckinridge} J, {Donahue} M, {Dressler} A, {Gaudi}
  BS, {Greene} T, {Guyon} O, {Hirata} C, {Kalirai} J, {Kasdin} NJ, {Moos} W,
  {Perlmutter} S, {Postman} M, {Rauscher} B, {Rhodes} J, {Wang} Y, {Weinberg}
  D, {Centrella} J, {Traub} W, {Baltay} C, {Colbert} J, {Bennett} D,
  {Kiessling} A, {Macintosh} B, {Merten} J, {Mortonson} M, {Penny} M, {Rozo} E,
  {Savransky} D, {Stapelfeldt} K, {Zu} Y, {Baker} C, {Cheng} E, {Content} D,
  {Dooley} J, {Foote} M, {Goullioud} R, {Grady} K, {Jackson} C, {Kruk} J,
  {Levine} M, {Melton} M, {Peddie} C, {Ruffa} J, {Shaklan} S (2013) {Wide-Field
  InfraRed Survey Telescope-Astrophysics Focused Telescope Assets WFIRST-AFTA
  Final Report}. ArXiv e-prints \eprint{1305.5422}

\bibitem[{{Spergel} et~al(2015){Spergel}, {Gehrels}, {Baltay}, {Bennett},
  {Breckinridge}, {Donahue}, {Dressler}, {Gaudi}, {Greene}, {Guyon}, {Hirata},
  {Kalirai}, {Kasdin}, {Macintosh}, {Moos}, {Perlmutter}, {Postman},
  {Rauscher}, {Rhodes}, {Wang}, {Weinberg}, {Benford}, {Hudson}, {Jeong},
  {Mellier}, {Traub}, {Yamada}, {Capak}, {Colbert}, {Masters}, {Penny},
  {Savransky}, {Stern}, {Zimmerman}, {Barry}, {Bartusek}, {Carpenter}, {Cheng},
  {Content}, {Dekens}, {Demers}, {Grady}, {Jackson}, {Kuan}, {Kruk}, {Melton},
  {Nemati}, {Parvin}, {Poberezhskiy}, {Peddie}, {Ruffa}, {Wallace}, {Whipple},
  {Wollack}, and {Zhao}}]{wfirst}
{Spergel} D, {Gehrels} N, {Baltay} C, {Bennett} D, {Breckinridge} J, {Donahue}
  M, {Dressler} A, {Gaudi} BS, {Greene} T, {Guyon} O, {Hirata} C, {Kalirai} J,
  {Kasdin} NJ, {Macintosh} B, {Moos} W, {Perlmutter} S, {Postman} M, {Rauscher}
  B, {Rhodes} J, {Wang} Y, {Weinberg} D, {Benford} D, {Hudson} M, {Jeong} WS,
  {Mellier} Y, {Traub} W, {Yamada} T, {Capak} P, {Colbert} J, {Masters} D,
  {Penny} M, {Savransky} D, {Stern} D, {Zimmerman} N, {Barry} R, {Bartusek} L,
  {Carpenter} K, {Cheng} E, {Content} D, {Dekens} F, {Demers} R, {Grady} K,
  {Jackson} C, {Kuan} G, {Kruk} J, {Melton} M, {Nemati} B, {Parvin} B,
  {Poberezhskiy} I, {Peddie} C, {Ruffa} J, {Wallace} JK, {Whipple} A, {Wollack}
  E, {Zhao} F (2015) {Wide-Field InfrarRed Survey Telescope-Astrophysics
  Focused Telescope Assets WFIRST-AFTA 2015 Report}. ArXiv e-prints
  \eprint{1503.03757}

\bibitem[{{Suyu}(2012)}]{Suyu12}
{Suyu} SH (2012) {Cosmography from two-image lens systems: overcoming the lens
  profile slope degeneracy}. ArXiv e-prints (12020287) \eprint{1202.0287}

\bibitem[{{Suyu} and {Halkola}(2010)}]{SuyuHalkola10}
{Suyu} SH, {Halkola} A (2010) {The halos of satellite galaxies: the companion
  of the massive elliptical lens SL2S J08544-0121}. \aap 524:A94,
  \doi{10.1051/0004-6361/201015481}, \eprint{1007.4815}

\bibitem[{{Suyu} et~al(2006){Suyu}, {Marshall}, {Hobson}, and
  {Blandford}}]{SuyuEtal06}
{Suyu} SH, {Marshall} PJ, {Hobson} MP, {Blandford} RD (2006) {A Bayesian
  analysis of regularized source inversions in gravitational lensing}. \mnras
  371:983--998, \doi{10.1111/j.1365-2966.2006.10733.x},
  \eprint{astro-ph/0601493}

\bibitem[{{Suyu} et~al(2009){Suyu}, {Marshall}, {Blandford}, {Fassnacht},
  {Koopmans}, {McKean}, and {Treu}}]{SuyuEtal09}
{Suyu} SH, {Marshall} PJ, {Blandford} RD, {Fassnacht} CD, {Koopmans} LVE,
  {McKean} JP, {Treu} T (2009) {Dissecting the Gravitational Lens B1608+656. I.
  Lens Potential Reconstruction}. \apj 691:277--298,
  \doi{10.1088/0004-637X/691/1/277}, \eprint{0804.2827}

\bibitem[{{Suyu} et~al(2010){Suyu}, {Marshall}, {Auger}, {Hilbert},
  {Blandford}, {Koopmans}, {Fassnacht}, and {Treu}}]{SuyuEtal10}
{Suyu} SH, {Marshall} PJ, {Auger} MW, {Hilbert} S, {Blandford} RD, {Koopmans}
  LVE, {Fassnacht} CD, {Treu} T (2010) {Dissecting the Gravitational lens
  B1608+656. II. Precision Measurements of the Hubble Constant, Spatial
  Curvature, and the Dark Energy Equation of State}. \apj 711:201--221,
  \doi{10.1088/0004-637X/711/1/201}, \eprint{0910.2773}

\bibitem[{{Suyu} et~al(2012){Suyu}, {Hensel}, {McKean}, {Fassnacht}, {Treu},
  {Halkola}, {Norbury}, {Jackson}, {Schneider}, {Thompson}, {Auger},
  {Koopmans}, and {Matthews}}]{SuyuEtal12a}
{Suyu} SH, {Hensel} SW, {McKean} JP, {Fassnacht} CD, {Treu} T, {Halkola} A,
  {Norbury} M, {Jackson} N, {Schneider} P, {Thompson} D, {Auger} MW, {Koopmans}
  LVE, {Matthews} K (2012) {Disentangling Baryons and Dark Matter in the Spiral
  Gravitational Lens B1933+503}. \apj 750:10, \doi{10.1088/0004-637X/750/1/10},
  \eprint{1110.2536}

\bibitem[{{Suyu} et~al(2013){Suyu}, {Auger}, {Hilbert}, {Marshall}, {Tewes},
  {Treu}, {Fassnacht}, {Koopmans}, {Sluse}, {Blandford}, {Courbin}, and
  {Meylan}}]{SuyuEtal13}
{Suyu} SH, {Auger} MW, {Hilbert} S, {Marshall} PJ, {Tewes} M, {Treu} T,
  {Fassnacht} CD, {Koopmans} LVE, {Sluse} D, {Blandford} RD, {Courbin} F,
  {Meylan} G (2013) {Two Accurate Time-delay Distances from Strong Lensing:
  Implications for Cosmology}. \apj 766:70, \doi{10.1088/0004-637X/766/2/70},
  \eprint{1208.6010}

\bibitem[{{Suyu} et~al(2014){Suyu}, {Treu}, {Hilbert}, {Sonnenfeld}, {Auger},
  {Blandford}, {Collett}, {Courbin}, {Fassnacht}, {Koopmans}, {Marshall},
  {Meylan}, {Spiniello}, and {Tewes}}]{SuyuEtal14}
{Suyu} SH, {Treu} T, {Hilbert} S, {Sonnenfeld} A, {Auger} MW, {Blandford} RD,
  {Collett} T, {Courbin} F, {Fassnacht} CD, {Koopmans} LVE, {Marshall} PJ,
  {Meylan} G, {Spiniello} C, {Tewes} M (2014) {Cosmology from Gravitational
  Lens Time Delays and Planck Data}. \apjl 788:L35,
  \doi{10.1088/2041-8205/788/2/L35}, \eprint{1306.4732}

\bibitem[{{Suyu} et~al(2017){Suyu}, {Bonvin}, {Courbin}, {Fassnacht}, {Rusu},
  {Sluse}, {Treu}, {Wong}, {Auger}, {Ding}, {Hilbert}, {Marshall}, {Rumbaugh},
  {Sonnenfeld}, {Tewes}, {Tihhonova}, {Agnello}, {Blandford}, {Chen},
  {Collett}, {Koopmans}, {Liao}, {Meylan}, and {Spiniello}}]{SuyuEtal17}
{Suyu} SH, {Bonvin} V, {Courbin} F, {Fassnacht} CD, {Rusu} CE, {Sluse} D,
  {Treu} T, {Wong} KC, {Auger} MW, {Ding} X, {Hilbert} S, {Marshall} PJ,
  {Rumbaugh} N, {Sonnenfeld} A, {Tewes} M, {Tihhonova} O, {Agnello} A,
  {Blandford} RD, {Chen} GCF, {Collett} T, {Koopmans} LVE, {Liao} K, {Meylan}
  G, {Spiniello} C (2017) {H0LiCOW - I. H$_{0}$ Lenses in COSMOGRAIL's
  Wellspring: program overview}. \mnras 468:2590--2604,
  \doi{10.1093/mnras/stx483}, \eprint{1607.00017}

\bibitem[{{Switzer} et~al(2013){Switzer}, {Masui}, {Bandura}, {Calin}, {Chang},
  {Chen}, {Li}, {Liao}, {Natarajan}, {Pen}, {Peterson}, {Shaw}, and
  {Voytek}}]{Switzer13}
{Switzer} ER, {Masui} KW, {Bandura} K, {Calin} LM, {Chang} TC, {Chen} XL, {Li}
  YC, {Liao} YW, {Natarajan} A, {Pen} UL, {Peterson} JB, {Shaw} JR, {Voytek} TC
  (2013) {Determination of z {$\sim$} 0.8 neutral hydrogen fluctuations using
  the 21 cm intensity mapping autocorrelation}. \mnras 434:L46--L50,
  \doi{10.1093/mnrasl/slt074}, \eprint{1304.3712}

\bibitem[{{Tagore} and {Keeton}(2014)}]{TagoreKeeton14}
{Tagore} AS, {Keeton} CR (2014) {Statistical and systematic uncertainties in
  pixel-based source reconstruction algorithms for gravitational lensing}.
  \mnras 445:694--710, \doi{10.1093/mnras/stu1671}, \eprint{1408.6297}

\bibitem[{{Takada} et~al(2014){Takada}, {Ellis}, {Chiba}, {Greene}, {Aihara},
  {Arimoto}, {Bundy}, {Cohen}, {Dor{\'e}}, {Graves}, {Gunn}, {Heckman},
  {Hirata}, {Ho}, {Kneib}, {F{\`e}vre}, {Lin}, {More}, {Murayama}, {Nagao},
  {Ouchi}, {Seiffert}, {Silverman}, {Sodr{\'e}}, {Spergel}, {Strauss}, {Sugai},
  {Suto}, {Takami}, and {Wyse}}]{Takada:2014}
{Takada} M, {Ellis} RS, {Chiba} M, {Greene} JE, {Aihara} H, {Arimoto} N,
  {Bundy} K, {Cohen} J, {Dor{\'e}} O, {Graves} G, {Gunn} JE, {Heckman} T,
  {Hirata} CM, {Ho} P, {Kneib} JP, {F{\`e}vre} OL, {Lin} L, {More} S,
  {Murayama} H, {Nagao} T, {Ouchi} M, {Seiffert} M, {Silverman} JD, {Sodr{\'e}}
  L, {Spergel} DN, {Strauss} MA, {Sugai} H, {Suto} Y, {Takami} H, {Wyse} R
  (2014) {Extragalactic science, cosmology, and Galactic archaeology with the
  Subaru Prime Focus Spectrograph}. \pasj 66:R1, \doi{10.1093/pasj/pst019},
  \eprint{1206.0737}

\bibitem[{{Tegmark} and {Zaldarriaga}(2009)}]{2009PhRvD..79h3530T}
{Tegmark} M, {Zaldarriaga} M (2009) {Fast Fourier transform telescope}. \prd
  79(8):083530, \doi{10.1103/PhysRevD.79.083530}, \eprint{0805.4414}

\bibitem[{{Tegmark} and {Zaldarriaga}(2010)}]{2010PhRvD..82j3501T}
{Tegmark} M, {Zaldarriaga} M (2010) {Omniscopes: Large area telescope arrays
  with only NlogN computational cost}. \prd 82(10):103501,
  \doi{10.1103/PhysRevD.82.103501}, \eprint{0909.0001}

\bibitem[{{Tewes} et~al(2013{\natexlab{a}}){Tewes}, {Courbin}, and
  {Meylan}}]{Tewes2013}
{Tewes} M, {Courbin} F, {Meylan} G (2013{\natexlab{a}}) {COSMOGRAIL: the
  COSmological MOnitoring of GRAvItational Lenses. XI. Techniques for time
  delay measurement in presence of microlensing}. \aap 553:A120,
  \doi{10.1051/0004-6361/201220123}, \eprint{1208.5598}

\bibitem[{{Tewes} et~al(2013{\natexlab{b}}){Tewes}, {Courbin}, {Meylan},
  {Kochanek}, {Eulaers}, {Cantale}, {Mosquera}, {Magain}, {Van Winckel},
  {Sluse}, {Cataldi}, {V{\"o}r{\"o}s}, and {Dye}}]{Tewes2013b}
{Tewes} M, {Courbin} F, {Meylan} G, {Kochanek} CS, {Eulaers} E, {Cantale} N,
  {Mosquera} AM, {Magain} P, {Van Winckel} H, {Sluse} D, {Cataldi} G,
  {V{\"o}r{\"o}s} D, {Dye} S (2013{\natexlab{b}}) {COSMOGRAIL: the COSmological
  MOnitoring of GRAvItational Lenses. XIII. Time delays and 9-yr optical
  monitoring of the lensed quasar RX J1131-1231}. \aap 556:A22,
  \doi{10.1051/0004-6361/201220352}, \eprint{1208.6009}

\bibitem[{{Tie} and {Kochanek}(2018)}]{TK18}
{Tie} SS, {Kochanek} CS (2018) {Microlensing makes lensed quasar time delays
  significantly time variable}. \mnras 473:80--90, \doi{10.1093/mnras/stx2348},
  \eprint{1707.01908}

\bibitem[{{Tihhonova} et~al(2017){Tihhonova}, {Courbin}, {Harvey}, {Hilbert},
  {Rusu}, {Fassnacht}, {Bonvin}, {Marshall}, {Meylan}, {Sluse}, {Suyu}, {Treu},
  and {Wong}}]{TihhonovaEtal17}
{Tihhonova} O, {Courbin} F, {Harvey} D, {Hilbert} S, {Rusu} CE, {Fassnacht} CD,
  {Bonvin} V, {Marshall} PJ, {Meylan} G, {Sluse} D, {Suyu} SH, {Treu} T, {Wong}
  KC (2017) {H0LiCOW VIII. A weak lensing measurement of the external
  convergence in the field of the lensed quasar HE$\,$0435$-$1223}. ArXiv
  e-prints \eprint{1711.08804}

\bibitem[{{Treu} et~al(2016){Treu}, {Brammer}, {Diego}, {Grillo}, {Kelly},
  {Oguri}, {Rodney}, {Rosati}, {Sharon}, {Zitrin}, {Balestra}, {Brada{\v c}},
  {Broadhurst}, {Caminha}, {Halkola}, {Hoag}, {Ishigaki}, {Johnson}, {Karman},
  {Kawamata}, {Mercurio}, {Schmidt}, {Strolger}, {Suyu}, {Filippenko}, {Foley},
  {Jha}, and {Patel}}]{TreuEtal16}
{Treu} T, {Brammer} G, {Diego} JM, {Grillo} C, {Kelly} PL, {Oguri} M, {Rodney}
  SA, {Rosati} P, {Sharon} K, {Zitrin} A, {Balestra} I, {Brada{\v c}} M,
  {Broadhurst} T, {Caminha} GB, {Halkola} A, {Hoag} A, {Ishigaki} M, {Johnson}
  TL, {Karman} W, {Kawamata} R, {Mercurio} A, {Schmidt} KB, {Strolger} LG,
  {Suyu} SH, {Filippenko} AV, {Foley} RJ, {Jha} SW, {Patel} B (2016)
  {''Refsdal'' Meets Popper: Comparing Predictions of the Re-appearance of the
  Multiply Imaged Supernova Behind MACSJ1149.5+2223}. \apj 817:60,
  \doi{10.3847/0004-637X/817/1/60}, \eprint{1510.05750}

\bibitem[{{Vanderriest} et~al(1989){Vanderriest}, {Schneider}, {Herpe},
  {Chevreton}, {Moles}, and {Wlerick}}]{Vanderriest1989}
{Vanderriest} C, {Schneider} J, {Herpe} G, {Chevreton} M, {Moles} M, {Wlerick}
  G (1989) {The value of the time delay Delta t(A, B) for the 'double' quasar
  0957+561 from optical photometric monitoring}. \aap 215:1--13

\bibitem[{{Vegetti} and {Koopmans}(2009)}]{VegettiKoopmans09}
{Vegetti} S, {Koopmans} LVE (2009) {Bayesian strong gravitational-lens
  modelling on adaptive grids: objective detection of mass substructure in
  Galaxies}. \mnras 392:945--963, \doi{10.1111/j.1365-2966.2008.14005.x},
  \eprint{0805.0201}

\bibitem[{{Vuissoz} et~al(2007){Vuissoz}, {Courbin}, {Sluse}, {Meylan},
  {Ibrahimov}, {Asfandiyarov}, {Stoops}, {Eigenbrod}, {Le Guillou}, {van
  Winckel}, and {Magain}}]{Vuissoz2007}
{Vuissoz} C, {Courbin} F, {Sluse} D, {Meylan} G, {Ibrahimov} M, {Asfandiyarov}
  I, {Stoops} E, {Eigenbrod} A, {Le Guillou} L, {van Winckel} H, {Magain} P
  (2007) {COSMOGRAIL: the COSmological MOnitoring of GRAvItational Lenses. V.
  The time delay in SDSS J1650+4251}. \aap 464:845--851,
  \doi{10.1051/0004-6361:20065823}, \eprint{astro-ph/0606317}

\bibitem[{{Vuissoz} et~al(2008){Vuissoz}, {Courbin}, {Sluse}, {Meylan},
  {Chantry}, {Eulaers}, {Morgan}, {Eyler}, {Kochanek}, {Coles}, {Saha},
  {Magain}, and {Falco}}]{Vuissoz2008}
{Vuissoz} C, {Courbin} F, {Sluse} D, {Meylan} G, {Chantry} V, {Eulaers} E,
  {Morgan} C, {Eyler} ME, {Kochanek} CS, {Coles} J, {Saha} P, {Magain} P,
  {Falco} EE (2008) {COSMOGRAIL: the COSmological MOnitoring of GRAvItational
  Lenses. VII. Time delays and the Hubble constant from WFI J2033-4723}. \aap
  488:481--490, \doi{10.1051/0004-6361:200809866}, \eprint{0803.4015}

\bibitem[{{Wallington} et~al(1996){Wallington}, {Kochanek}, and
  {Narayan}}]{WallingtonEtal96}
{Wallington} S, {Kochanek} CS, {Narayan} R (1996) {LensMEM: A Gravitational
  Lens Inversion Algorithm Using the Maximum Entropy Method}. \apj 465:64,
  \doi{10.1086/177401}

\bibitem[{{Walsh} et~al(1979){Walsh}, {Carswell}, and {Weymann}}]{Walsh1979}
{Walsh} D, {Carswell} RF, {Weymann} RJ (1979) {0957 + 561 A, B - Twin
  quasistellar objects or gravitational lens}. \nat 279:381--384,
  \doi{10.1038/279381a0}

\bibitem[{{Warren} and {Dye}(2003)}]{WarrenDye03}
{Warren} SJ, {Dye} S (2003) {Semilinear Gravitational Lens Inversion}. \apj
  590:673--682, \doi{10.1086/375132}

\bibitem[{{Wayth} and {Webster}(2006)}]{WaythWebster06}
{Wayth} RB, {Webster} RL (2006) {LENSVIEW: software for modelling resolved
  gravitational lens images}. \mnras 372:1187--1207,
  \doi{10.1111/j.1365-2966.2006.10922.x}, \eprint{arXiv:astro-ph/0609542}

\bibitem[{{Weinberg} et~al(2013){Weinberg}, {Mortonson}, {Eisenstein},
  {Hirata}, {Riess}, and {Rozo}}]{Weinberg:2013}
{Weinberg} DH, {Mortonson} MJ, {Eisenstein} DJ, {Hirata} C, {Riess} AG, {Rozo}
  E (2013) {Observational probes of cosmic acceleration}. \physrep 530:87--255,
  \doi{10.1016/j.physrep.2013.05.001}, \eprint{1201.2434}

\bibitem[{{Williams} and {Saha}(2000)}]{WilliamsSaha00}
{Williams} LLR, {Saha} P (2000) {Pixelated Lenses and $H_{0}$ from Time-Delay
  Quasars}. \aj 119:439--450, \doi{10.1086/301234}

\bibitem[{{Wong} et~al(2017){Wong}, {Suyu}, {Auger}, {Bonvin}, {Courbin},
  {Fassnacht}, {Halkola}, {Rusu}, {Sluse}, {Sonnenfeld}, {Treu}, {Collett},
  {Hilbert}, {Koopmans}, {Marshall}, and {Rumbaugh}}]{WongEtal17}
{Wong} KC, {Suyu} SH, {Auger} MW, {Bonvin} V, {Courbin} F, {Fassnacht} CD,
  {Halkola} A, {Rusu} CE, {Sluse} D, {Sonnenfeld} A, {Treu} T, {Collett} TE,
  {Hilbert} S, {Koopmans} LVE, {Marshall} PJ, {Rumbaugh} N (2017) {H0LiCOW -
  IV. Lens mass model of HE 0435-1223 and blind measurement of its time-delay
  distance for cosmology}. \mnras 465:4895--4913, \doi{10.1093/mnras/stw3077},
  \eprint{1607.01403}

\bibitem[{{Wucknitz}(2002)}]{Wucknitz02}
{Wucknitz} O (2002) {Degeneracies and scaling relations in general power-law
  models for gravitational lenses}. \mnras 332:951--961,
  \doi{10.1046/j.1365-8711.2002.05426.x}, \eprint{arXiv:astro-ph/0202376}

\bibitem[{{Wyithe} et~al(2007){Wyithe}, {Loeb}, and
  {Geil}}]{2007arXiv0709.2955W}
{Wyithe} S, {Loeb} A, {Geil} P (2007) {Baryonic Acoustic Oscillations in 21cm
  Emission: A Probe of Dark Energy out to High Redshifts}. ArXiv e-prints
  \eprint{0709.2955}

\bibitem[{Wyithe et~al(2008)Wyithe, Loeb, and Geil}]{WL08}
Wyithe S, Loeb A, Geil P (2008) Baryonic acoustic oscillations in 21cm
  emission: A probe of dark energy out to high redshifts. \mnras
  383:1195--1209, \eprint{0709.2955}

\bibitem[{Xu et~al(2015)Xu, Wang, and Chen}]{Xu:2014bya}
Xu Y, Wang X, Chen X (2015) {Forecasts on the Dark Energy and Primordial
  Non-Gaussianity Observations with the Tianlai Cylinder Array}. Astrophys J
  798(1):40, \doi{10.1088/0004-637X/798/1/40}, \eprint{1410.7794}

\bibitem[{{Yahalomi} et~al(2017){Yahalomi}, {Schechter}, and
  {Wambsganss}}]{YahalomiEtal17}
{Yahalomi} DA, {Schechter} PL, {Wambsganss} J (2017) {A Quadruply Lensed SN Ia:
  Gaining a Time-Delay...Losing a Standard Candle}. ArXiv e-prints
  \eprint{1711.07919}

\bibitem[{{Young} et~al(1981){Young}, {Deverill}, {Gunn}, {Westphal}, and
  {Kristian}}]{Young1981}
{Young} P, {Deverill} RS, {Gunn} JE, {Westphal} JA, {Kristian} J (1981) {The
  triple quasar Q1115+080A, B, C - A quintuple gravitational lens image}. \apj
  244:723--735, \doi{10.1086/158750}

\bibitem[{{Zwaan} et~al(2001){Zwaan}, {van Dokkum}, and
  {Verheijen}}]{Zwaan2001}
{Zwaan} MA, {van Dokkum} PG, {Verheijen} MAW (2001) {Hydrogen 21-Centimeter
  Emission from a Galaxy at Cosmological Distance}. Science 293:1800--1803,
  \doi{10.1126/science.1063034}, \eprint{arXiv:astro-ph/0109108}

\end{thebibliography}
\bibliographystyle{spbasic}

\end{document}